\documentclass[fleqn,usenatbib]{mnras}

\usepackage{newtxtext,newtxmath}

\usepackage[T1]{fontenc}
\usepackage{soul}

\def\lum{erg~s$^{-1}$}
\def\xa{XRT~161028}
\def\xb{XRT~140811}
\def\xc{XRT~030206}
\def\xd{XRT~151219}
\def\xe{XRT~151128}
\def\xf{XRT~160220}
\def\xg{XRT~110621}
\def\msun{M_\odot}

\def\arcsec{\hbox{$^{\prime\prime}$}}
\DeclareUnicodeCharacter{2009}{\,} 
\def\approxlt{\ifmmode \rlap{$<$}{}_{{}_{{}_{\textstyle\sim}}} \else%
$\rlap{$<$}{}_{{}_{{}_{\textstyle\sim}}}$\fi}

\DeclareRobustCommand{\VAN}[3]{#2}
\let\VANthebibliography\thebibliography
\def\thebibliography{\DeclareRobustCommand{\VAN}[3]{##3}\VANthebibliography}

\usepackage{graphicx}	
\usepackage{amsmath}
\usepackage{orcidlink}

\title[XMM-Newton FXTs]{XMM-{\it Newton}-discovered Fast X-ray Transients: Host galaxies and limits on contemporaneous detections of optical counterparts}

\author[]{
D. Eappachen\,\orcidlink{0000-0001-7841-0294}$^{1,2}$\thanks{E-mail: d.eappachen@sron.nl},
P.~G.~Jonker\,\orcidlink{0000-0001-5679-0695 }$^{2,1}$ 
J. Quirola-V\'asquez\,\orcidlink{0000-0001-8602-4641}$^{3,4,5}$,
D. Mata Sánchez\,\orcidlink{0000-0003-0245-9424}$^{6,7}$,
 A. Inkenhaag\,\orcidlink{0000-0001-5058-8111}$^{2,1}$,
\newauthor  A.J.~Levan$^{2}$,
M. Fraser\,\orcidlink{0000-0003-2191-1674}$^{8}$,
M.A.P. Torres\,\orcidlink{https://orcid.org/0000-0002-5297-2683}$^{6,7}$,
F.E.~Bauer\orcidlink{0000-0002-8686-8737}$^{3,4,9}$,
A. A. Chrimes\,\orcidlink{0000-0001-9842-6808}$^{2}$,
D. Stern \orcidlink{0000-0003-2686-9241}$^{10}$,
\newauthor M. J. Graham\orcidlink{0000-0002-3168-0139}$^{11}$,
S. J.~Smartt\orcidlink{0000-0002-8229-1731}$^{12}$,
K. W. Smith$^{12}$, 
M.E.~Ravasio\orcidlink{0000-0003-3193-4714}$^{2}$, 
A. I. Zabludoff\,\orcidlink{0000-0001-6047-8469}$^{13}$,
M. Yue\,$^{13,14}$\orcidlink{0000-0002-5367-8021},
\newauthor F. Stoppa\,\orcidlink{0000-0002-3424-8528}$^{2,15}$,
D. B. Malesani\orcidlink{0000-0002-7517-326X}$^{16,17}$,
N. C. Stone\,\orcidlink{0000-0002-4337-9458}$^{18}$,
S. Wen\,\orcidlink{0000-0002-0934-2686}$^{2}$\\
$^{1}$SRON, Netherlands Institute for Space Research, Niels Bohrweg 4, 2333 CA, Leiden, The Netherlands\\
$^{2}$Department of Astrophysics/IMAPP, Radboud University Nijmegen, P.O. Box 9010, 6500 GL, Nijmegen, The Netherlands\\
$^{3}$Instituto de Astrof\'isica, Pontificia Universidad Cat\'olica de Chile, Casilla 306, Santiago 22, Chile\\
$^{4}$Millennium Institute of Astrophysics (MAS), Nuncio Monse$\tilde{n}$or S\'otero Sanz 100, Providencia, Santiago, Chile\\
$^{5}$Observatorio Astron\'omico de Quito, Escuela Polit\'ecnica Nacional, 170136, Quito, Ecuador\\
$^{6}$Instituto de Astrof\'{i}sica de Canarias, E-38205 La Laguna, S/C de Tenerife, Spain\\
$^{7}$ Departamento de Astrof\'isica, Univ. de La Laguna, E-38206 La Laguna, Tenerife, Spain \\
$^{8}$School of Physics, O’Brien Centre for Science North, University College Dublin, Belfield, Dublin 4, Ireland\\
$^{9}$Space Science Institute, 4750 Walnut Street, Suite 205, Boulder, Colorado 80301, USA\\
$^{10}$Jet Propulsion Laboratory, California Institute of Technology, 4800 Oak Grove Drive, MS 169-224, Pasadena, CA 91109, USA \\
$^{11}$Division of Physics, Mathematics and Astronomy, California Institute of Technology, Pasadena, CA 91125, USA\\
$^{12}$Astrophysics Research Centre, School of Mathematics and Physics, Queen’s University Belfast, Belfast, UK\\
$^{13}$Steward Observatory, University of Arizona, 933 North Cherry Avenue, Tucson, AZ 85721, USA \\
$^{14}$MIT Kavli Institute for Astrophysics and Space Research, 77 Massachusetts Ave., Cambridge, MA 02139, USA\\
$^{15}$Department of Mathematics/IMAPP, Radboud University, P.O. Box 9010, NL-6500 GL Nĳmegen, The Netherlands \\
$^{16}$Cosmic Dawn Center (DAWN), Denmark \\
$^{17}$Niels Bohr Institute, University of Copenhagen, Jagtvej 128, 2200 Copenhagen N, Denmark \\
$^{18}$Racah Institute of Physics, The Hebrew University, Jerusalem, 91904, Israel\\
}

\date{Accepted XXX. Received YYY; in original form ZZZ}

\pubyear{2023}

\begin{document}
\label{firstpage}
\pagerange{\pageref{firstpage}--\pageref{lastpage}}
\maketitle

\begin{abstract}
Extragalactic fast X-ray transients (FXTs) are a 
class of soft (0.3-10 keV) X-ray transients lasting a few hundred seconds to several hours. Several progenitor mechanisms have been suggested to produce FXTs, including  supernova shock breakouts, binary neutron star mergers, or tidal disruptions involving an intermediate-mass black hole and a white dwarf. We present detailed host studies, including spectroscopic observations of the host galaxies of 7 {\it XMM-Newton}-discovered FXTs. 
The candidate hosts lie at redshifts 0.0928 $< z <$ 0.645 implying peak X-ray luminosities of 10$^{43}$ \lum $< L_X <$ 10$^{45}$ \lum\,and physical offsets of 1 kpc < $r_\mathrm{proj}$ < 22 kpc. These observations increase the number of FXTs with a spectroscopic redshift measurement by a factor of 2, although we note that one event is re-identified as a Galactic flare star. We infer host star formation rates and stellar masses by fitting the combined spectroscopic and archival photometric data. We also report on a contemporaneous optical counterpart search to the FXTs in Pan-STARRS and ATLAS by performing forced photometry at the position of the FXTs. We do not find any counterpart in our search. Given our constraints, including peak X-ray luminosities, optical limits, and host properties, we find that \xg{} is consistent with a SN SBO event. Spectroscopic redshifts of likely host galaxies for four events imply peak X-ray luminosities that are too high to be
consistent with SN SBOs, but we are unable to discard either the BNS or WD-IMBH TDE scenarios for these FXTs.
\end{abstract}




\section{Introduction}
\label{sec:intro}
Fast X-ray Transients (FXTs) are singular bursts of soft X-rays that last from minutes to hours, whose origin is still unknown. The majority (22 out of the 36) of the known FXTs have been discovered in  {\it Chandra} archival data (see \citealt{2013jonker}; \citealt{glennie}; \citealt{Irwin}; \citealt{Bauer}; \citealt{Xue2019};  \citealt{2021ATel14599....1L}; \citealt{2022A&A...663A.168Q}; \citealt{2022ApJ...927..211L}; \citealt{2023arXiv230301857E}; \citealt{Quirola2023}), while
12 have been found in archival {\it XMM-Newton} data (\citealt{Alp2020}).\footnote{Note that out of the 12 {\it XMM-Newton}-discovered FXTs, three are suspected to be due to flares by late-type stars (\citealt{Alp2020}).} 
In addition,
one FXT was discovered in eROSITA data \citep{2020ATel13416....1W} and one in {\it Swift}/XRT data (XRT~080109/SN2008D; \citealt{2008Natur.453..469S}). Earlier observatories are presumed to have detected some FXTs, but lacked sufficient spatial resolution to pinpoint them as extragalactic phenomena.

Among the 12 {\it XMM-Newton}-discovered FXTs, nine have candidate host galaxies, while for three no host has been identified to date. \citet{Alp2020} estimate the redshifts of those host galaxies to be between 0.1 and 1. Based on relatively limited photometric and spectroscopic information, they suggest that
most of these FXTs are due to a supernova shock breakout (SN SBO). Specifically, these SN SBOs are consistent with the expectations for blue supergiant core-collapse supernovae (CC-SNe).

An SBO is predicted to be the first electromagnetic emission from a supernova explosion (\citealt{2017hsn..book..967W}). This happens when the radiation-mediated shock crosses the surface of the star, thereby producing a bright X-ray/UV flash on a timescale of seconds to a fraction of an hour \citep{Falk1977,Klein1978,Matzner1999,Schawinski2008,Ganot2016,2017hsn..book..967W}. This is followed  by months-long SN emission, which is predominantly powered by radioactive decay and detectable at UV/optical/near-infrared (IR) wavelengths. The SBO emission carries information about the progenitor star, potentially enabling the inference of physical properties of the progenitor such as the stellar radius and mass-loss history. If the circumstellar medium (CSM) is dense enough, the breakout will occur at larger radii within the CSM, resulting in a longer-duration transient event.

The only confirmed SN SBO observed to date is XRT~080109, associated with SN~2008D, serendipitously discovered in {\it Swift} observations of SN~2007uy (\citealt{2008Natur.453..469S}). XRT~080109 was detected in one of the arms of the spiral galaxy NGC~2770 at a distance of 27 Mpc. The X-ray light curve is characterised  by a fast rise followed by an exponential decay with a peak X-ray luminosity of $\sim$ 6$\ \times \ 10^{43}$~\lum. Based on the multi-wavelength dataset, \citet{2008Natur.453..469S} shows that the observed properties are consistent with those of Wolf-Rayet stars ($R_\ast$ $\sim$ 10$^{11}$cm ), 
the favoured progenitors for type Ibc supernovae. 
More luminous SBOs have also been posited as an origin for some of the low luminosity long duration gamma-ray bursts (LGRBs) seen in the local Universe  (e.g., GRB 060218; \citealt{2006Natur.442.1008C}). 

Several other physical mechanisms have been suggested as the origin of FXTs; specifically a white dwarf (WD) - intermediate-mass black hole (IMBH) tidal disruption event (TDE)  and a binary neutron star (BNS) merger. In general, the X-ray light curves of TDEs involving a main sequence star and a supermassive black hole have a time scale ranging from months to years
(\citealt{2020SSRv..216...85S}). Instead, the relatively low mass of an IMBH compared to that of a supermassive black hole and the compactness of a WD compared to that of a main-sequence star implies that the time scale associated with the X-ray light curve of IMBH-WD TDEs is much shorter than that of an SMBH -- main-sequence TDE (\citealt{2009ApJ...695..404R}; \citealt{2020SSRv..216...39M}). A WD TDE remains a viable interpretation of at least some FXTs detected so far (e.g., XRT~000519; \citealt{2013jonker}). In order to be consistent with the observed volumetric rates of FXTs \citep[as reported by][]{Quirola2023}, one needs to assume that those IMBH-WD TDEs producing the X-ray emission are capable of launching jetted outflows. A beamed TDE emission would also naturally account for fairly high X-ray luminosities (\citealt{2016ApJ...819....3M}), similar to the ones observed in some FXTs. Indeed, there are also apparent similarities between these events and the sub-set of ultra-long duration gamma-ray bursts (\citealt{2014ApJ...781...13L}). 

A BNS merger is another possible progenitor for FXTs. Some BNS mergers could leave behind supramassive or even stable magnetars, which are expected to produce FXTs (\citealt{2006Sci...311.1127D}; \citealt{2008MNRAS.385.1455M}; \citealt{2013ApJ...763L..22Z}; \citealt{2014MNRAS.439.3916M}; \citealt{2017ApJ...835....7S}). For some FXTs like CDF-S~XT2 (\citealt{Xue2019}) and XRT~210423 (\citealt{2023arXiv230301857E}), the X-ray light curve shows a plateau and decay, similar to those seen in short gamma-ray bursts (SGRBs). This has been used as an argument in favour of a BNS origin. Since GW170817 it is known that SGRBs are caused by BNS mergers (\citealt{2017ApJ...848L..17C}). Converting the observed SGRB rate to an intrinsic rate that can be compared with the (nearly isotropic) FXT rate, a large beaming correction factor is required to reconcile the current observed FXTs volumetric density rate with that of SGRBs, providing potentially strong constraints on the opening angle of SGRB jets \citep{Quirola2023}.

While transient counterparts have been hard to find for FXTs (\citealt{Bauer}), we still expect progenitors of FXTs to be associated with a host. Host studies provide vital information to decipher the origin of the FXTs. For instance, finding the distance to the host galaxy helps constrain the energetics of the FXT (e.g., \citealt{2022MNRAS.514..302E}; \citealt{2023arXiv230301857E}). Understanding host environments can also play an important role in constraining the progenitor mechanism.

Host studies have been undertaken for a few of the {\it Chandra} FXTs. \citet{Bauer} finds a  dwarf galaxy host ($m_{r}$=27.5)  at $\sim$ z$_{\rm phot}$=0.4-3.2 that lies 0.13$^{\prime\prime}$ offset from the center of the X-ray uncertainty position of CDF-S XT1. \citet{Xue2019} reports that CDF-XT2 (\citealt{Xue2019}) lies at a projected distance of 3.3 $\pm$ 1.9 kpc from a $z_{\rm spec}$=0.74 star-forming galaxy. Deep Very Large Telescope (VLT) and Gran Telescopio Canarias (GTC) observations of XRT~000519 (\citealt{2013jonker})  reveal a candidate host galaxy to the northwest of the FXT position in the $g$-band image at $m_{g}$=26.29$\pm$0.09 mag (\citealt{2022MNRAS.514..302E}) which, if the host galaxy -- FXT association is real, rules out an SN SBO origin for this FXT. For XRT~170901, an elongated, possibly late-type, star-forming galaxy is reported by \citet{2022ApJ...927..211L} to fall within the 2$\sigma$ X-ray position. No clear host galaxy was found for XRT~030511 (\citealt{2022ApJ...927..211L}). For XRT~210423, \citet{2023arXiv230301857E} report a candidate host that lies within the 3$\sigma$ uncertainty region with a magnitude of 25.9$\pm$0.1 in the GTC/HiPERCAM $g_s$-filter. In addition, they reported two other candidate host galaxies; one with $z_{\rm spec}=1.5082 \pm 0.0001$ and offset 4.2$\pm$1\arcsec{} (37$\pm$9 kpc) from the FXT position and another one with $z_{\rm phot}=1.04^{+0.22}_{-0.14}$, offset by 3.6$\pm$1\arcsec{} (30$\pm$8 kpc). Irrespective of which of these three candidate host galaxies is the real host galaxy, if one of them is the host, the properties are consistent with a BNS merger origin for XRT~210423 (\citealt{2023arXiv230301857E}).

In this paper, we look in detail at the host properties of 7 of the {\it XMM-Newton}-discovered FXTs reported by \citet{Alp2020}, namely \xa{}, \xb{}, \xc{}, \xd{}, \xe{}, \xf{} and \xg{}. We also search for contemporaneous optical counterparts for those FXTs where such observations exist in either Pan-STARRS (\citealt{2016arXiv161205560C}), DECam (\citealt{2016MNRAS.460.1270D}) or ATLAS (\citealt{2018PASP..130f4505T}). 

Throughout the paper, we assume a flat $\Lambda$-CDM cosmology, with Hubble constant  H$_\mathrm{0}$= 67.4 $\pm$ 0.5 km s$^{-1}$Mpc$^{-1}$ and matter density parameter $\Omega_m$=0.315 $\pm$  0.007 (\citealt{2018arXiv180706209P}). Throughout, magnitudes are quoted in the AB system. We use the term "optical counterpart" for optical transient light associated with the FXT, and "candidate host" or "candidate host galaxy" for the host galaxy of the FXT. The uncertainties mentioned in the paper are at the 1$\sigma$ confidence level, unless mentioned otherwise.

\section{OBSERVATIONS AND ANALYSIS}
In order to determine the redshift of the candidate host galaxy, and thus distance to the FXT and the host properties, we obtained spectroscopic observations of candidate host galaxies to a subset of the FXTs reported in 
\citet{Alp2020}. A journal of the spectroscopic observations is given in Table~\ref{tab:spec}.

\begin{table*}
\small
\begin{center}
\caption{A journal of the spectroscopic observations of the candidate host of the {\it XMM-Newton}-discovered FXTs used in this paper. }
\label{tab:spec}
\begin{tabular}{llllllll}
\hline
Target &  Telescope/Instrument  & Date (UT) & Observations & {Grism} & Exposure time~[s] & Airmass & Seeing (arscec) \\
		\hline
		\xa &  GTC/OSIRIS &  2022 Sept 19 & 3 & R500R  & 600 & 1.1 & $\sim 0.7$\\
		\xb &  GTC/OSIRIS &  2022 Sept 19 & 3  & R500R  & 600 & 1.3 & $\sim 0.8$\\
		\xc &  GTC/OSIRIS &   2022 Sept 19 & 1 & R500R & 2650 & 1.3 & $\sim 0.9$\\
  \xd &  Keck/LRIS &   2022 Nov 25 & 2 & Blue/600 \& Red/400& 1200 & 1.4 & $\sim 0.9$\\
  \xe &  Keck/LRIS &   2022 Nov 25 & 2 & Blue/600 \& Red/400 & 1200 & 1.3 & $\sim 1.0$\\
  \xf{} & Magellan/LDSS3 & 2021 Aug 02 & 3 & VPH-All & 900& 1.3 & $\sim 1.2$ \\
  \xg{} & Magellan/LDSS3 & 2021 Aug 02 & 3 & VPH-All & 900 & 1.2 & $\sim 1.2$ \\
\hline
\end{tabular}
\end{center}
\end{table*}

\subsection{Astrometry}

We reviewed the FXTs reported by \citet{Alp2020} in the {\it XMM-Newton} serendipitous catalog (\citealt{2020A&A...641A.136W}). We have updated the Right Ascension (R.A.)~and Declination (Dec.) and the associated uncertainty region for the FXTs discussed in the paper, see Table \ref{tab:XMM_sere}. In the {\it XMM-Newton} serendipitous catalog X-ray detection astrometry is improved by cross-correlating X-ray detections with optical or infrared catalogues like USNO B1.0, 2MASS, or SDSS (DR8). For the FXT, \xg{} we use the coordinates and X-ray uncertainty reported by \citet{Novara2020} in which they cross-match the brightest sources detected in the XMM–Newton observation with the USNO B1 optical catalog (\citealt{2003AJ....125..984M}).

\begin{table*}
\small
\begin{center}
\caption{The updated the R.A. and Dec. and the associated X-ray uncertainty region of the FXTs obtained from the {\it XMM-Newton} serendipitous catalog.}

\label{tab:XMM_sere}
\begin{tabular}{cccc}
\hline
\multicolumn{1}{|p{1cm}|}{\centering FXT}
&\multicolumn{1}{|p{2.5cm}|}{\centering R.A. J2000 \\ ($^{\circ}$)}
&\multicolumn{1}{|p{2.5cm}|}{\centering Dec J2000  \\($^{\circ}$)}
&\multicolumn{1}{|p{3.5cm}|}{\centering X-ray uncertainty (1$\sigma$) \\ (${\arcsec{}}$)}\\
\hline
\xa & 263.23645 & 43.51225 &  1.1 \\ %
		\xb & 43.65351 & 41.07393 &   1 \\
            \xc & 29.28802 & 37.62769 &  0.8  \\
          \xd & 173.53129 & 0.87324 & 1.3  \\
            \xe & 167.07858 & -5.07474& 1.8 \\
            \xf{} & 204.19900 & -41.33699  & 1.6 \\
           $^{\ast}$\xg{} & 37.89542  & -60.62869 & 1.9 \\

\hline
\end{tabular}
\end{center}
\begin{flushright}
$^{\ast}$ From \citet{Novara2020}
\end{flushright}
\end{table*}

\subsection{Optical spectroscopic observations}
\subsubsection{GTC OSIRIS}
We obtained spectra of candidate host galaxies for the FXTs XRT~161028, XRT~140811 and XRT~030206 using the Optical System for Imaging and low-Intermediate-Resolution Integrated Spectroscopy (OSIRIS;  \citealt{2000SPIE.4008..623C}) instrument mounted on the 10.4--m {Gran Telescopio Canarias} (GTC). We used the R500R grism and one arcsec-wide slit (4800--10000 \AA{}  with a moderate resolving power of R $\approx$ 350 at $\lambda_c$=7165 \AA ). The spectrum was recorded on a mosaic of two Marconi CCDs. We undertook bias subtraction, flat field correction and spectral extraction using \textsc{PyRAF}  (\citealt{2012ascl.soft07011S}) and  wavelength (arc lamps Xe, Ne and HgAr) and flux calibration using {\sc molly}  (\citealt{2019ascl.soft07012M}). 

\subsubsection{Keck LRIS}
We observed \xd{} and \xe{} on UT 2022 November 25 using the Low-Resolution Imaging Spectrometer (LRIS; \citealt{1995PASP..107..375O}) on the Keck I telescope. For both sources, we obtained two 1200~s exposures using the 600 $\ell$ 
mm$^{-1}$ blue grism ($\lambda_{\rm blaze} =$
4000 \AA), the 400 $\ell$ mm$^{-1}$ red grating ($\lambda_{\rm blaze} =$ 8500 \AA), the 5600 \AA\, dichroic, and the 1 arcsec slit.  This instrument configuration covers the full optical window at moderate resolving power, $R \equiv \lambda / \Delta \lambda \approx 1500$ for objects filling the slit.  Because this was a half-night allocation with only a single standard observation, we used archival observations of standard stars from \citet{1990ApJ...358..344M} observed in January 2020 with the same instrument configuration for flux calibration. We reduced the spectra using standard techniques with IRAF (\citealt{1986SPIE..627..733T}).

\subsubsection{Magellan LDSS3 }
The Low Dispersion Survey Spectrograph (LDSS-3) is a spectrograph mounted on the 6.5--m Magellan telescope at the Las Campanas Observatory in Chile. We obtained spectra of the host candidates for XRT~160220 and XRT~110621 using LDSS-3. We used the VPH-All grism  and a 1 arcsec slit width (covering 4250--10000 \AA{} with a resolving power of  R $\approx$ 1150 at $\lambda_c$=7100 \AA) for our observations. We used \textsc{PyRAF} scripts (\citealt{2012ascl.soft07011S}) for cleaning and reducing the data and {\sc molly} (\citealt{2019ascl.soft07012M}) for wavelength (using the He, Ne, and Ar arc lamps) and flux calibration.

\subsection{Optical photometric observations}
In addition to our targeted observations to obtain galaxy redshifts, we also make use of archival imaging surveys to provide photometric measurements of the host galaxies, and for surveys where the temporal coverage matches the FXT event, we extract forced photometry around the time of the FXT detection. Optical photometry of seven candidate host galaxies is given in Table~\ref{tab:phot}.

\begin{table*}
\small
\begin{center}
\caption{Host photometry and position of the candidate FXT host galaxies}

\label{tab:phot}
\begin{tabular}{ccccccccc}
\hline
\multicolumn{1}{|p{1cm}|}{\centering Host galaxy of}
&\multicolumn{1}{|p{1.5cm}|}{\centering R.A.~and Dec \\J2000  ($^{\circ}$)}
&\multicolumn{1}{|p{1.5cm}|}{\centering Source}
&\multicolumn{1}{|p{1cm}|}{\centering $m_{u}$\\ (AB mag)   } 
&\multicolumn{1}{|p{1cm}|}{\centering $m_{g}$\\(AB mag)   } 
&\multicolumn{1}{|p{1cm}|}{\centering $m_{r}$\\ (AB mag)  }
&\multicolumn{1}{|p{1cm}|}{\centering $m_{i}$\\ (AB mag)   }
&\multicolumn{1}{|p{1cm}|}{\centering $m_{z}$\\ (AB mag)  } 
&\multicolumn{1}{|p{1cm}|}{\centering $m_{y}$\\ (AB mag)  } \\
\hline
\xa & 263.23707 43.51231 & SDSS17/Pan-STARRS  & 22.3 $\pm$  0.3  &21.19 $\pm$  0.06 & 20.83 $\pm$  0.04& 20.64 $\pm$  0.04  & 20.57 $\pm$  0.06  &  --\\ %
		\xb & 43.65365 41.07406 &  Pan-STARRS &  -- & 21.80 $\pm$  0.06 & 20.61 $\pm$  0.02 & 19.71 $\pm$  0.01 & 19.32 $\pm$  0.02 & 19.02  $\pm$  0.04   \\
            \xc & 29.28776 37.62768 & Pan-STARRS &   -- & 21.89 $\pm$  0.11 & 22.05  $\pm$  0.12 & 22.37  $\pm$   0.11 & 21.73 $\pm$  0.15 & --\\
            $^{\ast}$\xd & 173.53037 0.87409 & SDSS17/PS/VIKING & 22.2 $\pm$  0.4 & 22.4 $\pm$   0.2 & 21.6 $\pm$   0.1 & 21.05 $\pm$  0.05  & 20.82 $\pm$  0.06 & 20.4 $\pm$ 0.2  \\
            \xe & 167.07885 -5.07495 & Pan-STARRS & -- & 21.5 $\pm$ 0.1 & 20.30 $\pm$ 0.04 & 19.83 $\pm$ 0.03 & 19.47 $\pm$ 0.03 & 19.63 $\pm$ 0.09  \\
             $^{\ast\ast}$\xf{} & 204.19926 -41.33718  & DECam/VHS & -- & -- & -- & 22.25 $\pm$ 0.05 & 21.23 $\pm$ 0.04 & -- \\
            $^{\ast\ast\ast}$\xg{} & 37.89582 -60.62918 & GROND & -- & 19.58$\pm$0.01 & 19.58$\pm$0.01 & 19.53$\pm$0.01 & 19.67$\pm$0.02 & --\\
\hline
\end{tabular}
\end{center}
\vspace{-0.5cm}
\begin{flushright}
$^{\ast}$(AB mag) $m_{J}$=20.5 $\pm$ 0.2, 
$m_{H}$=20.2 $\pm$ 0.3,
$m_{K}$=19.9 $\pm$ 0.2 \\

$^{\ast\ast}$(AB mag) $m_{J}$=20.25 $\pm$ 0.28

$^{\ast\ast\ast}$
(AB mag)$m_{J}$=19.39 $\pm$ 0.07, 
$m_{H}$=19.63 $\pm$ 0.14,
$m_{K}$=19.67 $\pm$ 0.18;
From \citet{Novara2020} \\
\end{flushright}
\end{table*}

\subsubsection{Pan-STARRS and Blanco/DECam }
The Panoramic Survey Telescope and Rapid Response System (Pan-STARRS; \citealt{2016arXiv161205560C}) is a 1.8-meter telescope  located at Haleakala Observatory, Hawaii. It uses a 1.4 Gigapixel camera  to image the sky in five broadband filters ($g$, $r$, $i$, $z$, $y$). Pan-STARRS operations began on 2010 May 13. We obtained deep, stacked, Pan-STARRS images of the fields of the FXTs. Deep Pan-STARRS or Dark Energy Camera (DECam) images of the field of the 7 FXTs under study here are shown in Figure~\ref{fig:images}. For each FXT, we show the filter image where the host candidate is most clearly visible. Red circles show the 1$\sigma$ confidence error region of the FXT. The candidate host galaxies are marked with yellow lines. 
We used $\textsc{SExtractor}$ (\citealt{1996A&AS..117..393B}) to extract the R.A. and Dec. 
from the centre of the candidate host galaxies using the Pan-STARRS images. The positions of the host galaxies are then used to calculate the angular and physical offset between the centre of the candidate host galaxy and the centre of the FXT X-ray location for each FXT. We also performed forced photometry at the position of the FXTs in Pan-STARRS (\citealt{2020ApJS..251....3M}). We use the point spread function derived from the image to determine the upper limit on the photometry at the known, fixed, FXT position in the difference images.
The 5$\sigma$ limiting magnitudes of stacked Pan-STARRS images in the $g$, $r$, $i$, $z$ and $y$- filters are 23.3, 23.2, 23.1, 22.3 and 21.4 mag, respectively (\citealt{2016arXiv161205560C}). 

 We used deep images of the field of the FXTs \xf{} and \xg{} obtained using the DECam, mounted on the  Blanco 4--meter Telescope at Cerro Tololo Inter-American Observatory (\citealt{2016MNRAS.460.1270D}). The DECam fields of \xf{} and \xg{} are shown in Figure~\ref{fig:images}.
The 10$\sigma$ limiting magnitude for galaxies in Dark Energy Survey  utilizing  DECam is $g$ = 23.4, $r$~=~23.2, $i$ = 22.5, $z$ = 21.8, and $Y$ = 20.1 (\citealt{2018ApJS..235...33D}).

\subsubsection{ATLAS}
The Asteroid Terrestrial-impact Last Alert System (ATLAS) is a robotic astronomical survey (\citealt{2018PASP..130f4505T}) which maintains a quadruple 0.5m telescope system with two units in Hawaii and one each in Chile (El Sauce) and South Africa (Sutherland). ATLAS observes the whole visible sky several times every night to a limiting magnitude in the $o$-filter (5600-8200 \AA) of $\sim$ 19.5. We also obtained upper limits to the flux density at the position of FXTs using the ATLAS forced photometry server (\citealt{2020PASP..132h5002S}).
\subsubsection{ESO--2.2m/GROND}
The Gamma-ray Burst Optical/Near-infrared Detector (GROND; \citealt{2008PASP..120..405G}) is attached to the MPG/ESO--2.2~m telescope located at the La Silla Observatory. We adopt the GROND photometric data for \xg{} reported by \citealt{Novara2020}.

\subsubsection{SDSS}
The Sloan Digital Sky Survey (SDSS) uses a 2.5-m optical telescope dedicated to a multi-spectral imaging and a spectroscopic redshift survey. We use the SDSS17 (\citealt{2022ApJS..259...35A}) photometric data for some of the candidate hosts.

\subsection{Near-infrared photometric observations}
\subsubsection{VISTA}
The Visible and Infrared Survey Telescope for Astronomy (VISTA; \citealt{2006SPIE.6269E..0XD}), part of ESO’s Paranal Observatory, is a 4-meter-class telescope.
VIKING is the VISTA Kilo-Degree Infrared Galaxy Survey (\citealt{2013Msngr.154...32E}). We use photometric data from VIKING for \xd{}. The VISTA Hemisphere Survey (VHS; \citealt{2013Msngr.154...35M}) is imaging the entire southern hemisphere of the sky in at least two infrared filters ($J$ and $K_s$). We obtain the $J$-filter magnitude for \xf{} from VHS.

\begin{figure*}
    \centering
    \includegraphics[scale=0.30]{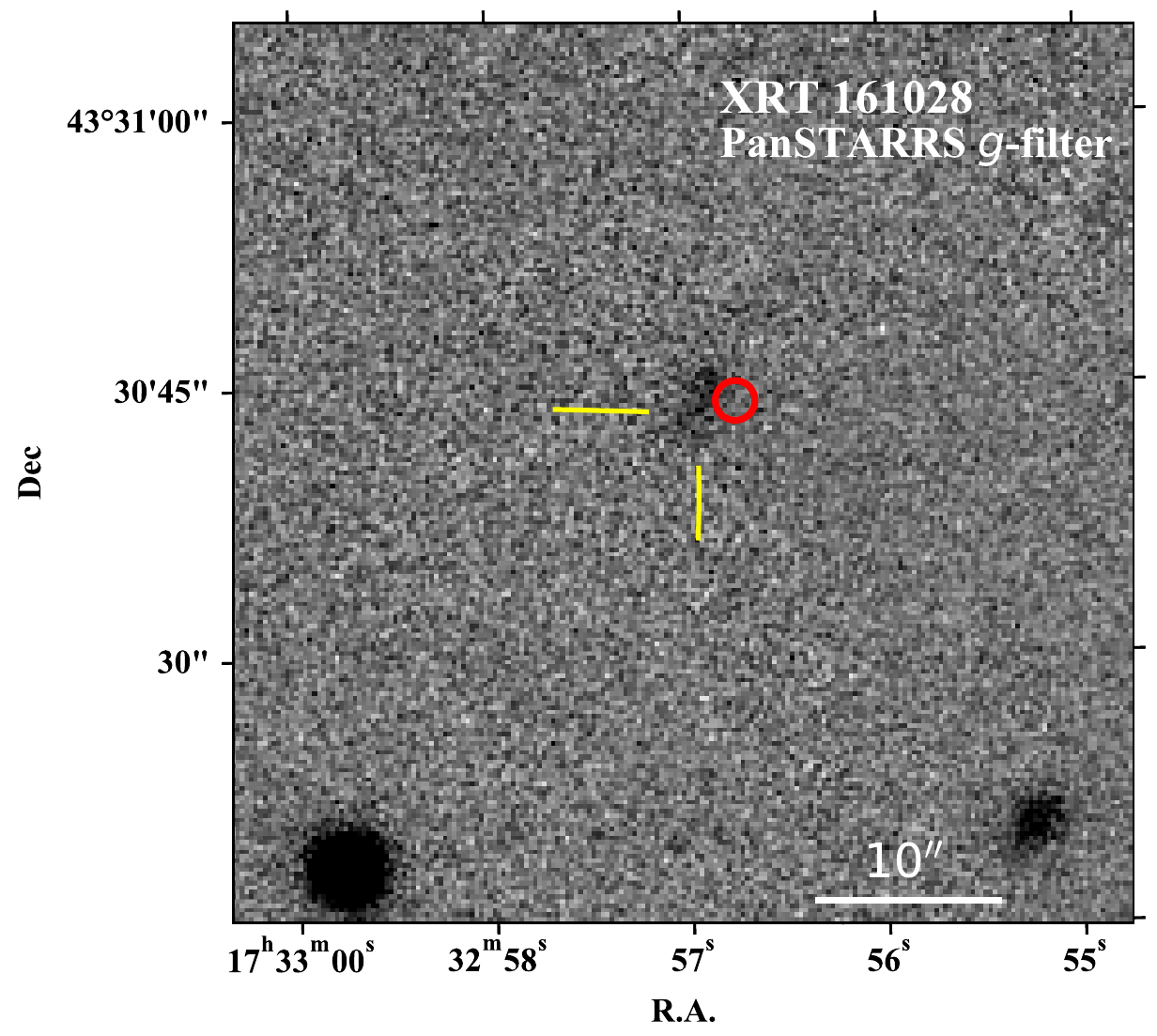}
    \vspace{-0.1cm}
	\includegraphics[scale=0.30]{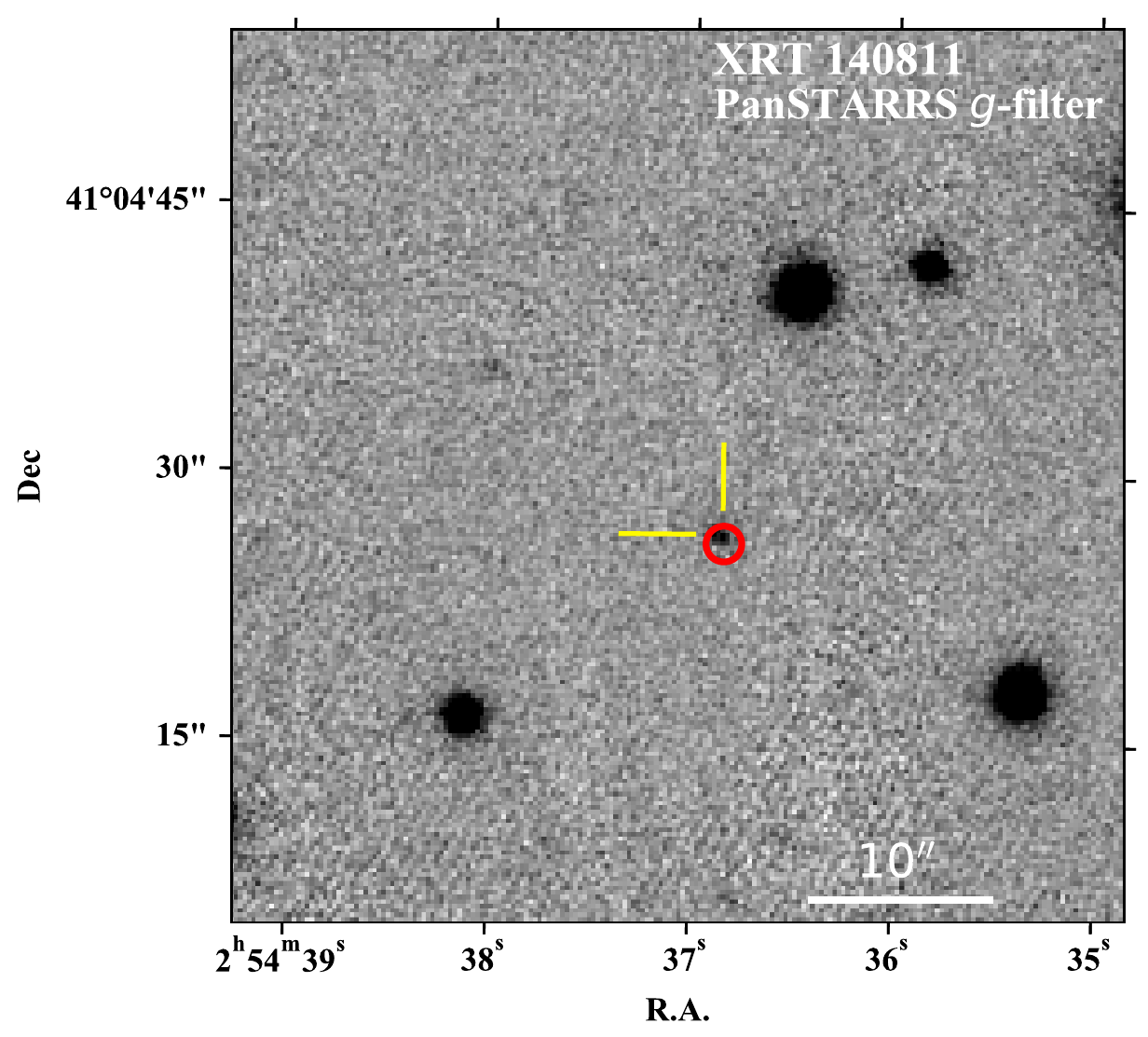}
    
    \includegraphics[scale=0.30]{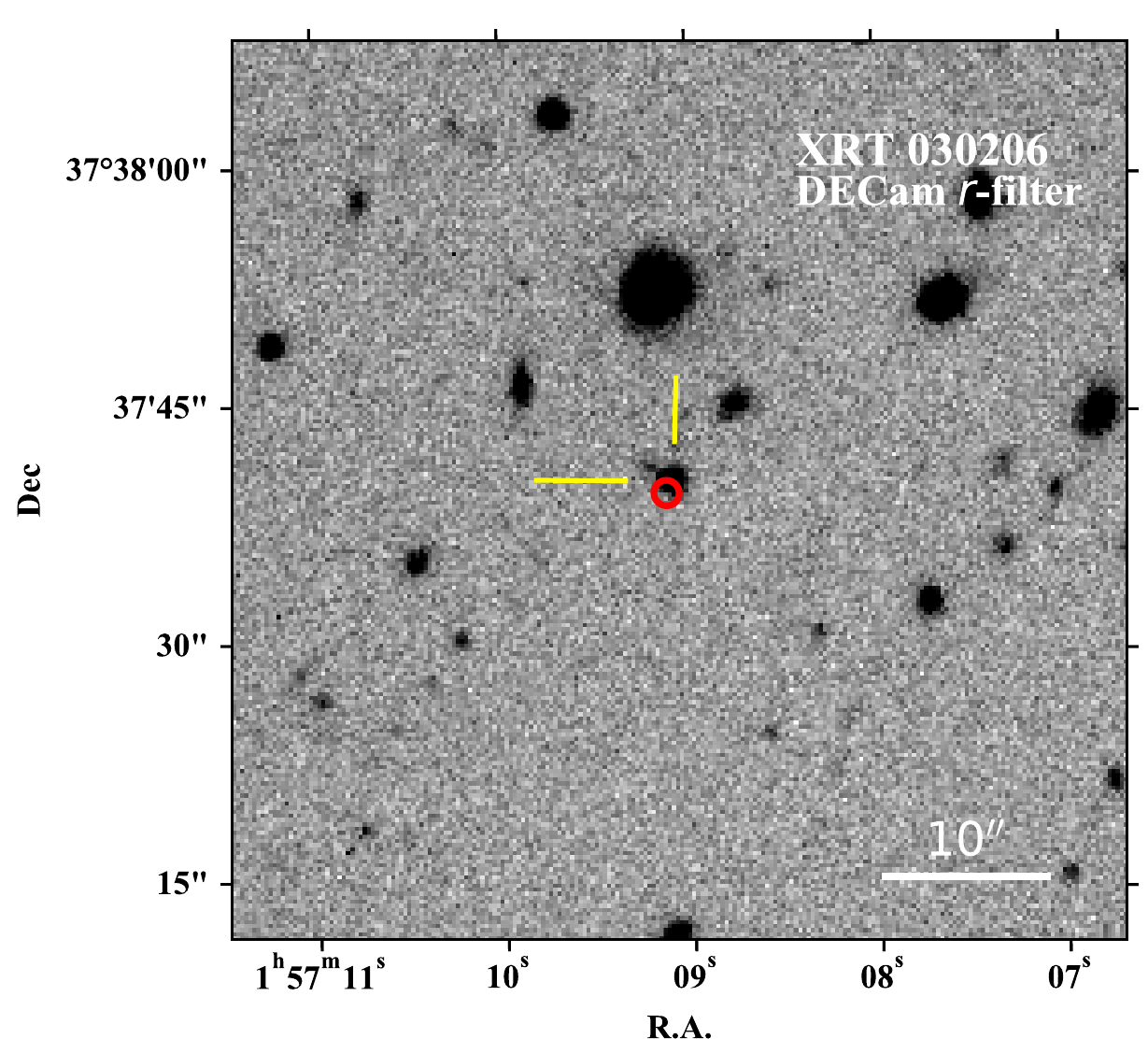}
\vspace{-0.1cm}
	\includegraphics[scale=0.30]{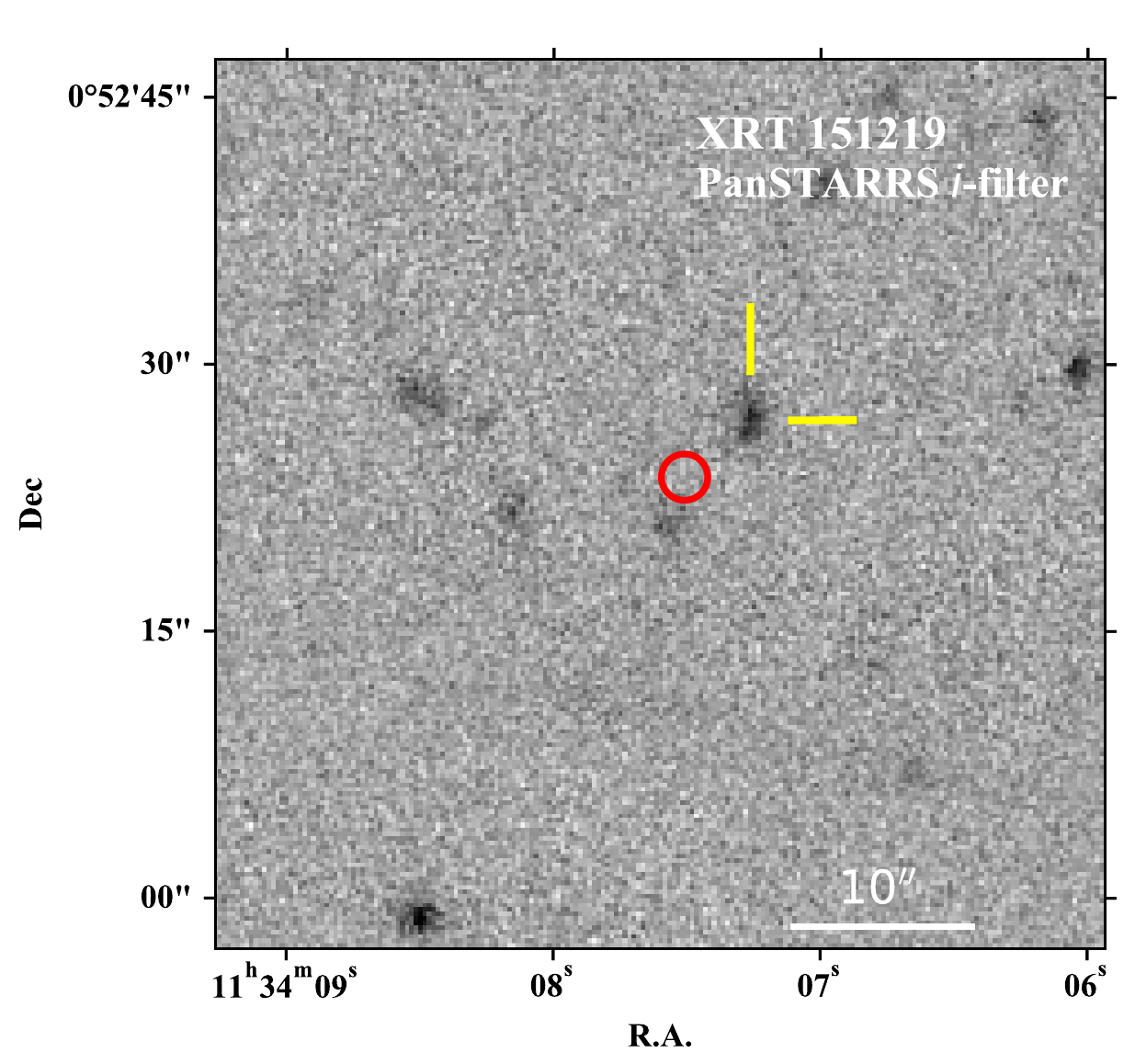}
        
        \includegraphics[scale=0.29]{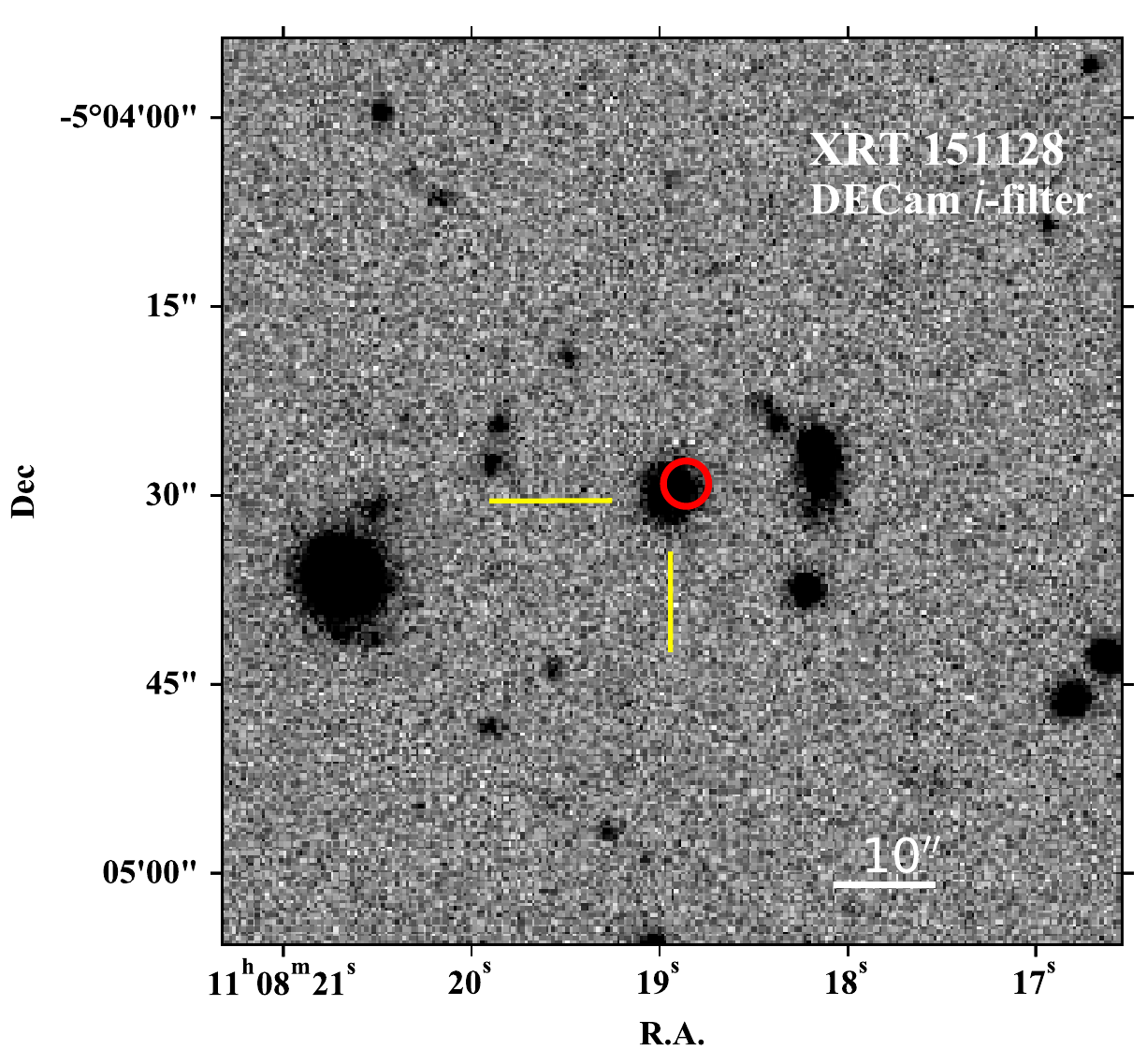}
        \vspace{-0.1cm}
         \includegraphics[scale=0.30]{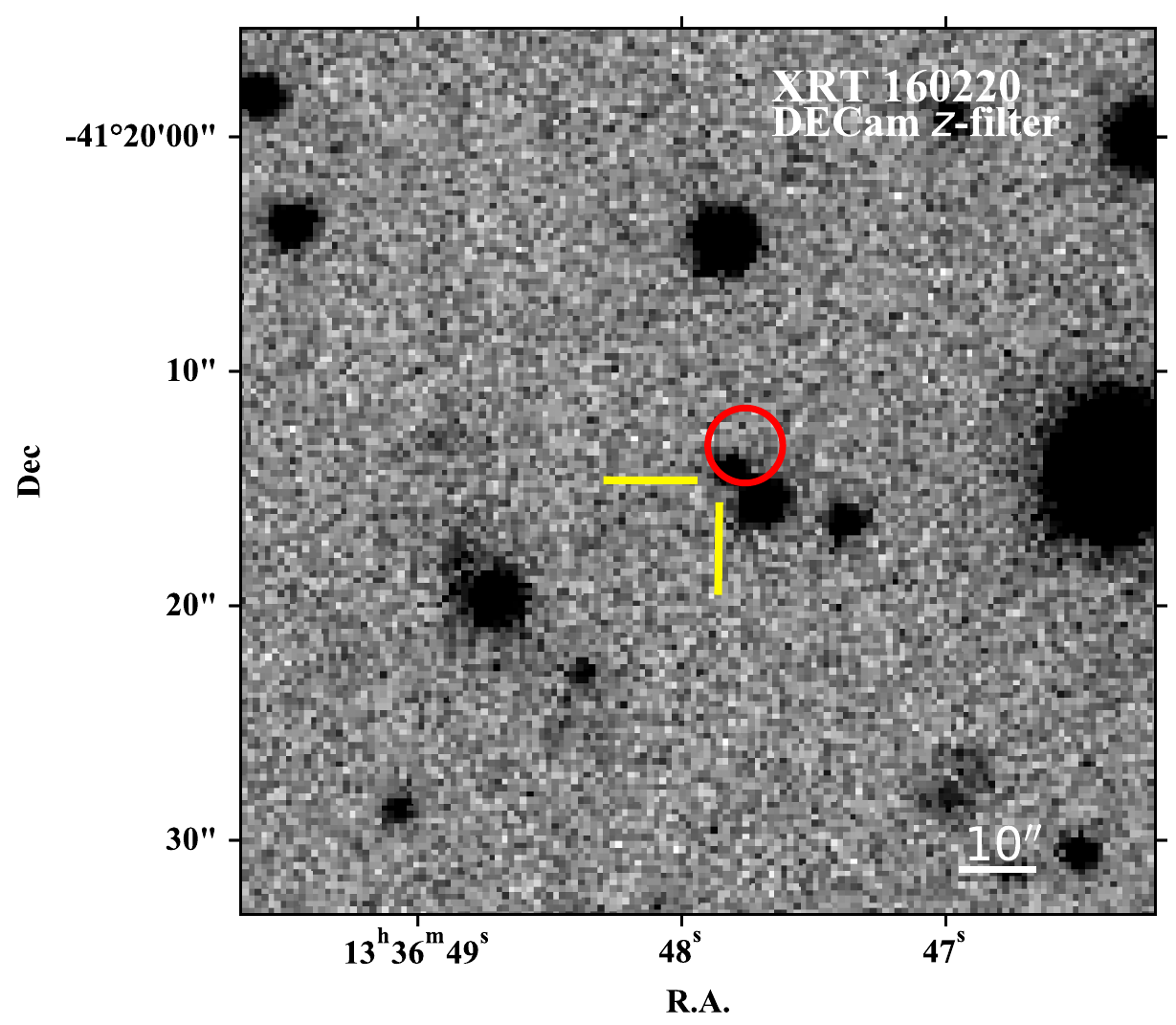}
          \vspace{-0.1cm}
           \hspace{6.2cm}
         \includegraphics[scale=0.30]{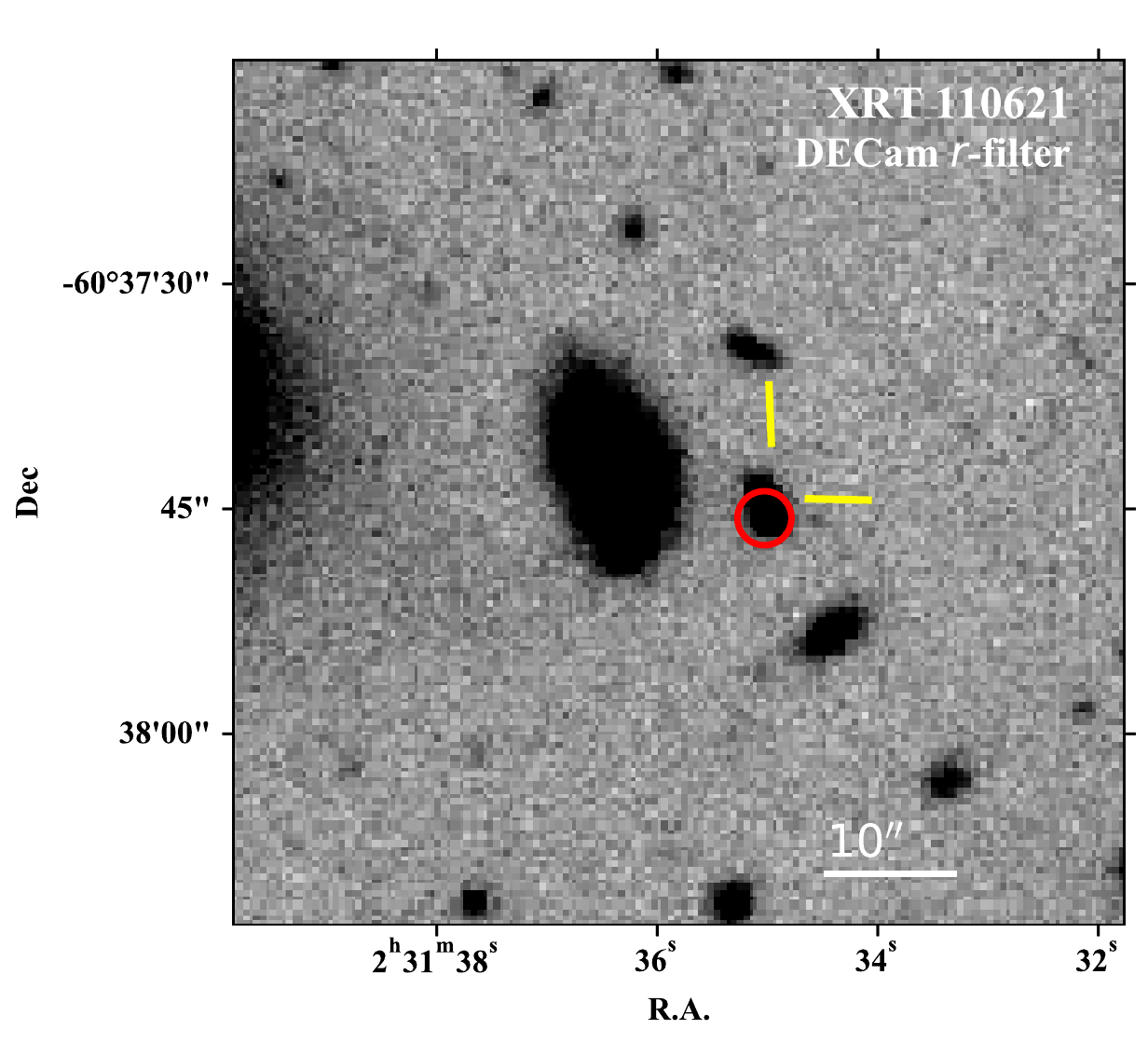}

        \caption{Deep, stacked, Pan-STARRS or DECam images of the field of seven of the twelve {\it XMM-Newton}-discovered FXTs reported by \citet{Alp2020}. The red circles show the $1\sigma$ X-ray positional uncertainty region of the FXT, while the candidate hosts are marked with yellow lines.}
         \label{fig:images}
\end{figure*}

\section{Results}

\subsection{Forced photometry}
\begin{figure*} 
    \centering
	\includegraphics[scale=0.45]{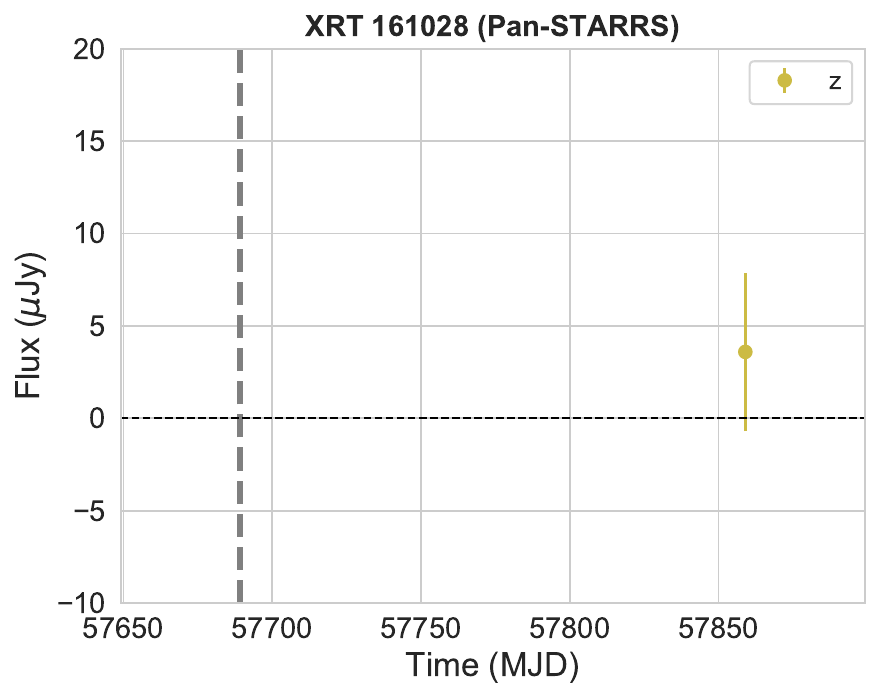}
      \vspace{0.1cm}
      \includegraphics[scale=0.45]{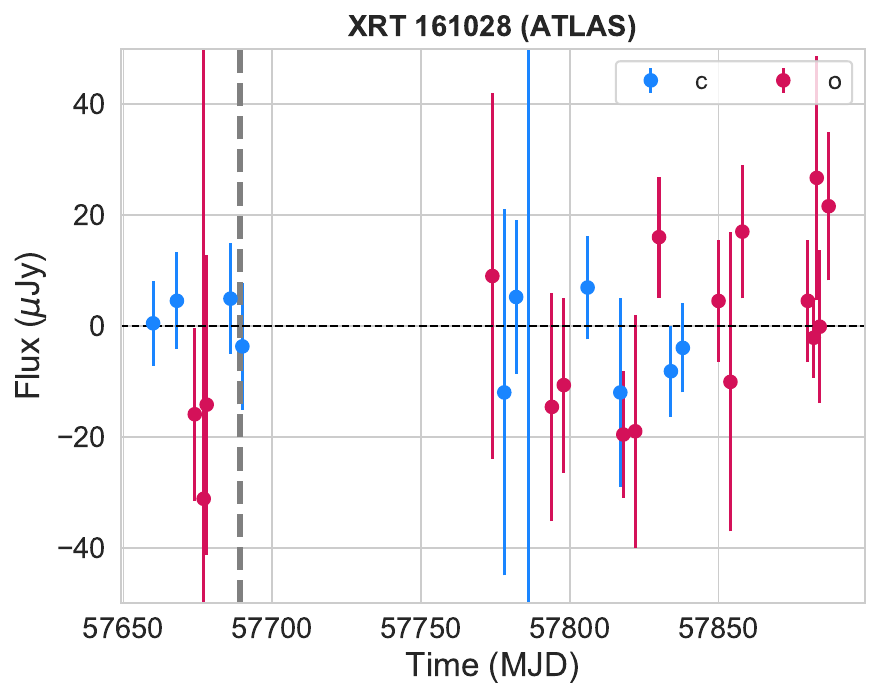}
       \vspace{0.1cm}
        \includegraphics[scale=0.45]{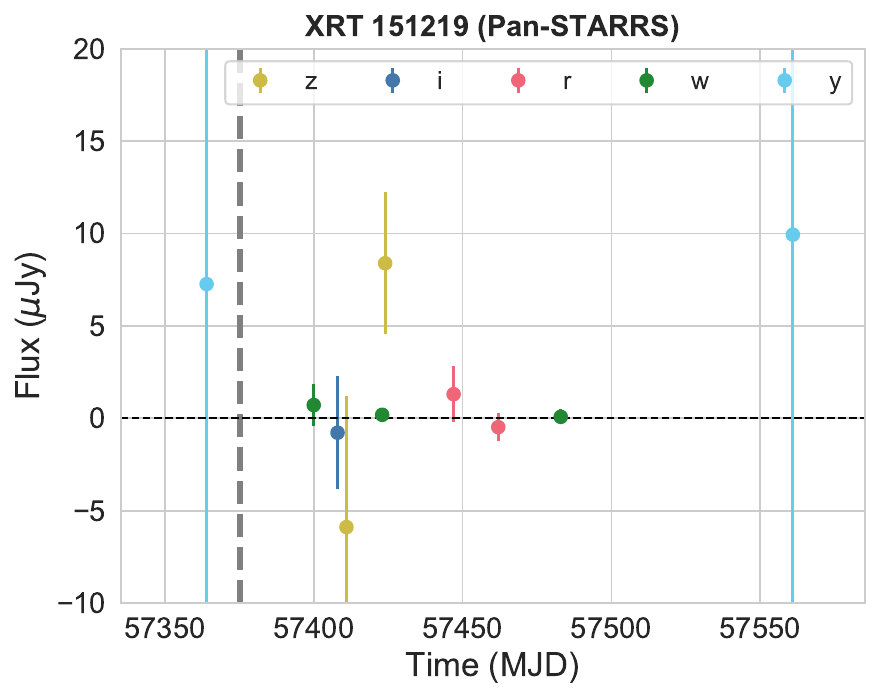}
        \vspace{0.1cm}
         \includegraphics[scale=0.45]{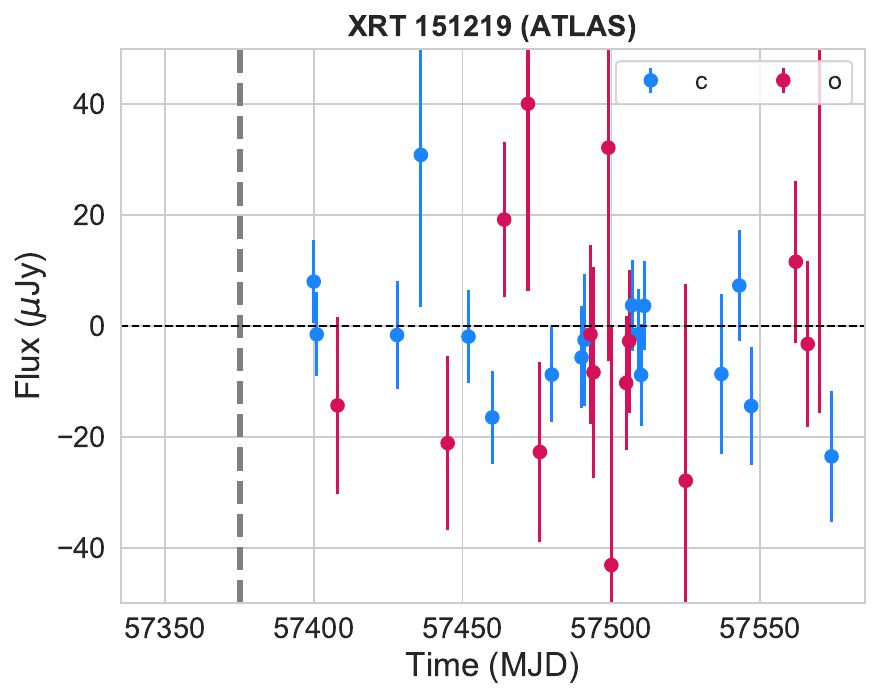}
        \vspace{0.1cm}
        \includegraphics[scale=0.45]{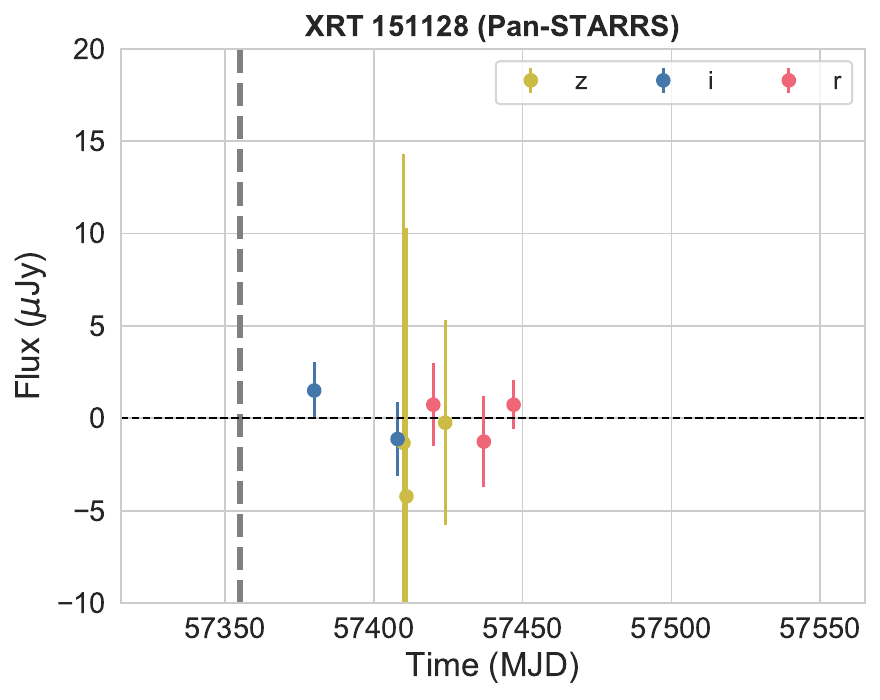}
        \vspace{0.1cm}
        \includegraphics[scale=0.45]{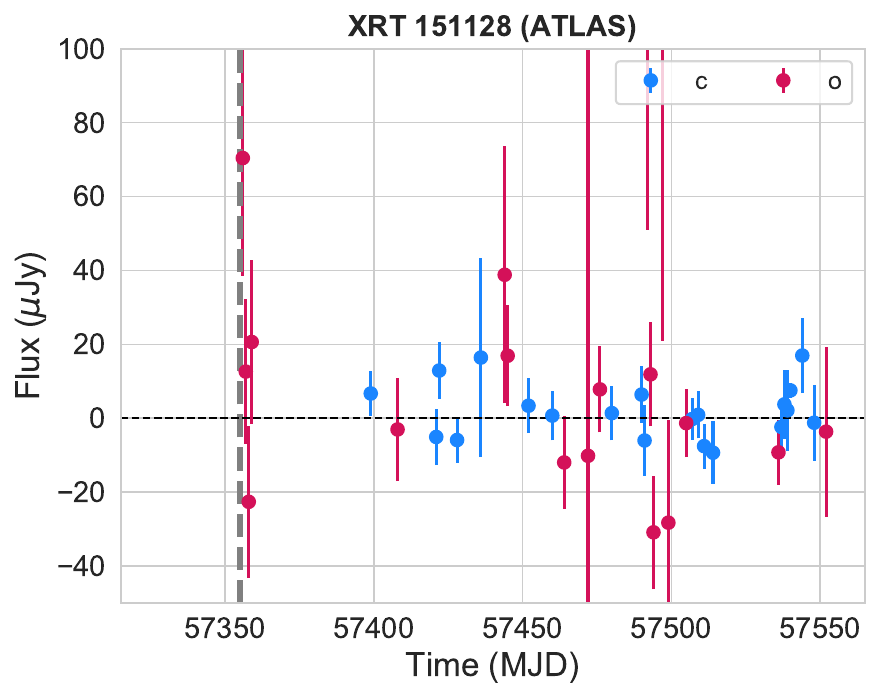}
        \vspace{0.1cm}
        \hspace{7.0cm}
       \includegraphics[scale=0.44]{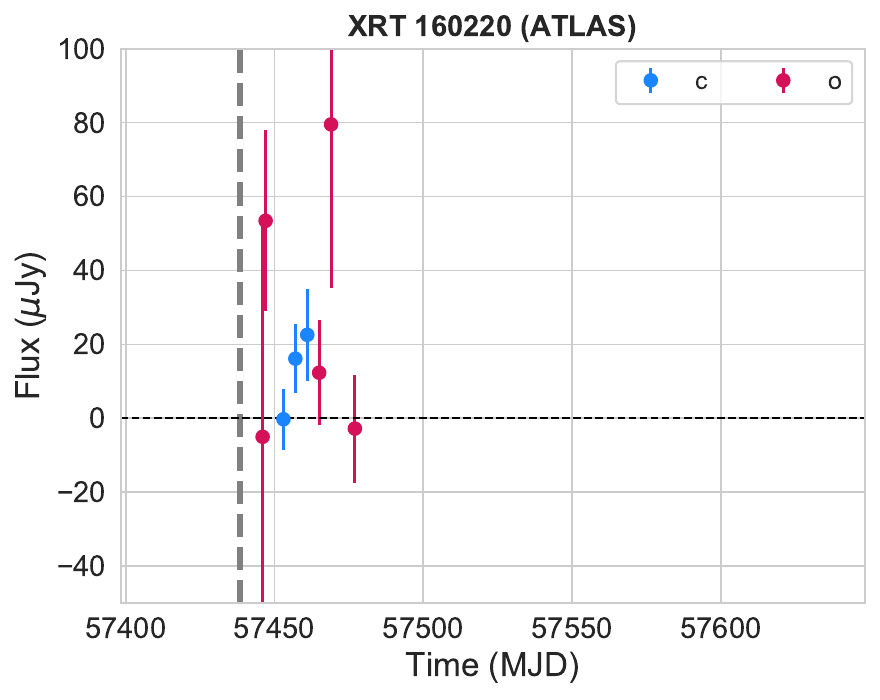}
        \vspace{0.1cm}
         \caption{ The results of the Pan-STARRS and ATLAS forced photometry starting 30 days before and lasting 200 days after the FXT for \xa{}, \xd{}, \xe{} and \xf{}. The X-axis shows time in Modified Julian Date (MJD), and the Y-axis shows the 1-day average flux densities and their uncertainties (in $\mu$Jy). Different filters are indicated by different colours. The vertical, dashed, grey line indicates the time the FXT occurred. No optical source was detected to 5$\sigma$ limits at the FXT location in any filters for any of these FXTs.}
 
\label{fig:FP}
\end{figure*}

We performed forced photometry using Pan-STARRS, and the ATLAS forced photometry servers for  FXTs  \xa{}, \xd{}, \xe{} and \xf{}.  Figure~\ref{fig:FP} shows the resulting light curves, with time in Modified Julian Date (MJD) and 1-day average fluxes and uncertainties (in $\mu$Jy) from 30 days before to 200 days after the date of occurrence of the FXT (indicated by the grey vertical dashed line). No contemporaneous counterpart (5$\sigma$ detection threshold) was found for any of these FXTs.
Different filters are indicated with different colours.  For \xb{}, \xc{} and \xg{}, no forced photometry results are available within 200 days of the event.

We derive upper limits by converting the flux density measurement uncertainty to an AB magnitude (we take 5 $\times$ the uncertainty in the flux density measurement for the upper limit calculation). The AB magnitude upper limit inferred from the forced photometry is given in Table~\ref{tab:LM}. The observation closest in time to the FXT is that in the $z$-filter for \xa{} in Pan-STARRS. It is $\sim$170 days after the burst.  In the case of \xd{}, there are multi-band observations of the field of the FXT $\sim$25 days after the transient. We also have Pan-STARRS observations within $\sim$25 days and ATLAS observations within a day after \xe{}.

\begin{figure*} 
    \centering
	\includegraphics[scale=0.45]{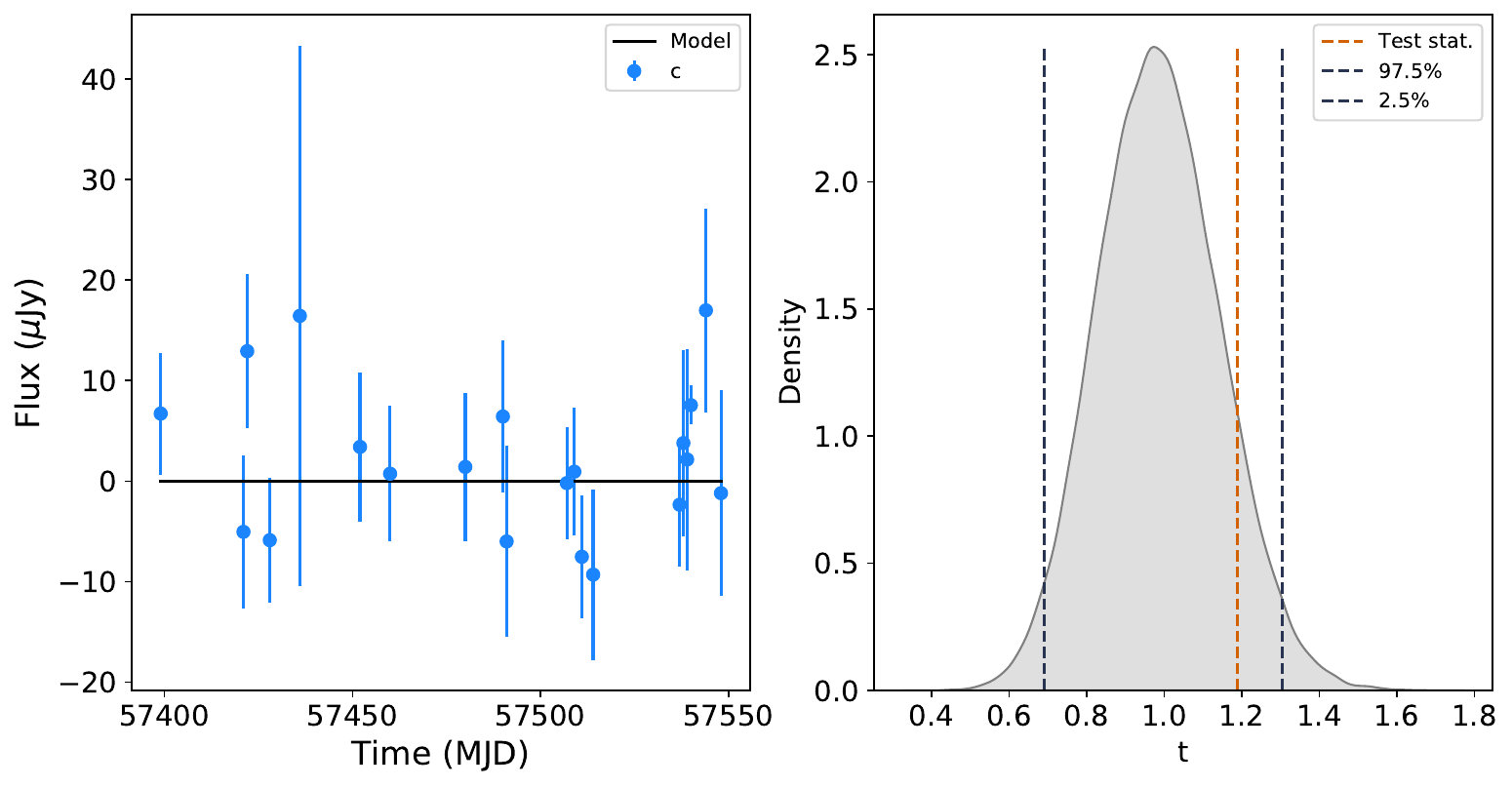}
        \caption{Results of the {\sc ConTEST} (\citealt{2023arXiv230209308S}) test performed on the ATLAS forced photometry data (using the $c$-filter observations) of \xd{}. {\it Left panel}: The flux density, relative uncertainties (blue dots and bars), and the constant at zero model (black line) are shown. {\it Right panel}: The test statistic (orange dashed line) is found to lie between the two critical values, that, for a significance level $\alpha$ = 0.05, are the 2.5 and 97.5 percentiles of the simulations' distances distribution. We find that the null-hypothesis of consistency at zero is not rejected.}
    \label{fig:stat}
\end{figure*}

We use the statistical test {\sc ConTEST} (\citealt{2023arXiv230209308S}), developed explicitly for astronomical data coupled with measurement uncertainties, to check for consistency between the observations and a constant model at zero. The null hypothesis of consistency is tested by calculating a distance measure, the test statistic $t$, between the observations and the model. This step is then followed by a simulation-based methodology where the model is assumed to be the ground truth, and a high number of samples are generated from it based on the observed uncertainties. Calculating the test statistic for each new simulated sample and comparing all of them to the test statistic of the original observations allows us to interpret whether the assumption of consistency has been violated or not.
For an observed test statistic $t$ from an unknown distribution $T$, the P-value is the probability of observing a test-statistic value as extreme as the one observed if the null hypothesis $H_0$ were true. For a two-tailed test, that is: 
$$P-value=2\times min\{P(t\geq T | H_0), P(t< T | H_0)\}.$$ The null hypothesis is not rejected when the P-value is greater than the significance level $\alpha$, which, in our case, is set to the standard value $\alpha$=0.05. 
In {\sc ConTEST}, the P-value is estimated as the proportion of test statistics of the simulated samples that have a $t$-value larger than that derived for the observation -- model combination. As shown in Figure~\ref{fig:stat}, another way to evaluate the result of the test is checking that the test statistic $t$ lies between the two critical values, that, for a significance level $\alpha$ = 0.05, are the 2.5 and 97.5 percentiles of the simulations’ distances distribution.
We performed this test considering different time periods of 0--100 and 0--200 days after the FXT onset in different filters. Due to the lack of data points in the forced photometry light curve, we only did this test in one setting (0--200 days) for the Pan-STARRS data, considering the data points from all the filters together. \footnote{This somewhat unrealistically tests for the presence of a source with an emission spectrum that is flat over the optical bands.} For all the cases, the null hypothesis is not rejected; hence, the model of zero flux density detected in the period 0-200 days after the FXT is consistent with the observations. Figure~\ref{fig:stat} shows an example of the output of the statistical test ConTEST performed on the ATLAS forced photometry data (using the $c$-filter data) of \xd{}. We find that the forced photometry light curves for \xa{}, \xd{}, \xe{} and \xf{} show no evidence for the detection of an optical source with a flux density above our upper limits.

\begin{table*}
\small
\begin{center}

\caption{The 5$\sigma$ upper limits (AB mag) inferred from forced photometry.}

\label{tab:LM}
\begin{tabular}{cccccccc}
\hline
\multicolumn{1}{|p{1cm}|}{\centering FXT}
&\multicolumn{1}{|p{1cm}|}{ \centering Time (MJD)} 
&\multicolumn{1}{|p{1cm}|}{\centering Pan-STARRS filters}
&\multicolumn{1}{|p{3cm}|}{\centering  Pan-STARRS \\5$\sigma$ upper limits \\(AB mag)}
&\multicolumn{1}{|p{1cm}|}{\centering $\sim$$\Delta$T$_{PS}$ (days)} 
&\multicolumn{1}{|p{1cm}|}{\centering ATLAS filters}
&\multicolumn{1}{|p{3cm}|}{\centering  ATLAS \\5$\sigma$ upper limits \\ (AB mag)}
&\multicolumn{1}{|p{1cm}|}{\centering $\sim$$\Delta$T$_{ATLAS}$ (days)}  \\

 \hline
\xa{} & 57689.1108 & $z$ & 20.6 & 170 & $c,o$ & 19.5,18.4 & 01, 85\\
\xd{} & 57375.1343 & $w^{\star}$, $z$ & 22.0, 20.7 & 25, 49 & $c$ & 20.0 & 25\\
\xe{} & 57354.9722 & $i,r$ & 21.7, 21.3 & 26, 66  & $o$ &  18.4 & 01\\
\xf{} & 57438.3113 & -- &  -- &  -- & $o,c$ & 18.7, 19.8 & 09, 15\\

 \hline
\end{tabular}
\end{center}
\begin{flushleft}
$^{\star}$A wide filter denoted by $w$, that essentially spans the $gri$ filters.
\end{flushleft}

\end{table*} 

\subsection{Identification, photometry, and spectroscopy of candidate host galaxies}
\label{sec:redshift}
For each of the FXTs listed below, we searched for a candidate host galaxy in either DECam or Pan-STARRS images. If a candidate host galaxy is found we determine its chance alignment probability in the following way: we determined the density of sources as bright as or brighter than the object in the filter under consideration in a square region, 
we provide the coordinate centres of this region, the size of the region, the measured source density, and the filter under consideration in Table~{\ref{tab:CA}}. Assuming Poisson statistics and  considering the area of the error region, we compute the probability of a chance alignment for each FXT (see Table~{\ref{tab:CA}}).

\subsubsection{\xa}

An extended object is found in Pan-STARRS images  ({\it top left} panel of Figure~\ref{fig:images}) at (R.A.,Dec.)=(263.$^{\circ}$23707, 43.$^{\circ}$51231). The centroid of this object is offset by $1.7 \pm 1.1$\arcsec{} from the X-ray position. The source has a kron magnitude of $r$ = 20.69 $\pm$ 0.05 and a $g$--$r$ colour of 0.36 $\pm$ 0.07.

\label{sec:redshift_161028}
The spectrum of the candidate host galaxy \xa{} (shown in the {\it top} panel of Figure~\ref{fig:spec}) exhibits the emission lines H$\alpha$ $\lambda$6564.6\footnote{ Rest wavelengths of the lines in vacuum are from https://classic.sdss.org/dr6/algorithms/linestable.html}, H$\beta$ $\lambda$4862.7, [O II] $\lambda$3728.5, [O III] $\lambda$4960.3, 5008.2 and absorption line Mg$\lambda$5176.7. We fitted multiple Gaussians to the emission and absorption line using the {\sc lmfit}\footnote{https://lmfit.github.io/lmfit-py/} package and obtained the best-fit central wavelengths and their associated errors. We determine that the redshift of the galaxy is $z_\mathrm{spec} = 0.326  \pm 0.004$.

\begin{figure} 
    \centering
	\includegraphics[scale=0.30]{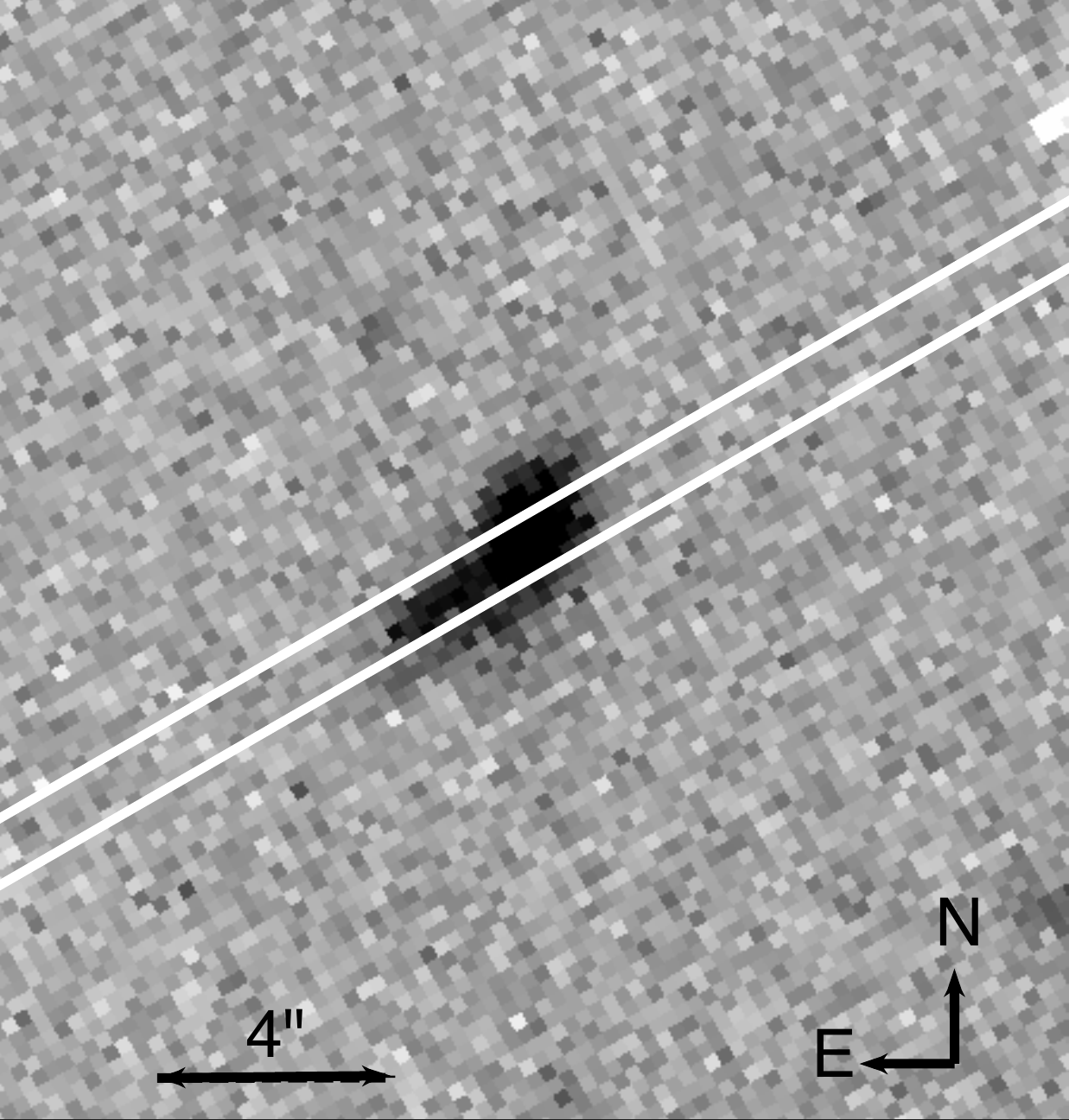}
        \caption{The GTC/OSIRIS the $r$-filter acquisition image of the candidate host of \xa{} (compare with the top left panel of Figure~\ref{fig:images}). The 1\arcsec{} slit is placed at a position angle of 120$^\circ$ North through East. This extended object is a blend of two (extended) sources as is also evident from the spectroscopic data.}
    \label{fig:acq_161028}
\end{figure}

In the GTC/OSIRIS spectroscopic acquisition image for the field of \xa{}, the host candidate appears to be an extended object (see Figure~\ref{fig:acq_161028}). However, upon the inspection of the spectroscopic data, we find that the source is a blend of two objects. We extracted both traces and found another candidate host at redshift $z_\mathrm{spec} = 0.645 \pm 0.007$. We identify the emission lines H$\beta$ $\lambda$4862.7, [O II] $\lambda$3728.5, [O III] $\lambda$5008.2 and  the absorption line Mg $\lambda$5176.7. These candidate hosts both lie within the 3$\sigma$ X-ray position.

\subsubsection{\xb}
The spectrum of the host candidate of \xb{} is consistent with the spectrum of a K3-star. Figure~\ref{fig:M2} shows the 5000 to 8500 \AA{} GTC/OSIRIS spectrum of the candidate counterpart of \xb{}  and K-star template spectrum from the ESO stellar library (\citealt{1998PASP..110..863P}). Therefore, we do not consider this object any further.

\subsubsection{\xc}
\xc{} is offset by 0.8 $\pm$ 0.8\arcsec{} from the centre of a candidate host galaxy at (R.A.,Dec)=(29$^\circ$.28776, 37$^\circ$.62768) ({\it second row-left} panel of Figure~\ref{fig:images}). The candidate host for \xc{} has $r$=22.05 $\pm$ 0.12.

The extracted 1D spectrum of the candidate host galaxy of \xc{} shows [O III] $\lambda$4960.3, 5008.2 and H$\alpha$ $\lambda$6564.6 emission lines.  We also identify extended emission in the 2D spectrum of the candidate host at the position of [O III] and H$\alpha$. We infer the spectroscopic redshift of the candidate host of \xc{} to be  $z_\mathrm{spec} = 0.281  \pm 0.003$. The residuals due to the subtraction of bright sky emission lines are present in the part of the spectrum we marked with a grey vertical band ({\it second} panel of Figure~\ref{fig:spec}). We visually inspected the 2D spectrum and determined that these are due to contamination from sky lines, particularly since there are no bright emission lines expected at these wavelengths (given the redshift derived from [OIII] and H$\alpha$).

\subsubsection{\xd}
In the Pan-STARRS $i$-filter image of \xd{} ({\it second row-right} panel in Figure~\ref{fig:images}), two candidate host galaxies can be identified. We consider the extended galaxy at an offset of 4.9 $\pm$ 1.3\arcsec{} as the probable  host (marked with yellow lines). 

For the candidate host galaxy of \xd{}, we determine the redshift using the emission lines H$\beta$ $\lambda$4862.7 and [O II] $\lambda$3728.5 to be $z_\mathrm{spec} = 0.584 \pm 0.009$. Other lines, such as [O III] $\lambda$4960.3, 5008.2, and MgII$\lambda$2799.1, can be identified in the spectrum ({\it third} panel of Figure~\ref{fig:spec}) and they support this redshift determination.

\subsubsection{\xe}
A possible host galaxy is detected in the DECam $i$-filter ({\it third row-left} panel in Figure~\ref{fig:images}) with an offset of 1.2 $\pm$ 1.8\arcsec{}. The photometry of \xe{} is given in  Table \ref{tab:phot}.

\begin{figure*} 
    \centering
	\includegraphics[scale=0.39]{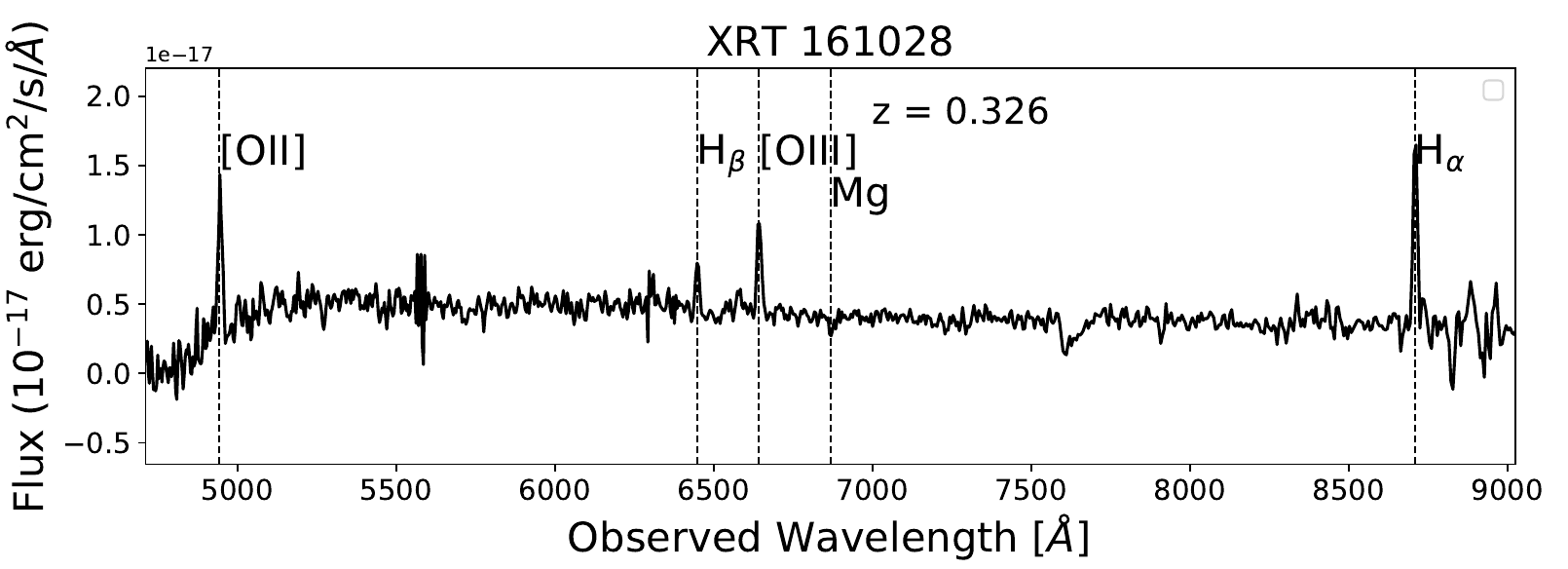}
        \vspace{0.1cm}
        \includegraphics[scale=0.40]{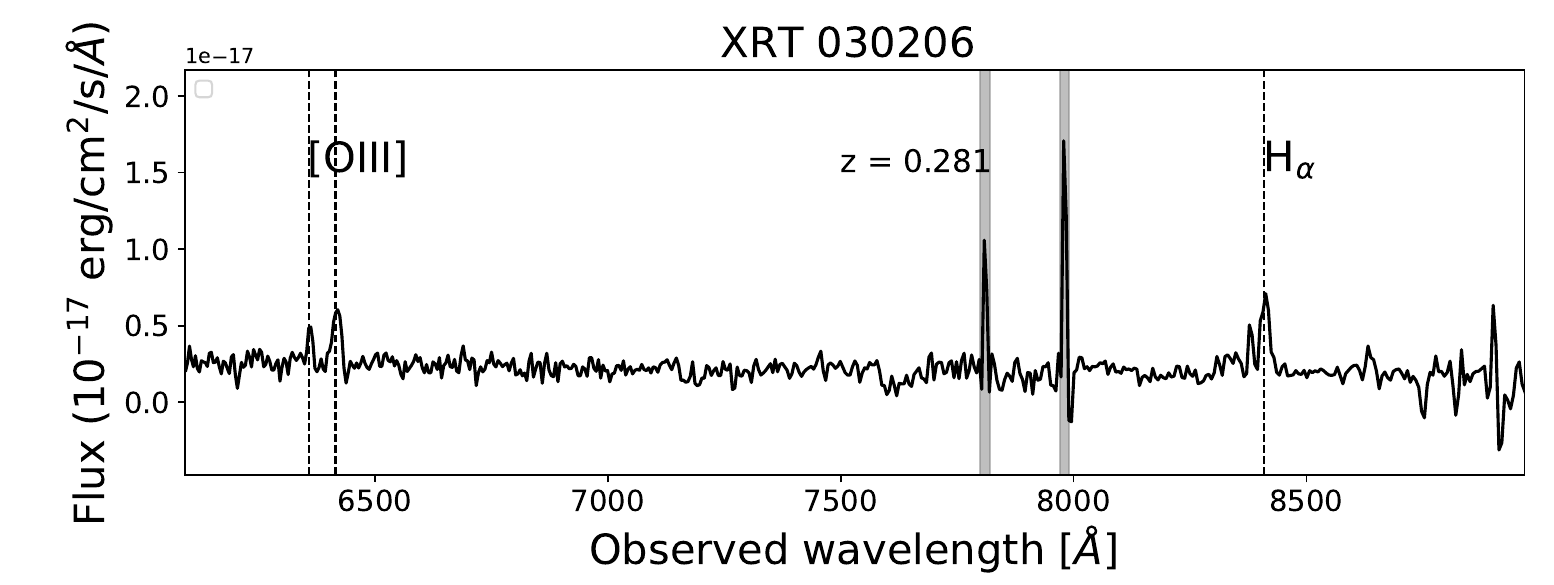}
        \vspace{0.1cm}
        \includegraphics[scale=0.36]{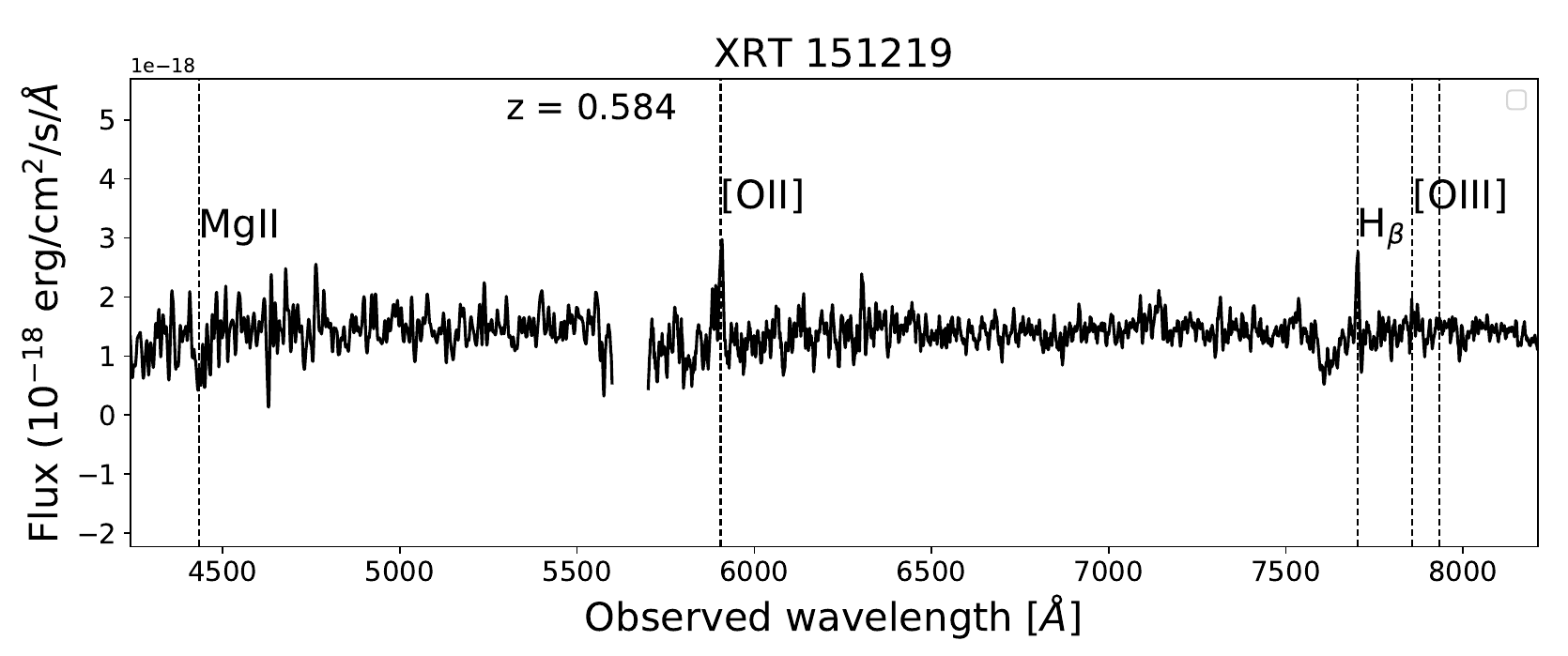}
        \vspace{0.1cm}
        \includegraphics[scale=0.42]{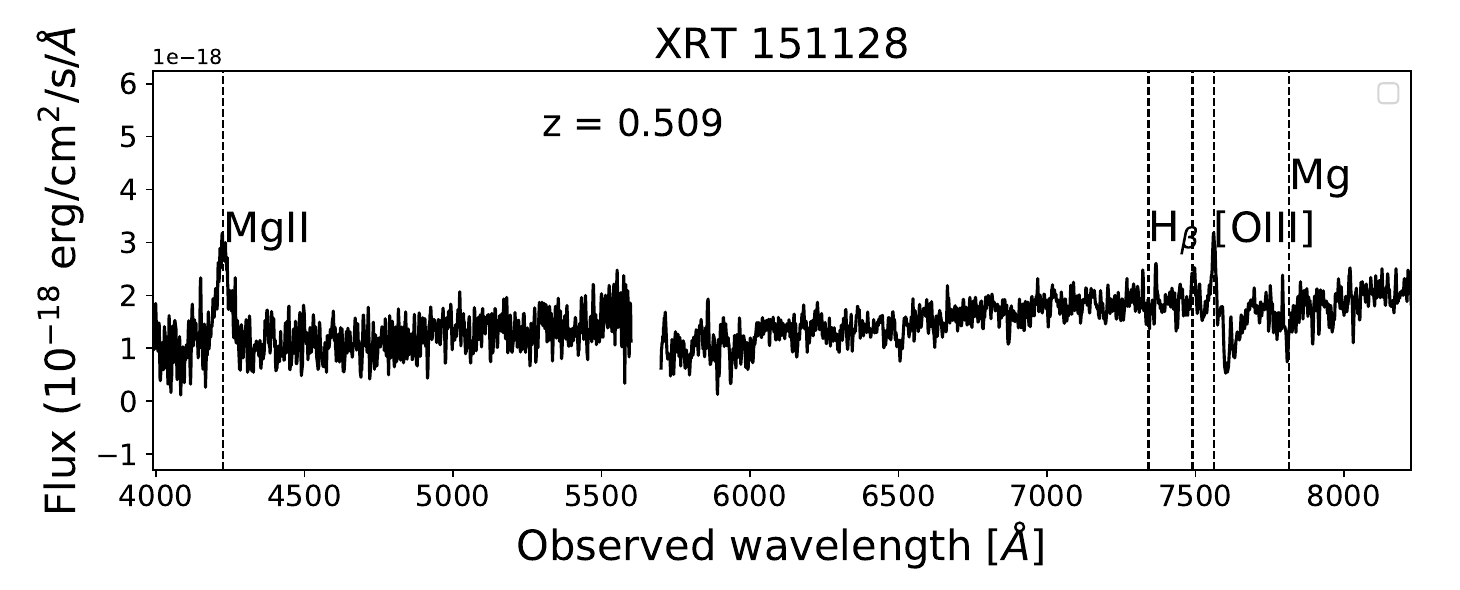}
        \vspace{0.1cm}
        \includegraphics[scale=0.39]{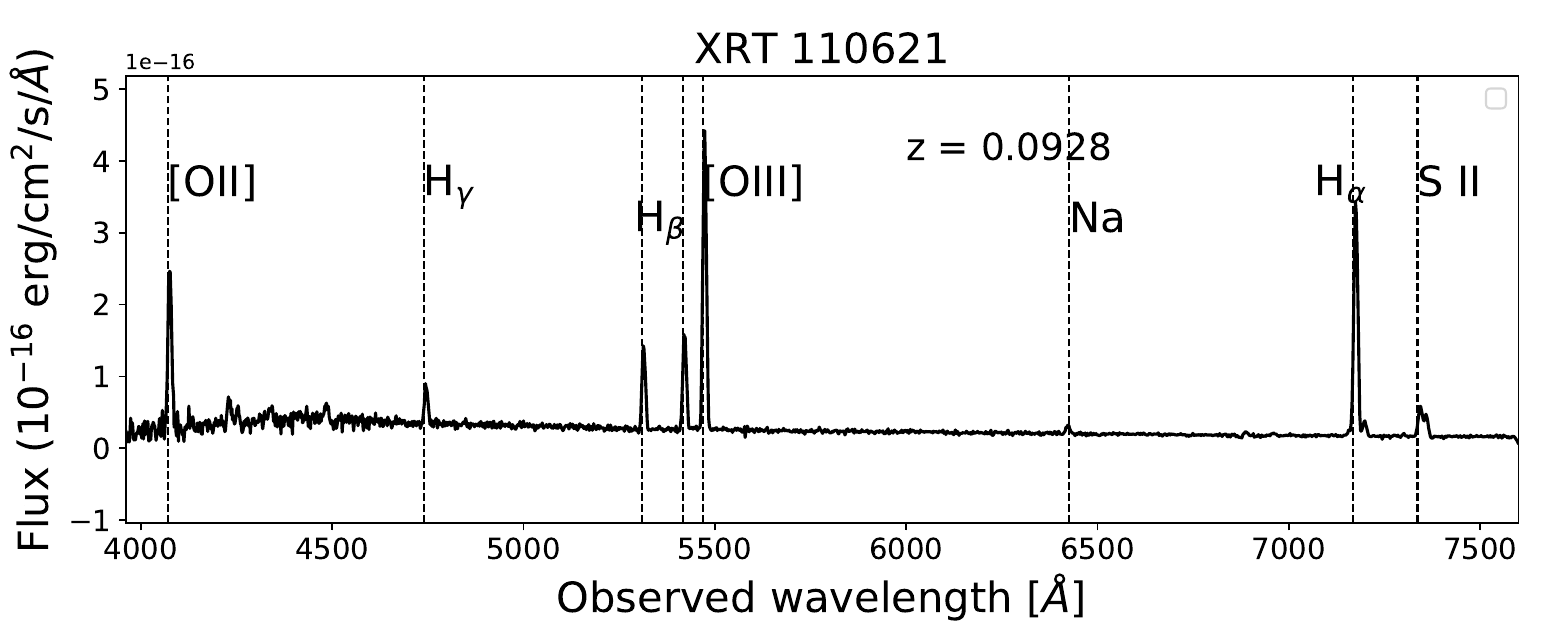}
        \vspace{0.1cm}
        
        \caption{The flux-calibrated spectra of five of the seven candidate hosts galaxies shown in Figure~\ref{fig:images}. For display purposes the boxcar smoothed spectrum (Box1DKernel with a width of 3 pixels) is shown for \xd{} and \xe{}.  Dashed vertical lines mark the location of important emission and absorption lines in each of the spectra. For \xc{}, there are sky residuals present in the spectrum towards the redder part of the spectrum (marked in grey).}
    \label{fig:spec}
\end{figure*}

The candidate host galaxy of \xe{} has a redshift of  $z_\mathrm{spec} = 0.509\pm 0.009$. The redshift was derived through fitting Gaussians to the emission line and the absorption lines [O III] $\lambda$5008.2, Mg$\lambda$5176.7 and MgII$\lambda$2799.1.  The flux calibrated spectrum of \xe{} is shown in the {\it fourth} panel of Figure~\ref{fig:spec}.

\subsubsection{\xf}
The $z$-filter DECam image of the field of \xf{} is shown in the {\it third row-right} panel of Figure~\ref{fig:images}. The DECam/VHS photometry of the host of \xf{} is given in Table \ref{tab:phot}. This host galaxy was identified within the 1$\sigma$ X-ray position by \citet{Alp2020}, marked by the yellow pointers in Figure \ref{fig:images}. The spectrum is dominated by an interloper star (the source adjacent to the host, within $\sim$2$\sigma$), and we were unable to extract the spectrum of the candidate host galaxy.

\subsubsection{\xg}

The DECam $r$-filter image of the field of \xg{} ({\it bottom} panel of Figure~\ref{fig:images}) shows a bright candidate host inside the 1.8\arcsec{} (1$\sigma$) X-ray position. We adopt the GROND photometry of the candidate host of \xg{} from \citet{Novara2020} (see Table \ref{tab:phot}).

We identify the emission lines 
H$\alpha$ $\lambda$6564.6, H$\beta$ $\lambda$4862.7, H$\gamma$ $\lambda$4341.68, [O III] $\lambda$4960.3, 5008.2, [O II] $\lambda$3727.1 and SII  $\lambda$6718.29 in the spectrum of the candidate host galaxy. Our spectroscopic redshift $z_\mathrm{spec} =  0.0928 \pm 0.0002$ is fully consistent with the redshift determined in \citet{Novara2020}.

\begin{figure} 
    \centering
	\includegraphics[scale=0.55]{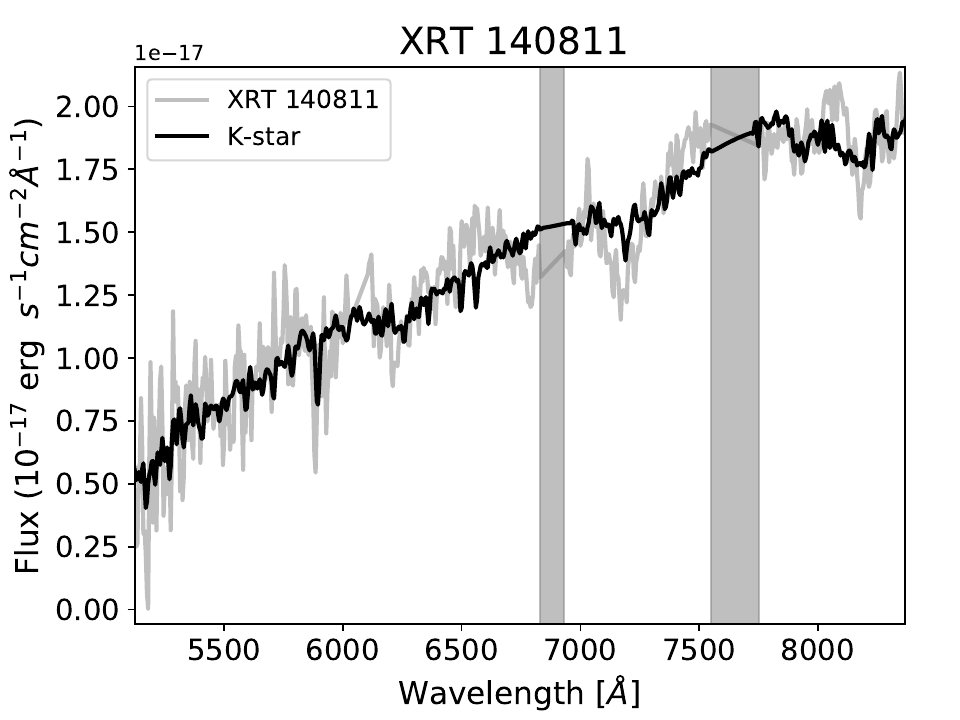}
        \caption{ The GTC/OSIRIS spectrum of the candidate host of \xb{} in the wavelength range 5000 to 8500 \AA{}. A K3V-star template spectrum obtained from the ESO stellar library is overplotted (the K3 spectral template is taken from \citealt{1998PASP..110..863P}). Given the good agreement, we deem it likely that \xb{} is not caused by an extra-galactic FXT but instead by a stellar flare. We manually masked the telluric regions (marked in grey) and a region affected by instrumental artefacts (6030-- 6100 \AA{}).}
    \label{fig:M2}
\end{figure}

\begin{table*}
\small
\begin{center}

\caption{Chance alignment probabilities.}

\label{tab:CA}
\begin{tabular}{cccccccc}
\hline
\multicolumn{1}{|p{1cm}|}{\centering Candidate host of}
&\multicolumn{1}{|p{2.5cm}|}{ \centering Coordinate centres of the region used for  density determination (R.A.~and Dec in $^{\circ}$)} 
&\multicolumn{1}{|p{1.5cm}|}{\centering Filter}
&\multicolumn{1}{|p{2cm}|}{\centering Region considered (\arcsec{})}
&\multicolumn{1}{|p{2cm}|}{\centering Source density \\  (\# / arcsec$^{-2}$)}
&\multicolumn{1}{|p{1.5cm}|}{\centering Radius of interest (\arcsec{})}
&\multicolumn{1}{|p{1.5cm}|}{\centering Chance alignment probability (per cent)}
&\multicolumn{1}{|p{2cm}|}{\centering Positional Uncertainty (1$\sigma$)
(\arcsec{})} \\

 \hline
\xa{} & 263.$^{\circ}$23650, 43.$^{\circ}$51246& Pan-STARRS $g$   & 70 $\times$ 70 &0.0008 & 1.7  & 0.7  & 1.1 \\
\xb{} & 43.$^{\circ}$65379, 41.$^{\circ}$07421 & Pan-STARRS $g$  & 70 $\times$ 70 & 0.002 & 1 & 0.6 & 1 \\
\xc{} & 29.$^{\circ}$2877890 37.$^{\circ}$6276687 & DECam $r$ & 57 $\times$ 57 &  0.002 & 0.8 & 0.4 & 0.8 \\
\xd{} & 173.$^{\circ}$53100, 0.$^{\circ}$87342 & Pan-STARRS $i$ & 70 $\times$ 70 &0.0008 & 4.9 & 5.8 & 1.3 \\
\xe{} & 167.$^{\circ}$07797, -5.$^{\circ}$07520 &  DECam $i$ & 70 $\times$ 70 & 0.0008 & 1.8 & 0.8 & 1.8 \\
\xf{}& 204.$^{\circ}$19986, -41$^{\circ}$.33725 & DECam $z$  & 38 $\times$ 38 & 0.007 &1.6 &  5.2 & 1.6 \\
\xg{} & 37.$^{\circ}$89449 -60.$^{\circ}$62715 & DECam $r$ &  70 $\times$ 70 & 0.0004  &1.9  & 0.5 & 1.9 \\

 \hline
\end{tabular}
\end{center}

\end{table*}

\subsection{Fitting the Spectral Energy Distribution (SED)}

We derived the star formation rate (SFR) and galaxy mass ($M_*$) of the candidate host galaxies using {\it BAGPIPES} (Bayesian Analysis of Galaxies for Physical Inference and Parameter EStimation; \citealt{2018MNRAS.480.4379C}). Taking into account the star formation history and the transmission function of neutral/ionized ISM  for broadband photometry and spectra, and using the \textsc{MultiNest} sampling algorithm, \texttt{BAGPIPES} fits stellar population models to the multi-band photometric data. Posterior probability distributions for the host galaxy redshift ($z$), age (mass-weighted age; Age$_{MW}$), extinction by dust ($A_V$),  specific star formation rate (sSFR) and metallicity ($Z$) can also be used as free parameters. As priors, we used  $A_V$ values between 0.0 and 2.0 mag (flat), an exponentially declining star formation history function, and the parametrization developed by \citet{2000ApJ...533..682C} to account for the effect of dust attenuation on the SEDs. 

\begin{figure*} 
    \centering
        \includegraphics[scale=0.37]{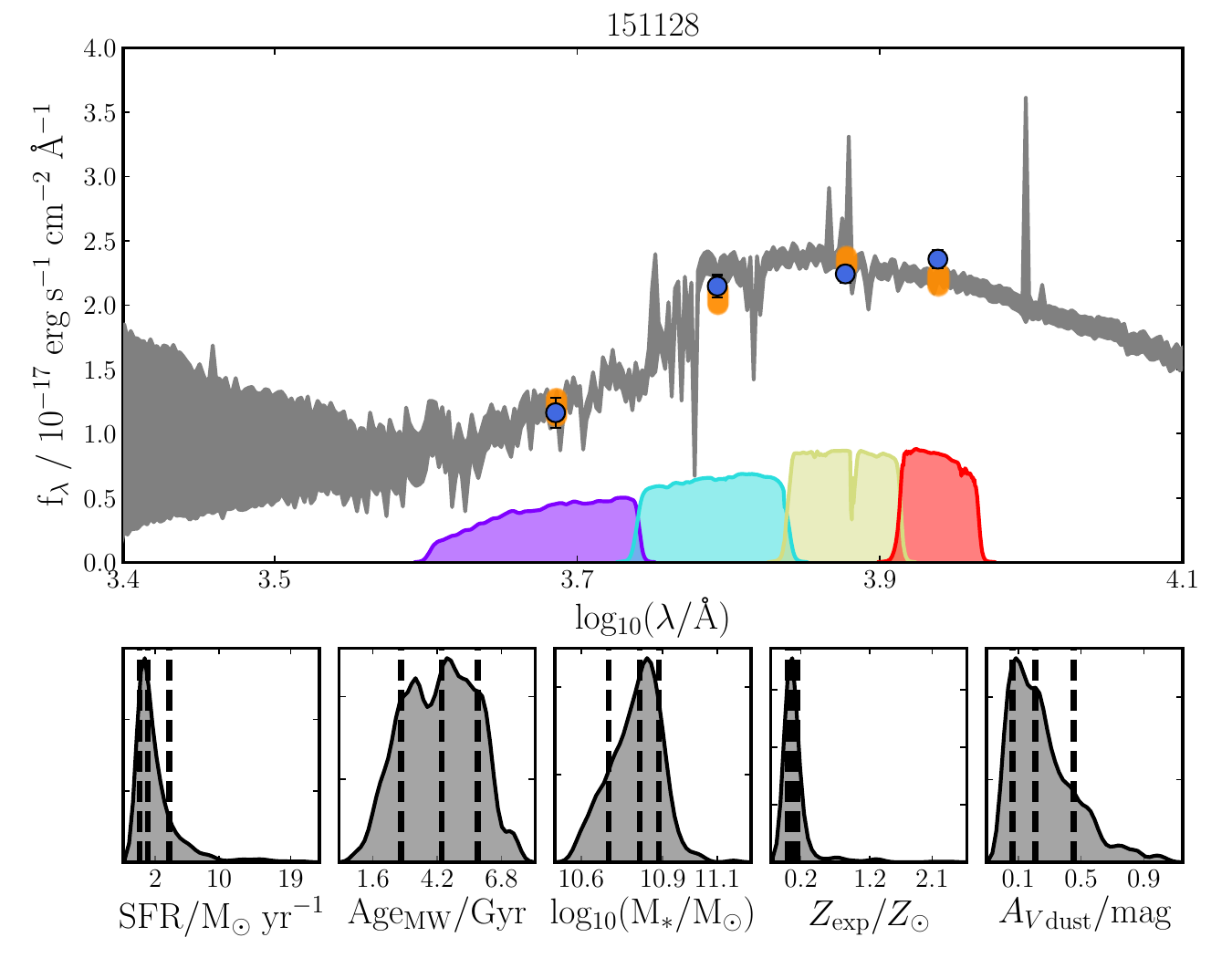}
        \vspace{0.2cm}
        \includegraphics[scale=0.37]{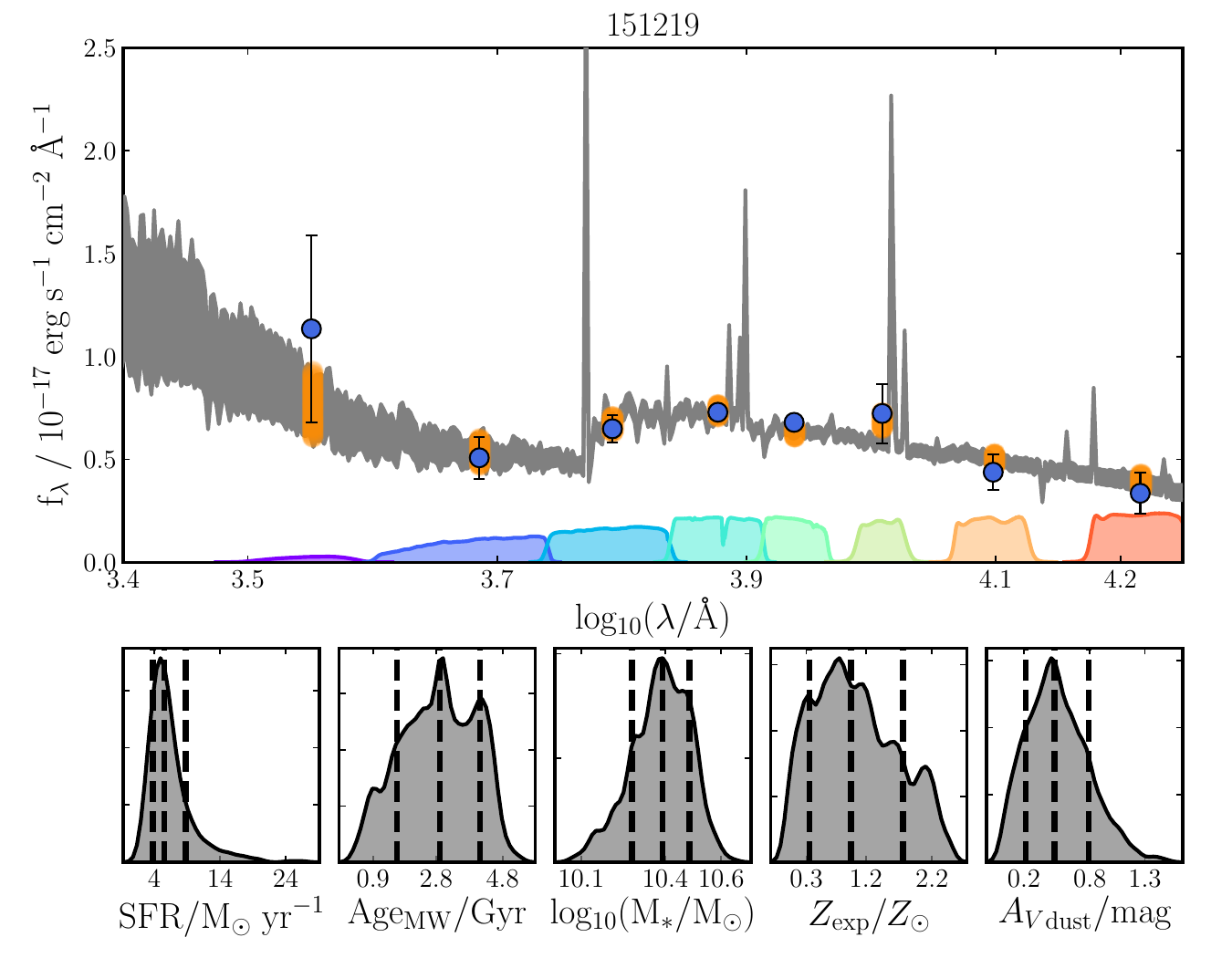}
        \vspace{0.2cm}
        \includegraphics[scale=0.37]{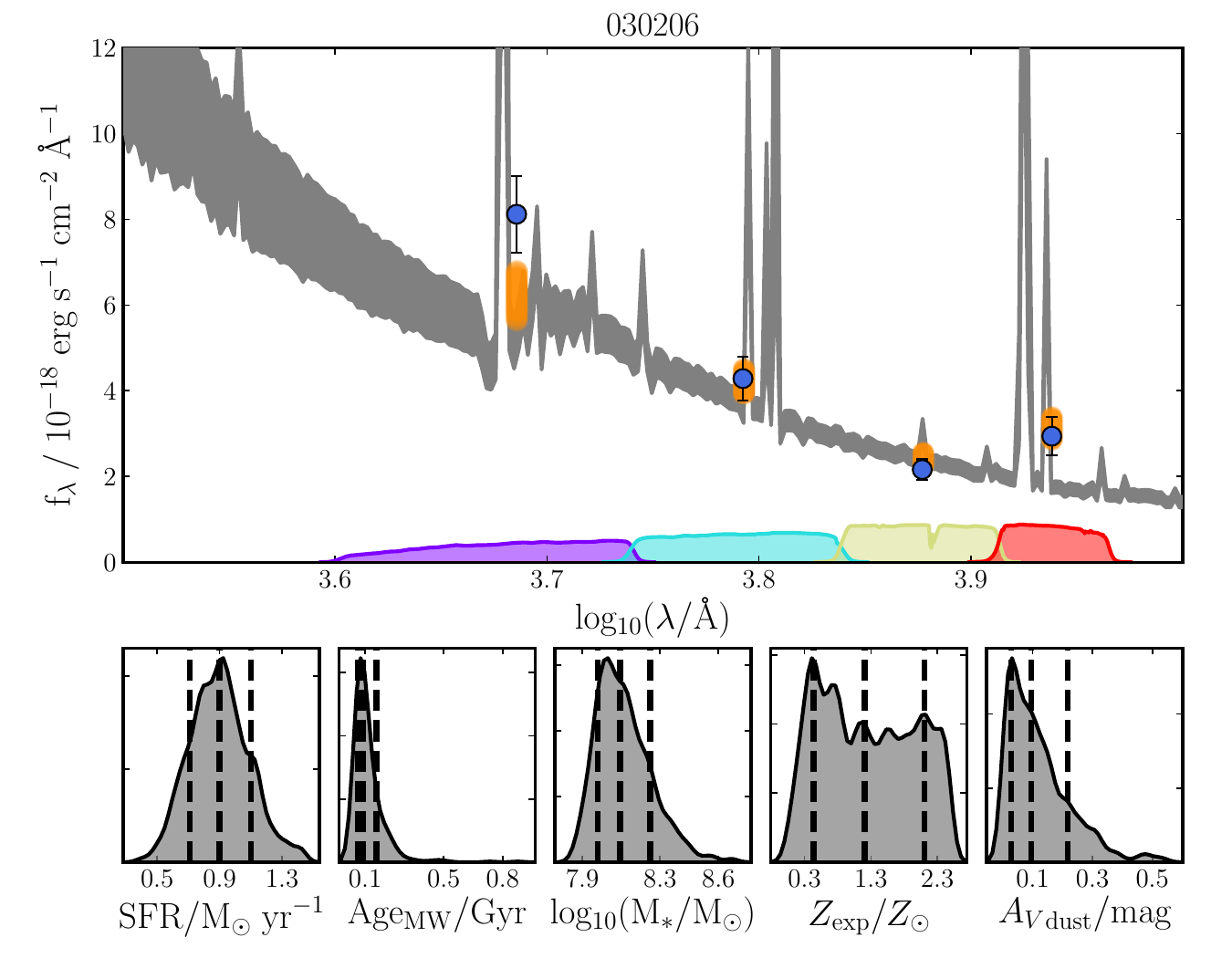}
        \vspace{0.2cm}
        \includegraphics[scale=0.37]{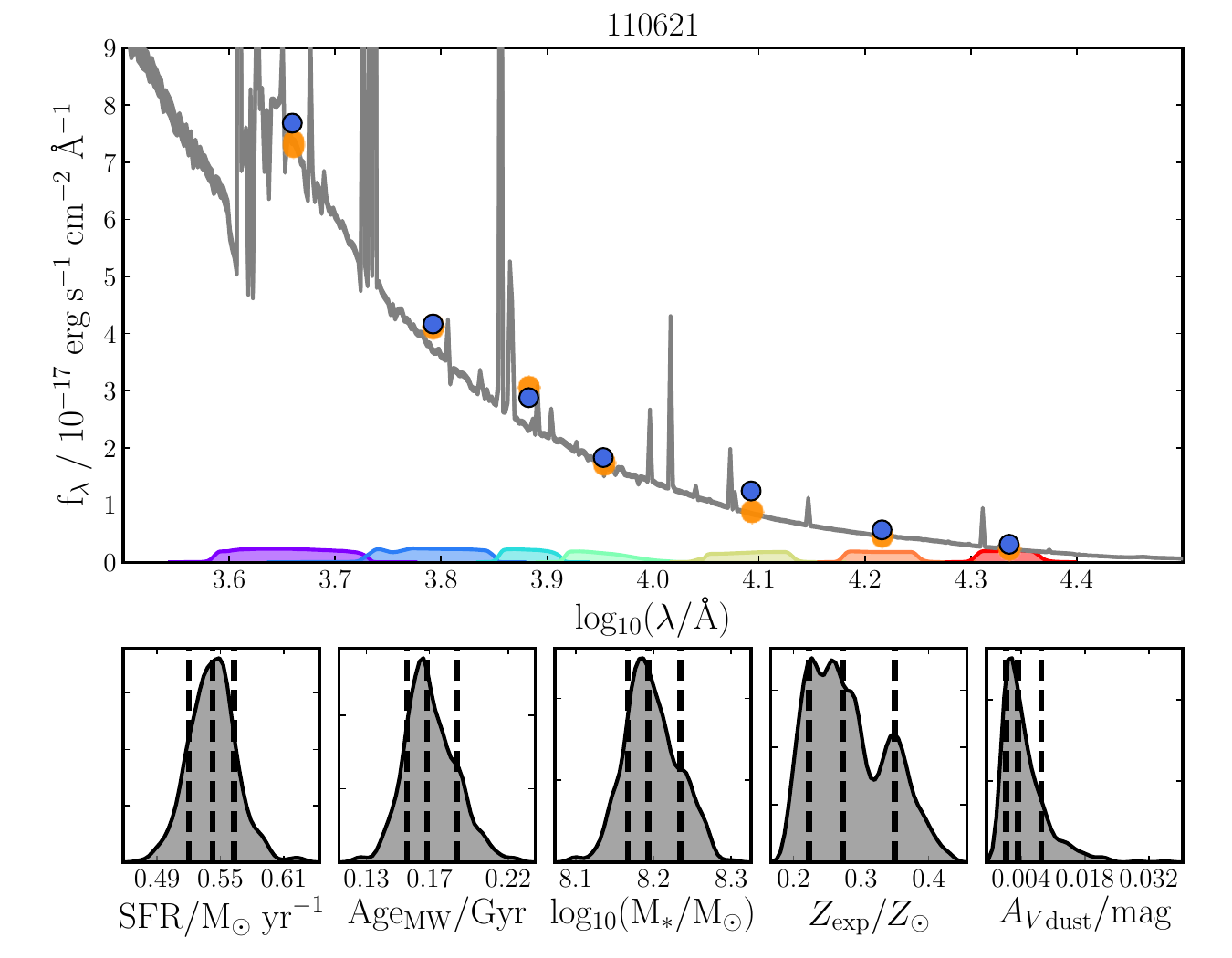}
        \vspace{0.2cm}
        \caption{Best fitting SED models obtained using the \texttt{BAGPIPES} package. The 16th to 84th percentile range for the posterior probability for the spectrum and photometry (shaded gray and orange, respectively) is shown. The input, observed, photometric data and their 1$\sigma$ uncertainties are given by the blue markers. The wavelength range covered by the photometric filters is marked by the coloured bars at the bottom of the plot. \textit{Bottom panels:} Posterior probability distributions for the four fitted parameters (SFR, age, galaxy stellar mass and metallicity). The 16th, 50th, and 84th percentile posterior values are indicated by vertical dashed black lines, from left to right in each sub-plot.}
    \label{fig:sed}
\end{figure*}

Figure~\ref{fig:sed} shows the 16th to 84th percentile range for the posterior probability distribution for best-fit template spectrum and derived broadband photometry (shaded \textit{grey} and \textit{orange}) for four of the host galaxies. The \textit{blue} points indicate our input photometric data for \xc{}, \xd{}, \xe{} and \xg{}. We obtained the posterior probability distribution for the fitted parameters using a flat prior for the redshift given the spectroscopic redshifts and their uncertainties (see Table~\ref{tab:results}). The posterior probability  distribution of the other fitted parameters is shown in the bottom panels of Figure~\ref{fig:sed}.

\begin{table*}
\small
\begin{center}
\caption{The inferred host properties of the FXT candidate host galaxies. We obtain the redshift either from our spectroscopic observations or the photometric redshift through our SED fitting. Host parameters, along with the peak X-ray luminosity of the FXT, assuming it is at the distance of the candidate host, are given.}
\label{tab:results}
\begin{tabular}{ccccccc}
\hline
\multicolumn{1}{|p{1.5cm}|}{\centering Target}
&\multicolumn{1}{|p{1.5cm}|}{\centering z$_{spec}$ } 
&\multicolumn{1}{|p{1cm}|}{\centering Angular offset (\arcsec) }
&\multicolumn{1}{|p{1.5cm}|}{\centering Physical offset (kpc) } 
&\multicolumn{1}{|p{1.7cm}|}{\centering  SFR ($\msun yr^{-1}$) } 
&\multicolumn{1}{|p{1.5cm}|}{\centering  Log($M_{*}[\msun]$) }
&\multicolumn{1}{|p{2.5cm}|}{\centering Peak X-ray luminosity of the FXT (\lum)  } \\

\hline

		\xa$^{\star}$&  0.326  $\pm$   0.004 &  -- & -- & -- & -- &  6 $\times$ 10$^{44}$\\[1mm] 
		 \xa{}$^{\star}$ &  0.645 $\pm$ 0.0007  &  -- & -- & --& -- & 3 $\times$ 10$^{45}$  \\[1mm] 
            \xc & 0.281  $\pm$   0.003   &  0.8 $\pm$ 0.8 & 3.5$\pm$ 3.5 & 0.9 $\pm$ 0.3 & 8.1 $\pm$ 0.2 & 5 $\times$ 10$^{44}$ \\[1mm] 
            \xd & 0.584 $\pm$ 0.009 &    4.9 $\pm$ 1.3  & 33.3 $\pm$ 8.9 & 5.3 $\pm$ 3.4 & 10.4 $\pm$ 0.1 & 2 $\times$ 10$^{45}$   \\[1mm] 
            \xe & 0.51 $\pm$ 0.01 &  1.2 $\pm$ 1.8 & 22.1 $\pm$ 7.6 & 11.3 $\pm$ 2.6 & 10.8 $\pm$ 0.1 & 3 $\times$ 10$^{44}$ \\
            [1mm] 
            \xg{} & 0.0928 $\pm$ 0.0002 &  0.6 $\pm$ 1.8 & 1.1 $\pm$ 3.2 &  $0.54 \pm 0.02$ & 8.19 $\pm$ 0.05 & 2 $\times$ 10$^{43}$ \\
            [1mm] 
\hline
\end{tabular}
\end{center}
\begin{flushleft}
$^{\star}$ We provide two entries for \xa{}, one for each of the two candidate hosts within  the 2$\sigma$ X-ray position (see Section~\ref{sec:redshift_161028}).
\end{flushleft}

\end{table*}

\section{Discussion} 

We obtained a spectrum of the candidate host of \xb{} and found it to be consistent with that of a K star (see Figure \ref{fig:M2}). Therefore, we conclude that \xb{} might be an K-dwarf flare.
\citet{Alp2020} also mention \xb{} as a possible misidentified Galactic foreground source. Given the source density of objects as bright as or brighter than the candidate host around \xb{}, there is only a 0.6 per cent chance to find an object close to the FXT position by chance, re-enforcing that the X-ray flare was likely due to a flare on the K star.
\citet{2012ApJ...753...30S} shows the dusty torus of an AGN will typically produce a W1-W2 $\geq$ 0.8, where W1 and W2 are Wide-Field Infrared Survey Explorer (WISE) filters at 3.4 and 4.6 $\mu$m, respectively. We calculate the WISE colours W1-W2 $<$ 0.23, which suggests that the candidate host is unlikely to be an AGN.
To compute the ratio $\log(L_X/L_{\rm bol}){=}\log(F_X/F_{\rm bol})$, we normalize stellar synthetic models of dwarf stars \citep[taken from][]{Phillips2020} to the photometric detections of XRT~140811 in the $r-$band, and compute bolometric fluxes by integrating the normalized models at optical/NIR wavelengths. We obtained $\log(F_X/F_{\rm bol})=-3.42$ for \xb{}, where $L_X$ and $L_{\rm bol}$ are the X-ray flare and average (non-flare) bolometric luminosities. Dwarf stars typically exhibit ratios not larger than $\log(L_X/L_{\rm bol}){\lesssim}-3.0$ (the dwarf star flare saturation limit; e.g., \citealt{Garcia2008,DeLuca2020}). This check also confirms that \xb{} is likely due to a stellar flare.

Therefore, out of the 7 FXTs discussed in this paper and 12 reported by \citet{Alp2020}, in total, we exclude \xb{} as an FXT through our observations, reducing the total FXT count to 11. Among the remaining six FXTs observed by us, we do not have sufficient spectroscopic/photometric information on  \xf{}. We discuss the five other FXTs in detail in the next Sections.

We calculated the offset between the centre of the candidate host galaxy and the centre of the FXT localisation uncertainty region (see  Table~\ref{tab:results}).
If we assume the X-ray uncertainty follows a Gaussian distribution, the distances from the origin for an offset source with Gaussian positional uncertainty follow a Ricean distribution (e.g., \citealt{ 2002AJ....123.1111B}). The uncertainty in the offsets are dominated by the uncertainty in the  localisation of {\it XMM-Newton} FXTs. Hence, the  distribution of  the candidate host offsets, considering the 1$\sigma$ uncertainty, does not give much information for comparison with the cumulative distribution of galactocentric offset in kpc seen in SGRBs (\citealt{2022arXiv220601763F}), LGRBs (\citealt{2017MNRAS.467.1795L}), superluminous supernovae  (SL-SNe; \citealt{2021ApJS..255...29S}), Type Ia supernovae ( \citealt{2020ApJ...901..143U}) and core-collapse supernovae (\citealt{2012ApJ...759..107K}, \citealt{2021ApJS..255...29S}).

\begin{figure*} 
    \centering
	\includegraphics[scale=0.60]{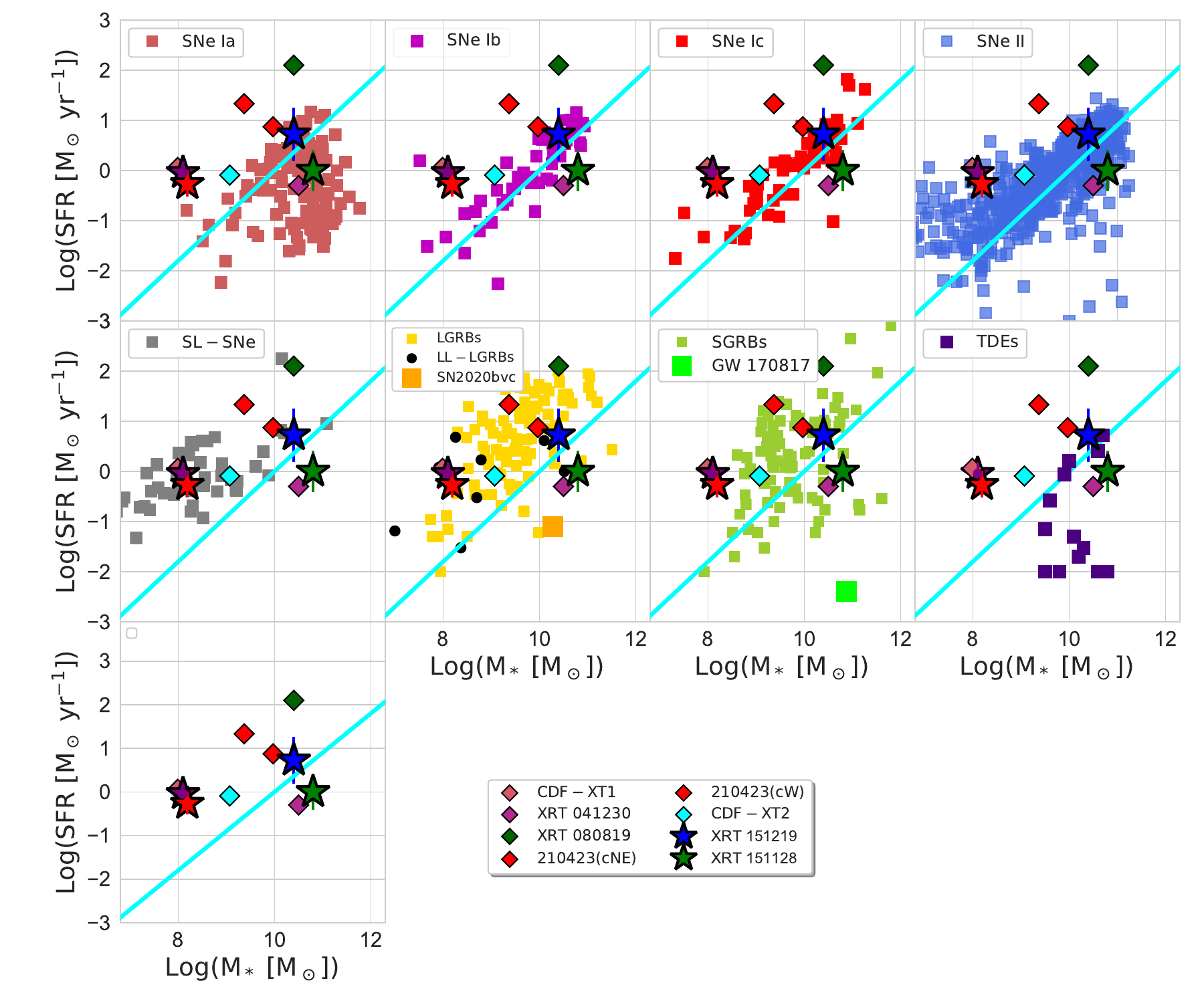}
       \caption{ M$_{*}$ and SFR of the candidate host galaxies (coloured stars) compared with host galaxies of other transient events.  Different panels show the host properties of the previously reported FXTs (\citealt{2022A&A...663A.168Q}; \citealt{Bauer}; \citealt{Xue2019}; \citealt{2023arXiv230301857E}), of LGRBs, SGRBs, low-luminosity LGRBs (see the references in figure 13 of \citealt{2022A&A...663A.168Q}; \citealt{2022arXiv220601763F}), SN Ia, SN Ib, SN Ic and SN II (\citealt{2014A&A...572A..38G}, \citealt{2021ApJS..255...29S}), GW\,170817(\citealt{Im2017}), and TDEs (\citealt{2020SSRv..216...32F}). The solid cyan lines show the best-fit local galaxy main
sequence relation from \citet{2010ApJ...721..193P}.  }
\label{fig:sfr}
\end{figure*}

Figure~\ref{fig:sfr} shows the stellar mass and SFR values of the candidate host galaxies compared with host galaxies of other FXT candidates (\citealt{2022A&A...663A.168Q}), CDFS-XT1(\citealt{Bauer}), CDFS-XT2 (\citealt{Xue2019}) and  XRT~210423 (\citealt{2023arXiv230301857E}). The host galaxies of the other transient  events, such as LGRBs, SGRBs, and low-luminosity LGRBs (see the references in figure 13 of \citealt{2022A&A...663A.168Q}; \citealt{2022arXiv220601763F}), SN Ia, SN Ib, SN Ic and SN II, SL-SNe, (\citealt{2014A&A...572A..38G}, \citealt{2021ApJS..255...29S}), TDEs (\citealt{2020SSRv..216...32F}) and GW\,170817 (\citealt{Im2017}) are denoted by coloured squares.

The determination of redshift/distance to the hosts of the FXTs allows us to convert the observed FXT (peak) flux to (peak) luminosity. The peak luminosity of the FXT is a powerful tool in distinguishing between the various proposed progenitor models of the FXTs. The progenitor models include BNS mergers ($L_{\rm X,peak}\approx 10^{44}$--10$^{51}$~erg~s$^{-1}$; \citealt{2014ARA&A..52...43B}), WD-IMBH TDEs ($L_{\rm{X,peak}} \approxlt 10^{48}$ erg s$^{-1}$; \citealt{2020SSRv..216...39M}; beamed emission), and SN SBO ($L_{\rm{X,peak}} \approxlt 10^{44}$ erg s$^{-1}$; \citealt{2008Natur.453..469S}; \citealt{2017hsn..book..967W}; \citealt{2022ApJ...933..164G}). We discuss if the {\it XMM-Newton}-discovered FXTs studied here are due to a SN SBO as suggested by \citet{Alp2020}. Using our inferred host properties, we also discuss other progenitor models for the FXTs.

For an SN SBO, we expect associated SNe emission to become detectable after some rise time delay.  The rise times ($t_r$) for various SNe are as follows: Type II SNe have median rise times of 7.5 $\pm$ 0.3 days in $g$ (\citealt{2015MNRAS.451.2212G}), whereas Type Ia SNe have a  mean rise time of 18.98 $\pm$ 0.54 days (\citealt{2015MNRAS.446.3895F}) and Type Ic SNe have an average $r$-filter rise time of 15 $\pm$ 6 days (\citealt{2019A&A...621A..71T}).  Type Ia SNe have a peak absolute magnitude ranging between -18
and -20 in the $B$-filter, while for Type Ic, it ranges between -15 and -20 mag in the $B$-filter. For  Type II SNe, the distribution of peak absolute magnitude is wider, varying from -14 to -21 in the $B$-filter (\citealt{2014AJ....147..118R}).

\begin{figure*} 
    \centering

     \includegraphics[scale=0.56]{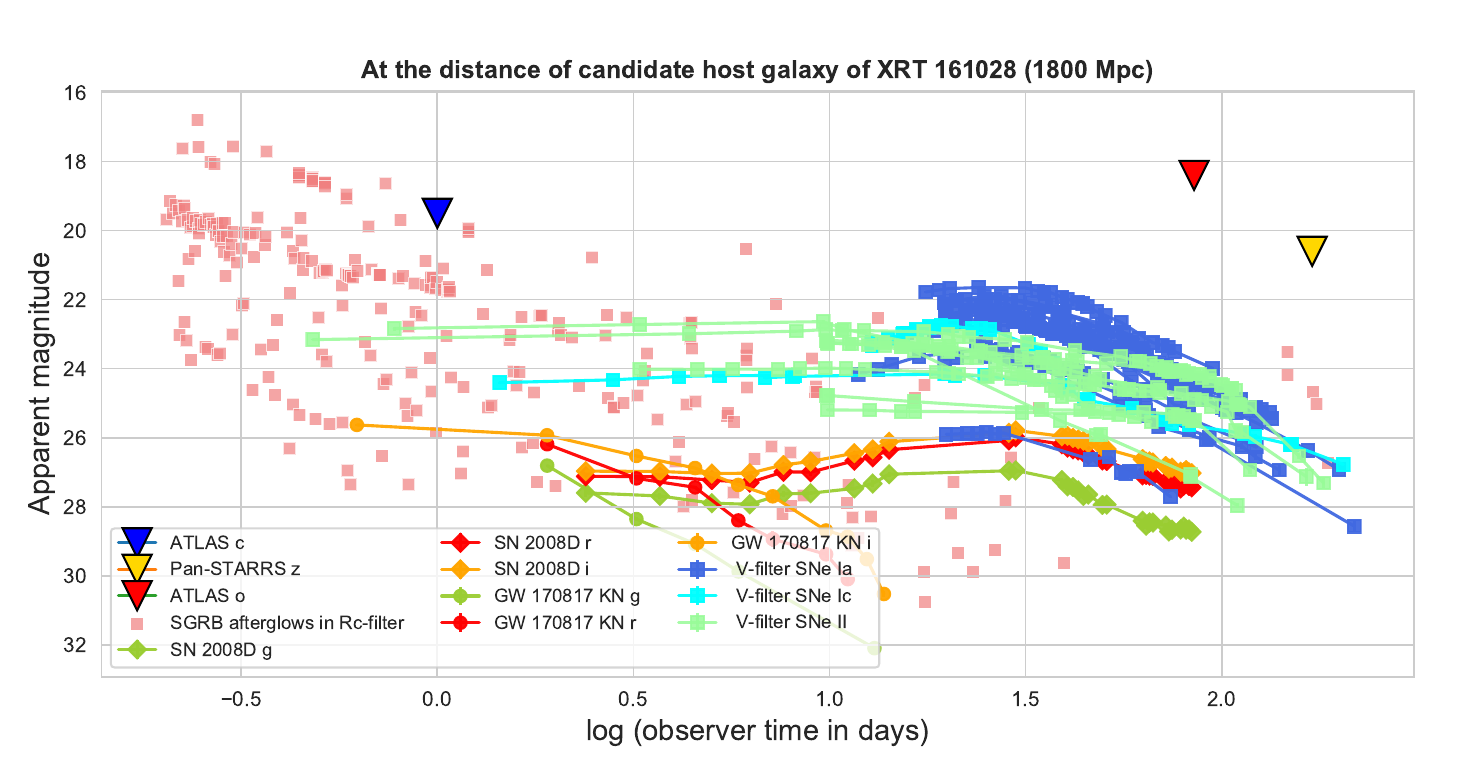}
        \vspace{0.1cm}
	\includegraphics[scale=0.55]{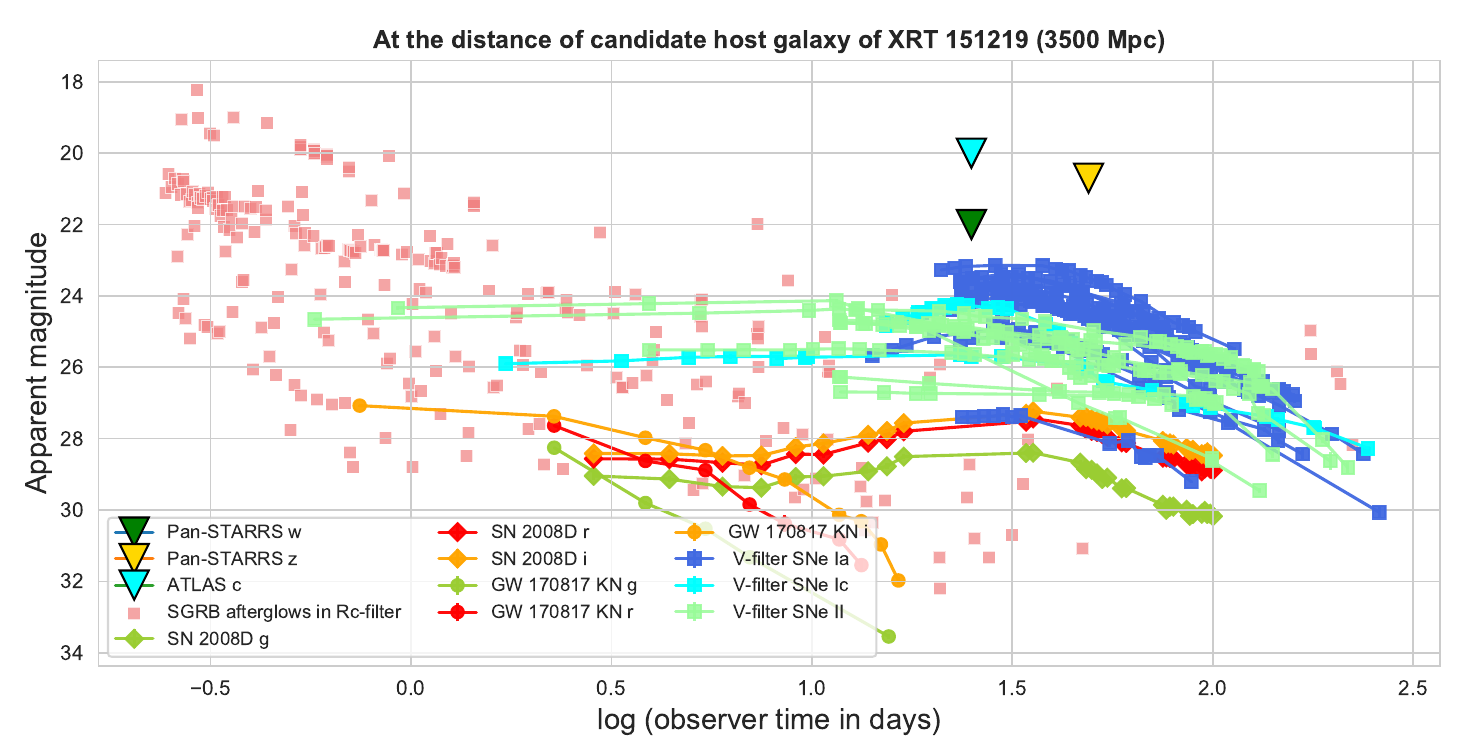}
        \vspace{0.1cm}
    
     \includegraphics[scale=0.60]{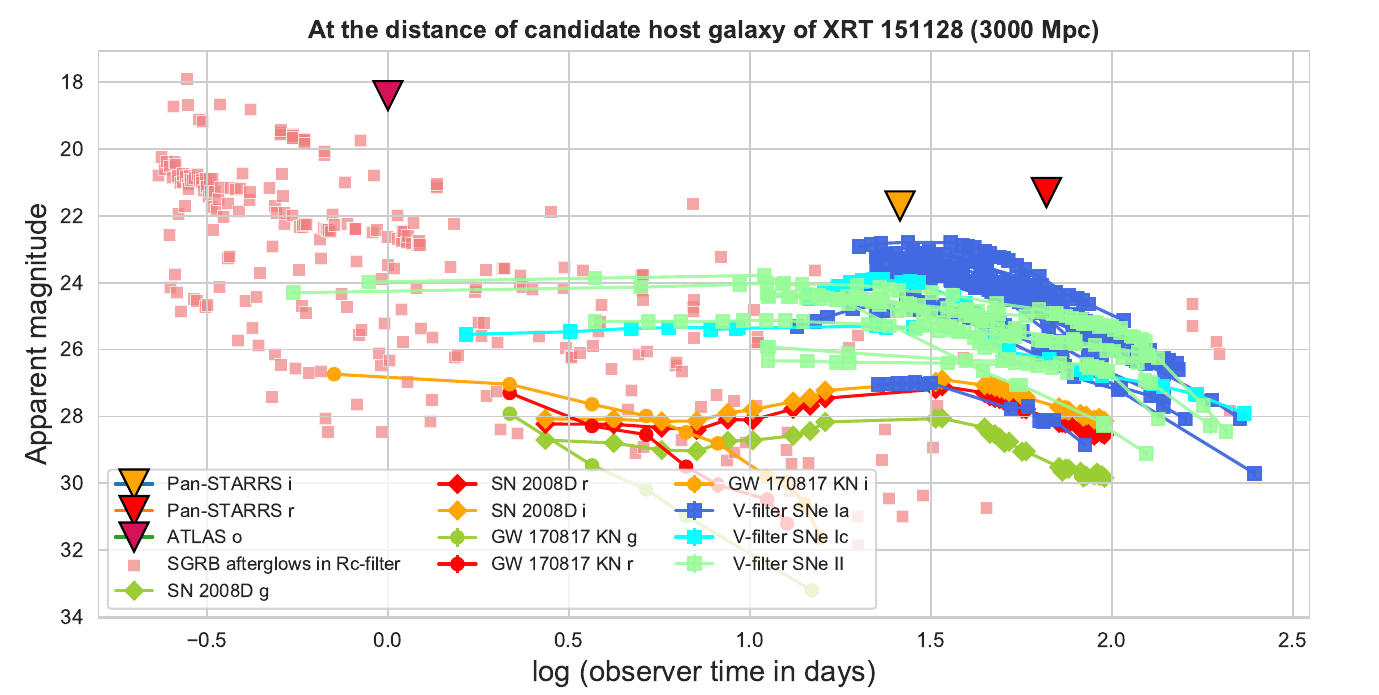}
        \vspace{0.1cm}
        \caption{We compare our optical limits on contemporaneous optical counterpart emission to three FXTs (shown with coloured triangles) with the light curve of  optical signals related to different progenitor models for FXTs. We compare with the light curve of the kilonova associated with GW~170817 observed in different filters (GW~170817 KN, g/r/i; \citealt{2017ApJ...848L..17C}; shown in circles), with different filter light curves of the supernova associated with the SBO, SN~2008D (\citealt{2008Natur.453..469S}; diamond symbols), with the light curves of a sample of Type II SNe (\citealt{2017ApJS..233....6H}), with a sample of Type Ia SNe (\citealt{2006AJ....131..527J}; \citealt{2008AJ....135.1598M}), and with the two Type Ic SNe, SN~2006aj (\citealt{2006ApJ...645L..21M}) and SN~1997ef (\citealt{2000ApJ...534..660I}; coloured squares). Observations in different filters are marked using different colours. Unfortunately, our forced photometry results on the detection of a contemporaneous optical counterpart do not rule out any of the optical signals for the progenitor scenarios under discussion in this work.}
    \label{fig:OL}
\end{figure*}

In Figure~\ref{fig:OL} we compare our limits
on contemporaneous optical counterparts for three of the FXTs, \xa{}, \xd{} and \xe{} (see Table~\ref{tab:LM}) with the light curves of GW~170817 \citep[][$g$ and $r$-filters]{2017ApJ...848L..17C}, SN~2008D \citep[][$g$, $r$ and $i$ filters]{2008Natur.453..469S}, a sample of Type II SNe \citep[][$V$-filter]{2017ApJS..233....6H}, a sample of Type Ia SNe (\citealt{2006AJ....131..527J}; \citealt{2008AJ....135.1598M}, $V$-filter), and the light curves of two Type Ic SNe, SN~2006aj (\citealt{2006ApJ...645L..21M}) and SN~1997ef (\citealt{2000ApJ...534..660I}, both $V$-filter). The comparison samples are first moved into their rest frames before being shifted to the redshift of each FXT. We make no $K$-correction, but note that the difference in rise times resulting from the $K$-correction at these redshifts will not be large. Because the explosion time is unknown for the comparison SNe, the SNe light curves are placed such that the peak occurs at time $t_r$ after the FXT.

The clinching evidence for an FXT and BNS merger association would be the detection of a kilonova signal associated with an FXT. Kilonovae are faint, fast-evolving sources powered by radioactive decay from the lanthanides that are formed in the neutron-rich material ejected during the BNS merger (\citealt{2012ApJ...746...48M}). Lanthanide material has very high opacities (\citealt{2013ApJ...775...18B}); therefore, they emit little in the optical on timescales beyond a few days and so are expected to be very red sources after a few days. On timescales of weeks, emission from the atypical Type Ia SN that may accompany the white dwarf disruption is expected to be rather blue with absolute $u$ to $i$-filter magnitudes ranging from $M_U$=-16\ to\ -19\ and $M_I$=-17\ to\ -18\ (see figure 4 in \citealt{2016ApJ...819....3M}). The light curve peaks on a time scale of $\sim$1--2 weeks (rest frame).

\subsection{\xa{}}

\begin{figure} 
    \centering
	\includegraphics[scale=0.50]{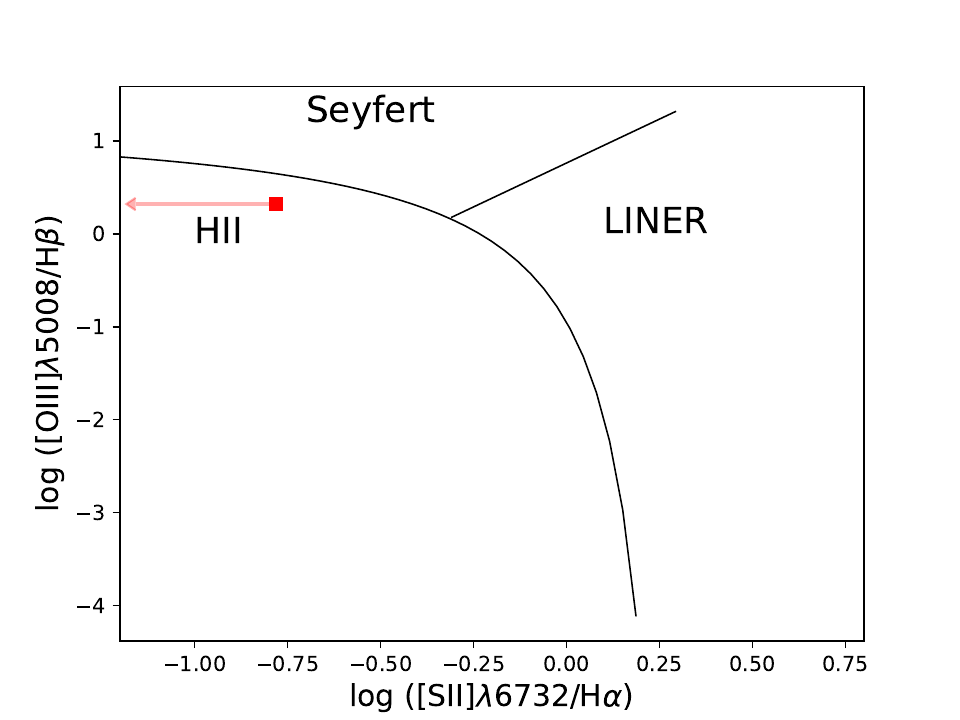}
        \caption{Baldwin-Phillips-Terlevich (BPT) diagram (\citealt{1981PASP...93....5B})
using the flux measurements for the emission lines detected in the spectrum of cN. We used the lines  H$\alpha$, H$\beta$, [OIII] and SII for calculating the position of cN in the BPT diagram. The position of cN falls within the area in the BPT diagram occupied by H~II regions. The size of the symbol is larger than the size of the measurement error in the emission line flux ratio $\log({\rm [OIII]/H_\beta})$.}
    \label{fig:BPT}
\end{figure}

We find that the apparent candidate host is a blend of two sources (see Section~\ref{sec:redshift_161028}). The spectroscopic redshift of the first candidate host galaxy (cN) is $z_\mathrm{spec}$ = 0.326 $\pm$ 0.004, which implies a luminosity distance of $\sim$1300 Mpc considering the cosmology\footnote{We use https://www.astro.ucla.edu/wright/CosmoCalc.html} mentioned in the Section~\ref{sec:intro}. If \xa{} is associated with this candidate host galaxy, it had a peak luminosity (0.3--10 keV) of 6 $\times$ 10$^{44}$ \lum. This peak X-ray luminosity is too bright for an SN SBO origin. 
Our second candidate host (cZ) has a redshift $z_\mathrm{spec}$ = 0.645 $\pm$ 0.007. At this redshift, the FXT had a peak X-ray luminosity of  3 $\times$ 10$^{45}$ \lum,  firmly ruling out an SN SBO association. There is only a 0.7 per cent chance of finding an object as bright as or brighter than the candidate host so close to \xa{} X-ray position by chance. Therefore, we consider it likely that the FXT is associated with either cN or cZ. For cN, using the emission lines H$\alpha$, H$\beta$, [OIII] and upper limits on the presence of the SII lines, we calculate the position of
the cN in a Baldwin-Phillips-Terlevich (BPT) diagram (\citealt{1981PASP...93....5B}). The position of cN falls within the area in the BPT diagram occupied by H~II regions (see Figure \ref{fig:BPT}; \citealt{2001ApJ...556..121K}; \citealt{2003MNRAS.346.1055K}; \citealt{2006MNRAS.372..961K}).

No transient optical light is associated with \xa{} in Pan-STARRS and ATLAS data to 5$\sigma$ limits. Figure~\ref{fig:OL} shows the optical limits obtained from the forced photometry compared with the light curves of GW~170817, SN200D, CC-SNe, Type~Ia and Type~Ic SNe, considering the distance of cN (luminosity distance of $\sim$ 1800 Mpc). We show the magnitude upper limits in ATLAS~$c$ (approximately $g$ + $r$-filters), $o$ (approximately $r$ + $i$-filters) and Pan-STARRS $z$.  If any of these transients was at the redshift of cN, our upper limits are such that our observations would not have yielded in a detection of such a transient. In most cases, we are comparing similar filters. Although the $z$-filter is significantly redder than the comparison filters, even assuming an extremely blue transient (e.g., AT2018cow; \citealt{2018ApJ...865L...3P}), $r-z$  is only $\sim$ 0.5. Therefore, the $z$-filter limit is also unconstraining.

\subsection{\xc{}} 
\xc{} has a peak X-ray luminosity (0.3--10 keV) of 5 $\times$ 10$^{44}$ \lum\,at a redshift of $z_\mathrm{spec}$ = 0.281 $\pm$ 0.003. This is above the peak X-ray luminosity of an SN SBO.
The inferred SFR and  M$_*$ for the candidate host of \xc{} is similar to that of the host population of Type II SNe and SL-SNe.

The candidate host of the FXT \xc{} has a chance alignment probability of 0.4 per cent. 
The peak X-ray luminosity and the inferred SFR and  M$_*$ for the candidate host of \xc{} suggest that BNS merger origin for this FXT is plausible, while the SN SBO scenario seems unlikely considering the peak X-ray luminosity.

\subsection{\xd{}} 
\xd{} is associated with the candidate host galaxy at a redshift of $z_\mathrm{spec}$= 0.584 $\pm$ 0.009 (luminosity distance of $\sim$3500 Mpc), it would have a peak X-ray luminosity (0.3--10 keV) of 2 $\times$ 10$^{45}$ \lum. An SBO origin of the \xd{} is therefore ruled out (\citealt{2008Natur.453..469S}; \citealt{2017hsn..book..967W}; \citealt{2022ApJ...933..164G}), but such a peak luminosity is consistent with that predicted for BNS mergers (\citealt{2014ARA&A..52...43B}). The host parameters such as SFR and M$_*$ (blue star in Figure~\ref{fig:sfr}) obtained through the SED fitting for \xd{} are consistent with that of a SGRB or SN host.

For \xd{}, we find no contemporaneous transient counterpart  to the 5$\sigma$ limit in the Pan-STARRS $w$ (approximately $g$ + $r$ + $i$-filters) and $z$ and ATLAS $c$-filter data. The {\it middle panel} of Figure~\ref{fig:OL} shows the optical limits in comparison with the other transient light curves if they were at the distance of the candidate host galaxy. The upper limits in all the filters are unconstraining.

Given the chance alignment probability of 5.8\%, we also consider a scenario where the candidate host galaxy is close to the FXT position by chance. Assuming the distance of the candidate host galaxy at $\sim$3500 Mpc or further, we derive an upper absolute magnitude limit of $>$ -19.5 from the limiting magnitude $m_r$= 23.2 of the Pan-STARRS image. If at the distance of the candidate host galaxy, we cannot discard a companion dwarf galaxy host (these absolute magnitudes are similar to that of the Large Magellanic Cloud; \citealt{2022NatAs...6..911B}).   
Therefore, we cannot exclude a BNS merger or a WD TDE origin, while the SN SBO is ruled out due to the high peak X-ray luminosity.

 However, we find an X-ray flare in {\it XMM-Newton} serendipitous catalog within the 3$\sigma$ X-ray region of \xd{} centred at (R.A.,Dec.)=(173.$^{\circ}$53113, 0.$^{\circ}$87498) with a 1$\sigma$ positional uncertainty of 3.6\arcsec{} on 2017 June 5. This X-ray flare has a flux of  $\sim$ 4.2 $\times$ 10$^{-14}$ erg cm$^{-2}$ s$^{-1}$. \citet{Alp2020} also indicates the presence of quiescent emission before and after the \xd{} observation. Hence an active galactic nucleus (AGN) variability scenario for \xd{} also cannot be ruled out.

\subsection{\xe{}} 
The spectroscopic redshift of the candidate host galaxy of \xe{} is $z_\mathrm{spec}$ = 0.509 $\pm$ 0.009 such that the FXT has a peak X-ray luminosity (0.3--10 keV) of   3 $\times$ 10$^{44}$ \lum, which is slightly above the peak luminosity predicted by the SN SBO models (\citealt{2017hsn..book..967W}; \citealt{2022ApJ...933..164G}). The broad emission lines of the candidate host suggest it is an AGN.  The large
[OIII]/H$_\mathrm{\beta}$ ratio is also indicative of AGN activity (\citealt{1981PASP...93....5B}). We calculate the WISE colours W1-W2 = 0.16 $\pm$ 0.14 and W2-W3 = 3.54 $\pm$ 0.37 (W3 band at 12$\mu$m). There are a variety of reasons why an apparently active galaxy might have inactive mid-infrared colours. Foremost, it is well established that only higher luminosity AGN will have the torus emission dominate over the galaxy stellar emission,
causing redder mid-infrared colours (e.g., \citealt{2002ApJ...579L..71M}, \citealt{2010ApJ...708..584E}). In addition, due to light travel time, a variable AGN that has turned off will not have all of its AGN diagnostic features respond simultaneously (e.g., \citealt{2022ApJ...936..162S}). Finally, \citet{2019ApJ...876...50L} note that different torus dust properties, including an absence of a torus, might explain the failure of WISE colour selection to identify an AGN. The candidate host galaxy occupies the same SFR vs.~M$_*$ parameter space as most SNe~Ia, SNe~II host but very few SGRB hosts fall in this region of that parameter space.

We did not detect any counterpart (5$\sigma$) for \xe{} in Pan-STARRS  and ATLAS data to a magnitude limit of 21.7 ($i$-filter) and 21.3 ($r$-filter), 26 and 66 days  after the FXT, respectively (see Table \ref{tab:LM}). We also have an upper limit from ATLAS $o$-filter observations a day after the FXT. However, at the redshift of $z$=0.51, our upper limits are not deep enough to find a counterpart associated with the FXT \xe{} for the models we consider here (see  the {\it bottom} panel of Figure~\ref{fig:OL}).

We compute the probability of chance alignment of the candidate host galaxy with respect to the X-ray position to be 0.8 per cent. We also consider a scenario where the candidate host is not associated with the FXT. The 5$\sigma$ stack limiting magnitudes of Pan-STARRS in the $i$-filter is 23.1 mag, which implies a limit of $-$19.3 on the absolute magnitude in the $i$-filter. At this high redshift, we cannot completely discard that there is an unseen dwarf galaxy at the position of the FXT which is the actual host galaxy, as this absolute magnitude limit  would still render the Large Magellanic Cloud undetected (\citealt{2022NatAs...6..911B}). The host properties for \xe{} do not allow us to discard the BNS or the WD-IMBH TDE scenario for \xe{}.

\subsection{\xg{}}
For \xg{} our observations confirm the host candidate properties reported by \citet{Alp2020} and \citet{Novara2020}. We find that \xg{} has a peak (0.3--10 keV) X-ray luminosity of 2 $\times$ 10$^{43}$ \lum\, if at the $z_\mathrm{spec}$ = 0.0928 $\pm$ 0.0002 of the candidate host galaxy. 

From our SED fitting using {\it BAGPIPES} to the 7-filter GROND photometry taken from \cite{Novara2020}, we derive a SFR and M$_*$ of $0.54 \pm 0.02$ $\msun$ yr$^{-1}$ and $\sim$ 1.5 $\times$ 10$^{8}$  $\msun$, respectively for the host of \xg. These values are similar to those for the hosts of Type II SNe and SL-SNe (see the red star in Figure~\ref{fig:sfr}). The inferred SFR and M$_*$ are similar to the values reported by \citet{Novara2020} who infer an SFR and  M$_*$ of $\sim$ 1 $\msun$ yr$^{-1}$ and $\sim$ (2--3) $\times$ 10$^{8}$ $\msun$ using Galaxy-Observed-Simulated SED Interactive Program  and $\sim$ 0.1--1 $\msun$ yr$^{-1}$ and $\sim$ (1.5--2) $\times$ 10$^{8}$ $\msun$ using {\it LePHARE} "photometric analysis for redshift estimate code" (see \citealt{Novara2020}). We do not have contemporaneous Pan-STARRS photometric data for this FXT. The peak X-ray luminosity is fainter than typically found in SGRBs, furthermore, the candidate host SFR and stellar mass $M_*$ (Figure \ref{fig:sfr}) instead favours a SN SBO origin. \citet{Novara2020} rules out the possibility of of an possibility of a flare from an AGN for \xg{}.

\section{Conclusions}
We present detailed host studies and a contemporaneous optical counterpart search for 7 of the 12 {\it XMM-Newton} FXTs reported by \citet{Alp2020}. Of the FXTs discussed in this
work, one (\xb{}) is likely due to a flare from a Galactic late-type star. \xg{} is consistent with being due to an SN SBO event. Spectroscopic redshifts for the likely host galaxies for the other four events imply peak X-ray luminosities that are too high to be
consistent with SN SBO events.  We are unable to discard either the BNS or WD-IMBH TDE scenarios for those four events. For \xd{}, we also cannot rule out the AGN flare scenario.

\section*{Acknowledgements}
We thank the anonymous referee for the helpful comments on this manuscript. DE acknowledges discussions with Shubham Srivastava. DMS acknowledges support from the Consejería de Economía, Conocimiento y Empleo del Gobierno de Canarias and the European Regional Development Fund (ERDF) under grant with reference ProID2021010132 (ACCISI/FEDER, UE); as well as support from the Spanish Ministry of Science and Innovation via an Europa Excelencia grant (EUR2021-122010. The work of DS was carried out
at the Jet Propulsion Laboratory, California Institute of Technology, under a contract with NASA. We acknowledge support from 
ANID - Millennium Science Initiative Program - ICN12\_009 (FEB, JQ-V), CATA-BASAL - FB210003 (FEB), and FONDECYT Regular - 1190818 (FEB) and 1200495 (FEB). AI and MER acknowledge support by NWO under grant number 184.034.002.

The Pan-STARRS1 Surveys (PS1) and the PS1 public science archive have been made possible through contributions by the Institute for Astronomy, the University of Hawaii, the Pan-STARRS Project Office, the Max-Planck Society and its participating institutes, the Max Planck Institute for Astronomy, Heidelberg and the Max Planck Institute for Extraterrestrial Physics, Garching, The Johns Hopkins University, Durham University, the University of Edinburgh, the Queen's University Belfast, the Harvard-Smithsonian Center for Astrophysics, the Las Cumbres Observatory Global Telescope Network Incorporated, the National Central University of Taiwan, the Space Telescope Science Institute, the National Aeronautics and Space Administration under Grant No. NNX08AR22G issued through the Planetary Science Division of the NASA Science Mission Directorate, the National Science Foundation Grant No. AST-1238877, the University of Maryland, Eotvos Lorand University (ELTE), the Los Alamos National Laboratory, and the Gordon and Betty Moore Foundation.

This work has made use of data from the Asteroid Terrestrial-impact Last Alert System (ATLAS) project. The Asteroid Terrestrial-impact Last Alert System (ATLAS) project is primarily funded to search for near earth asteroids through NASA grants NN12AR55G, 80NSSC18K0284, and 80NSSC18K1575; byproducts of the NEO search include images and catalogs from the survey area. This work was partially funded by Kepler/K2 grant J1944/80NSSC19K0112 and HST GO-15889, and STFC grants ST/T000198/1 and ST/S006109/1. The ATLAS science products have been made possible through the contributions of the University of Hawaii Institute for Astronomy, the Queen’s University Belfast, the Space Telescope Science Institute, the South African Astronomical Observatory, and The Millennium Institute of Astrophysics (MAS), Chile.

SDSS-IV acknowledges support and 
resources from the Center for High 
Performance Computing  at the 
University of Utah. The SDSS 
website is www.sdss.org. SDSS-IV is managed by the 
Astrophysical Research Consortium 
for the Participating Institutions 
of the SDSS Collaboration including 
the Brazilian Participation Group, 
the Carnegie Institution for Science, 
Carnegie Mellon University, Center for 
Astrophysics | Harvard \& 
Smithsonian, the Chilean Participation 
Group, the French Participation Group, 
Instituto de Astrof\'isica de 
Canarias, The Johns Hopkins 
University, Kavli Institute for the 
Physics and Mathematics of the 
Universe (IPMU) / University of 
Tokyo, the Korean Participation Group, 
Lawrence Berkeley National Laboratory, 
Leibniz Institut f\"ur Astrophysik 
Potsdam (AIP),  Max-Planck-Institut 
f\"ur Astronomie (MPIA Heidelberg), 
Max-Planck-Institut f\"ur 
Astrophysik (MPA Garching), 
Max-Planck-Institut f\"ur 
Extraterrestrische Physik (MPE), 
National Astronomical Observatories of 
China, New Mexico State University, 
New York University, University of 
Notre Dame, Observat\'ario 
Nacional / MCTI, The Ohio State 
University, Pennsylvania State 
University, Shanghai 
Astronomical Observatory, United 
Kingdom Participation Group, 
Universidad Nacional Aut\'onoma 
de M\'exico, University of Arizona, 
University of Colorado Boulder, 
University of Oxford, University of 
Portsmouth, University of Utah, 
University of Virginia, University 
of Washington, University of 
Wisconsin, Vanderbilt University, 
and Yale University. This research has made use of the CfA Supernova Archive, which is funded in part by the National Science Foundation through grant AST 0907903.

\section*{Data Availability}
All data will be made available in a reproduction package uploaded to Zenodo.



\bibliographystyle{mnras}
\bibliography{example} 

\begin{thebibliography}{}
\makeatletter
\relax
\def\mn@urlcharsother{\let\do\@makeother \do\$\do\&\do\#\do\^\do\_\do\%\do\~}
\def\mn@doi{\begingroup\mn@urlcharsother \@ifnextchar [ {\mn@doi@} {\mn@doi@[]}}
\def\mn@doi@[#1]#2{\def\@tempa{#1}\ifx\@tempa\@empty \href {http://dx.doi.org/#2} {doi:#2}\else \href {http://dx.doi.org/#2} {#1}\fi \endgroup}
\def\mn@eprint#1#2{\mn@eprint@#1:#2::\@nil}
\def\mn@eprint@arXiv#1{\href {http://arxiv.org/abs/#1} {{\tt arXiv:#1}}}
\def\mn@eprint@dblp#1{\href {http://dblp.uni-trier.de/rec/bibtex/#1.xml} {dblp:#1}}
\def\mn@eprint@#1:#2:#3:#4\@nil{\def\@tempa {#1}\def\@tempb {#2}\def\@tempc {#3}\ifx \@tempc \@empty \let \@tempc \@tempb \let \@tempb \@tempa \fi \ifx \@tempb \@empty \def\@tempb {arXiv}\fi \@ifundefined {mn@eprint@\@tempb}{\@tempb:\@tempc}{\expandafter \expandafter \csname mn@eprint@\@tempb\endcsname \expandafter{\@tempc}}}

\bibitem[\protect\citeauthoryear{{Abdurro'uf} et~al.,}{{Abdurro'uf} et~al.}{2022}]{2022ApJS..259...35A}
{Abdurro'uf} et~al., 2022, \mn@doi [\apjs] {10.3847/1538-4365/ac4414}, \href {https://ui.adsabs.harvard.edu/abs/2022ApJS..259...35A} {259, 35}

\bibitem[\protect\citeauthoryear{{Alp} \& {Larsson}}{{Alp} \& {Larsson}}{2020}]{Alp2020}
{Alp} D.,  {Larsson} J.,  2020, \mn@doi [\apj] {10.3847/1538-4357/ab91ba}, \href {https://ui.adsabs.harvard.edu/abs/2020ApJ...896...39A} {896, 39}

\bibitem[\protect\citeauthoryear{{Baldwin}, {Phillips}  \& {Terlevich}}{{Baldwin} et~al.}{1981}]{1981PASP...93....5B}
{Baldwin} J.~A.,  {Phillips} M.~M.,   {Terlevich} R.,  1981, \mn@doi [\pasp] {10.1086/130766}, \href {https://ui.adsabs.harvard.edu/abs/1981PASP...93....5B} {93, 5}

\bibitem[\protect\citeauthoryear{{Barnes} \& {Kasen}}{{Barnes} \& {Kasen}}{2013}]{2013ApJ...775...18B}
{Barnes} J.,  {Kasen} D.,  2013, \mn@doi [\apj] {10.1088/0004-637X/775/1/18}, \href {https://ui.adsabs.harvard.edu/abs/2013ApJ...775...18B} {775, 18}

\bibitem[\protect\citeauthoryear{{Bauer} et~al.,}{{Bauer} et~al.}{2017}]{Bauer}
{Bauer} F.~E.,  et~al., 2017, \mn@doi [\mnras] {10.1093/mnras/stx417}, \href {https://ui.adsabs.harvard.edu/abs/2017MNRAS.467.4841B} {467, 4841}

\bibitem[\protect\citeauthoryear{{Belokurov} \& {Evans}}{{Belokurov} \& {Evans}}{2022}]{2022NatAs...6..911B}
{Belokurov} V.,  {Evans} N.~W.,  2022, \mn@doi [Nature Astronomy] {10.1038/s41550-022-01740-w}, \href {https://ui.adsabs.harvard.edu/abs/2022NatAs...6..911B} {6, 911}

\bibitem[\protect\citeauthoryear{{Berger}}{{Berger}}{2014}]{2014ARA&A..52...43B}
{Berger} E.,  2014, \mn@doi [\araa] {10.1146/annurev-astro-081913-035926}, \href {https://ui.adsabs.harvard.edu/abs/2014ARA&A..52...43B} {52, 43}

\bibitem[\protect\citeauthoryear{{Bertin} \& {Arnouts}}{{Bertin} \& {Arnouts}}{1996}]{1996A&AS..117..393B}
{Bertin} E.,  {Arnouts} S.,  1996, \mn@doi [Astronomy and Astrophysics Supplement Series] {10.1051/aas:1996164}, \href {https://ui.adsabs.harvard.edu/\#abs/1996A&AS..117..393B} {117, 393}

\bibitem[\protect\citeauthoryear{{Bloom}, {Kulkarni}  \& {Djorgovski}}{{Bloom} et~al.}{2002}]{2002AJ....123.1111B}
{Bloom} J.~S.,  {Kulkarni} S.~R.,   {Djorgovski} S.~G.,  2002, \mn@doi [\aj] {10.1086/338893}, \href {https://ui.adsabs.harvard.edu/abs/2002AJ....123.1111B} {123, 1111}

\bibitem[\protect\citeauthoryear{{Calzetti}, {Armus}, {Bohlin}, {Kinney}, {Koornneef}  \& {Storchi-Bergmann}}{{Calzetti} et~al.}{2000}]{2000ApJ...533..682C}
{Calzetti} D.,  {Armus} L.,  {Bohlin} R.~C.,  {Kinney} A.~L.,  {Koornneef} J.,   {Storchi-Bergmann} T.,  2000, \mn@doi [\apj] {10.1086/308692}, \href {https://ui.adsabs.harvard.edu/abs/2000ApJ...533..682C} {533, 682}

\bibitem[\protect\citeauthoryear{{Campana} et~al.,}{{Campana} et~al.}{2006}]{2006Natur.442.1008C}
{Campana} S.,  et~al., 2006, \mn@doi [\nat] {10.1038/nature04892}, \href {https://ui.adsabs.harvard.edu/abs/2006Natur.442.1008C} {442, 1008}

\bibitem[\protect\citeauthoryear{{Carnall}, {McLure}, {Dunlop}  \& {Dav{\'e}}}{{Carnall} et~al.}{2018}]{2018MNRAS.480.4379C}
{Carnall} A.~C.,  {McLure} R.~J.,  {Dunlop} J.~S.,   {Dav{\'e}} R.,  2018, \mn@doi [\mnras] {10.1093/mnras/sty2169}, \href {https://ui.adsabs.harvard.edu/abs/2018MNRAS.480.4379C} {480, 4379}

\bibitem[\protect\citeauthoryear{{Cepa} et~al.,}{{Cepa} et~al.}{2000}]{2000SPIE.4008..623C}
{Cepa} J.,  et~al., 2000, in {Iye} M.,  {Moorwood} A.~F.,  eds,  Society of Photo-Optical Instrumentation Engineers (SPIE) Conference Series Vol. 4008, Optical and IR Telescope Instrumentation and Detectors. pp 623--631, \mn@doi{10.1117/12.395520}

\bibitem[\protect\citeauthoryear{{Chambers} et~al.,}{{Chambers} et~al.}{2016}]{2016arXiv161205560C}
{Chambers} K.~C.,  et~al., 2016, \mn@doi [arXiv e-prints] {10.48550/arXiv.1612.05560}, \href {https://ui.adsabs.harvard.edu/abs/2016arXiv161205560C} {p. arXiv:1612.05560}

\bibitem[\protect\citeauthoryear{{Cowperthwaite} et~al.,}{{Cowperthwaite} et~al.}{2017}]{2017ApJ...848L..17C}
{Cowperthwaite} P.~S.,  et~al., 2017, \mn@doi [\apjl] {10.3847/2041-8213/aa8fc7}, \href {https://ui.adsabs.harvard.edu/abs/2017ApJ...848L..17C} {848, L17}

\bibitem[\protect\citeauthoryear{{Dai}, {Wang}, {Wu}  \& {Zhang}}{{Dai} et~al.}{2006}]{2006Sci...311.1127D}
{Dai} Z.~G.,  {Wang} X.~Y.,  {Wu} X.~F.,   {Zhang} B.,  2006, \mn@doi [Science] {10.1126/science.1123606}, \href {https://ui.adsabs.harvard.edu/abs/2006Sci...311.1127D} {311, 1127}

\bibitem[\protect\citeauthoryear{{Dalton} et~al.,}{{Dalton} et~al.}{2006}]{2006SPIE.6269E..0XD}
{Dalton} G.~B.,  et~al., 2006, in {McLean} I.~S.,  {Iye} M.,  eds,  Society of Photo-Optical Instrumentation Engineers (SPIE) Conference Series Vol. 6269, Society of Photo-Optical Instrumentation Engineers (SPIE) Conference Series. p. 62690X, \mn@doi{10.1117/12.670018}

\bibitem[\protect\citeauthoryear{{Dark Energy Survey Collaboration} et~al.,}{{Dark Energy Survey Collaboration} et~al.}{2016}]{2016MNRAS.460.1270D}
{Dark Energy Survey Collaboration} et~al., 2016, \mn@doi [\mnras] {10.1093/mnras/stw641}, \href {https://ui.adsabs.harvard.edu/abs/2016MNRAS.460.1270D} {460, 1270}

\bibitem[\protect\citeauthoryear{{De Luca} et~al.,}{{De Luca} et~al.}{2020}]{DeLuca2020}
{De Luca} A.,  et~al., 2020, \mn@doi [\aap] {10.1051/0004-6361/201937163}, \href {https://ui.adsabs.harvard.edu/abs/2020A&A...634L..13D} {634, L13}

\bibitem[\protect\citeauthoryear{{Drlica-Wagner} et~al.,}{{Drlica-Wagner} et~al.}{2018}]{2018ApJS..235...33D}
{Drlica-Wagner} A.,  et~al., 2018, \mn@doi [\apjs] {10.3847/1538-4365/aab4f5}, \href {https://ui.adsabs.harvard.edu/abs/2018ApJS..235...33D} {235, 33}

\bibitem[\protect\citeauthoryear{{Eappachen} et~al.,}{{Eappachen} et~al.}{2022}]{2022MNRAS.514..302E}
{Eappachen} D.,  et~al., 2022, \mn@doi [\mnras] {10.1093/mnras/stac1194}, \href {https://ui.adsabs.harvard.edu/abs/2022MNRAS.514..302E} {514, 302}

\bibitem[\protect\citeauthoryear{{Eappachen} et~al.,}{{Eappachen} et~al.}{2023}]{2023arXiv230301857E}
{Eappachen} D.,  et~al., 2023, \mn@doi [arXiv e-prints] {10.48550/arXiv.2303.01857}, \href {https://ui.adsabs.harvard.edu/abs/2023arXiv230301857E} {p. arXiv:2303.01857}

\bibitem[\protect\citeauthoryear{{Eckart}, {McGreer}, {Stern}, {Harrison}  \& {Helfand}}{{Eckart} et~al.}{2010}]{2010ApJ...708..584E}
{Eckart} M.~E.,  {McGreer} I.~D.,  {Stern} D.,  {Harrison} F.~A.,   {Helfand} D.~J.,  2010, \mn@doi [\apj] {10.1088/0004-637X/708/1/584}, \href {https://ui.adsabs.harvard.edu/abs/2010ApJ...708..584E} {708, 584}

\bibitem[\protect\citeauthoryear{{Edge}, {Sutherland}, {Kuijken}, {Driver}, {McMahon}, {Eales}  \& {Emerson}}{{Edge} et~al.}{2013}]{2013Msngr.154...32E}
{Edge} A.,  {Sutherland} W.,  {Kuijken} K.,  {Driver} S.,  {McMahon} R.,  {Eales} S.,   {Emerson} J.~P.,  2013, The Messenger, \href {https://ui.adsabs.harvard.edu/abs/2013Msngr.154...32E} {154, 32}

\bibitem[\protect\citeauthoryear{{Falk} \& {Arnett}}{{Falk} \& {Arnett}}{1977}]{Falk1977}
{Falk} S.~W.,  {Arnett} W.~D.,  1977, \mn@doi [\apjs] {10.1086/190440}, \href {https://ui.adsabs.harvard.edu/abs/1977ApJS...33..515F} {33, 515}

\bibitem[\protect\citeauthoryear{{Firth} et~al.,}{{Firth} et~al.}{2015}]{2015MNRAS.446.3895F}
{Firth} R.~E.,  et~al., 2015, \mn@doi [\mnras] {10.1093/mnras/stu2314}, \href {https://ui.adsabs.harvard.edu/abs/2015MNRAS.446.3895F} {446, 3895}

\bibitem[\protect\citeauthoryear{{Fong} et~al.,}{{Fong} et~al.}{2022}]{2022arXiv220601763F}
{Fong} W.-f.,  et~al., 2022, arXiv e-prints, \href {https://ui.adsabs.harvard.edu/abs/2022arXiv220601763F} {p. arXiv:2206.01763}

\bibitem[\protect\citeauthoryear{{French}, {Wevers}, {Law-Smith}, {Graur}  \& {Zabludoff}}{{French} et~al.}{2020}]{2020SSRv..216...32F}
{French} K.~D.,  {Wevers} T.,  {Law-Smith} J.,  {Graur} O.,   {Zabludoff} A.~I.,  2020, \mn@doi [\ssr] {10.1007/s11214-020-00657-y}, \href {https://ui.adsabs.harvard.edu/abs/2020SSRv..216...32F} {216, 32}

\bibitem[\protect\citeauthoryear{{Galbany} et~al.,}{{Galbany} et~al.}{2014}]{2014A&A...572A..38G}
{Galbany} L.,  et~al., 2014, \mn@doi [\aap] {10.1051/0004-6361/201424717}, \href {https://ui.adsabs.harvard.edu/abs/2014A&A...572A..38G} {572, A38}

\bibitem[\protect\citeauthoryear{{Ganot} et~al.,}{{Ganot} et~al.}{2016}]{Ganot2016}
{Ganot} N.,  et~al., 2016, \mn@doi [\apj] {10.3847/0004-637X/820/1/57}, \href {https://ui.adsabs.harvard.edu/abs/2016ApJ...820...57G} {820, 57}

\bibitem[\protect\citeauthoryear{{Garc{\'\i}a-Alvarez}, {Drake}, {Kashyap}, {Lin}  \& {Ball}}{{Garc{\'\i}a-Alvarez} et~al.}{2008}]{Garcia2008}
{Garc{\'\i}a-Alvarez} D.,  {Drake} J.~J.,  {Kashyap} V.~L.,  {Lin} L.,   {Ball} B.,  2008, \mn@doi [\apj] {10.1086/587611}, \href {https://ui.adsabs.harvard.edu/abs/2008ApJ...679.1509G} {679, 1509}

\bibitem[\protect\citeauthoryear{{Glennie}, {Jonker}, {Fender}, {Nagayama}  \& {Pretorius}}{{Glennie} et~al.}{2015}]{glennie}
{Glennie} A.,  {Jonker} P.~G.,  {Fender} R.~P.,  {Nagayama} T.,   {Pretorius} M.~L.,  2015, \mn@doi [\mnras] {10.1093/mnras/stv801}, \href {https://ui.adsabs.harvard.edu/\#abs/2015MNRAS.450.3765G} {450, 3765}

\bibitem[\protect\citeauthoryear{{Goldberg}, {Jiang}  \& {Bildsten}}{{Goldberg} et~al.}{2022}]{2022ApJ...933..164G}
{Goldberg} J.~A.,  {Jiang} Y.-F.,   {Bildsten} L.,  2022, \mn@doi [\apj] {10.3847/1538-4357/ac75e3}, \href {https://ui.adsabs.harvard.edu/abs/2022ApJ...933..164G} {933, 164}

\bibitem[\protect\citeauthoryear{{Gonz{\'a}lez-Gait{\'a}n} et~al.,}{{Gonz{\'a}lez-Gait{\'a}n} et~al.}{2015}]{2015MNRAS.451.2212G}
{Gonz{\'a}lez-Gait{\'a}n} S.,  et~al., 2015, \mn@doi [\mnras] {10.1093/mnras/stv1097}, \href {https://ui.adsabs.harvard.edu/abs/2015MNRAS.451.2212G} {451, 2212}

\bibitem[\protect\citeauthoryear{{Greiner} et~al.,}{{Greiner} et~al.}{2008}]{2008PASP..120..405G}
{Greiner} J.,  et~al., 2008, \mn@doi [\pasp] {10.1086/587032}, \href {https://ui.adsabs.harvard.edu/abs/2008PASP..120..405G} {120, 405}

\bibitem[\protect\citeauthoryear{{Hicken} et~al.,}{{Hicken} et~al.}{2017}]{2017ApJS..233....6H}
{Hicken} M.,  et~al., 2017, \mn@doi [\apjs] {10.3847/1538-4365/aa8ef4}, \href {https://ui.adsabs.harvard.edu/abs/2017ApJS..233....6H} {233, 6}

\bibitem[\protect\citeauthoryear{{Im} et~al.,}{{Im} et~al.}{2017}]{Im2017}
{Im} M.,  et~al., 2017, \mn@doi [\apjl] {10.3847/2041-8213/aa9367}, \href {https://ui.adsabs.harvard.edu/abs/2017ApJ...849L..16I} {849, L16}

\bibitem[\protect\citeauthoryear{{Irwin} et~al.,}{{Irwin} et~al.}{2016}]{Irwin}
{Irwin} J.~A.,  et~al., 2016, \mn@doi [\nat] {10.1038/nature19822}, \href {https://ui.adsabs.harvard.edu/abs/2016Natur.538..356I} {538, 356}

\bibitem[\protect\citeauthoryear{{Iwamoto} et~al.,}{{Iwamoto} et~al.}{2000}]{2000ApJ...534..660I}
{Iwamoto} K.,  et~al., 2000, \mn@doi [\apj] {10.1086/308761}, \href {https://ui.adsabs.harvard.edu/abs/2000ApJ...534..660I} {534, 660}

\bibitem[\protect\citeauthoryear{{Jha} et~al.,}{{Jha} et~al.}{2006}]{2006AJ....131..527J}
{Jha} S.,  et~al., 2006, \mn@doi [\aj] {10.1086/497989}, \href {https://ui.adsabs.harvard.edu/abs/2006AJ....131..527J} {131, 527}

\bibitem[\protect\citeauthoryear{{Jonker} et~al.,}{{Jonker} et~al.}{2013}]{2013jonker}
{Jonker} P.~G.,  et~al., 2013, \mn@doi [\apj] {10.1088/0004-637X/779/1/14}, \href {https://ui.adsabs.harvard.edu/\#abs/2013ApJ...779...14J} {779, 14}

\bibitem[\protect\citeauthoryear{{Kauffmann} et~al.,}{{Kauffmann} et~al.}{2003}]{2003MNRAS.346.1055K}
{Kauffmann} G.,  et~al., 2003, \mn@doi [\mnras] {10.1111/j.1365-2966.2003.07154.x}, \href {https://ui.adsabs.harvard.edu/abs/2003MNRAS.346.1055K} {346, 1055}

\bibitem[\protect\citeauthoryear{{Kelly} \& {Kirshner}}{{Kelly} \& {Kirshner}}{2012}]{2012ApJ...759..107K}
{Kelly} P.~L.,  {Kirshner} R.~P.,  2012, \mn@doi [\apj] {10.1088/0004-637X/759/2/107}, \href {https://ui.adsabs.harvard.edu/abs/2012ApJ...759..107K} {759, 107}

\bibitem[\protect\citeauthoryear{{Kewley}, {Dopita}, {Sutherland}, {Heisler}  \& {Trevena}}{{Kewley} et~al.}{2001}]{2001ApJ...556..121K}
{Kewley} L.~J.,  {Dopita} M.~A.,  {Sutherland} R.~S.,  {Heisler} C.~A.,   {Trevena} J.,  2001, \mn@doi [\apj] {10.1086/321545}, \href {https://ui.adsabs.harvard.edu/abs/2001ApJ...556..121K} {556, 121}

\bibitem[\protect\citeauthoryear{{Kewley}, {Groves}, {Kauffmann}  \& {Heckman}}{{Kewley} et~al.}{2006}]{2006MNRAS.372..961K}
{Kewley} L.~J.,  {Groves} B.,  {Kauffmann} G.,   {Heckman} T.,  2006, \mn@doi [\mnras] {10.1111/j.1365-2966.2006.10859.x}, \href {https://ui.adsabs.harvard.edu/abs/2006MNRAS.372..961K} {372, 961}

\bibitem[\protect\citeauthoryear{{Klein} \& {Chevalier}}{{Klein} \& {Chevalier}}{1978}]{Klein1978}
{Klein} R.~I.,  {Chevalier} R.~A.,  1978, \mn@doi [\apjl] {10.1086/182740}, \href {https://ui.adsabs.harvard.edu/abs/1978ApJ...223L.109K} {223, L109}

\bibitem[\protect\citeauthoryear{{LaMassa}, {Georgakakis}, {Vivek}, {Salvato}, {Ananna}, {Urry}, {MacLeod}  \& {Ross}}{{LaMassa} et~al.}{2019}]{2019ApJ...876...50L}
{LaMassa} S.~M.,  {Georgakakis} A.,  {Vivek} M.,  {Salvato} M.,  {Ananna} T.~T.,  {Urry} C.~M.,  {MacLeod} C.,   {Ross} N.,  2019, \mn@doi [\apj] {10.3847/1538-4357/ab108b}, \href {https://ui.adsabs.harvard.edu/abs/2019ApJ...876...50L} {876, 50}

\bibitem[\protect\citeauthoryear{{Levan} et~al.,}{{Levan} et~al.}{2014}]{2014ApJ...781...13L}
{Levan} A.~J.,  et~al., 2014, \mn@doi [\apj] {10.1088/0004-637X/781/1/13}, \href {https://ui.adsabs.harvard.edu/abs/2014ApJ...781...13L} {781, 13}

\bibitem[\protect\citeauthoryear{{Lin}, {Irwin}  \& {Berger}}{{Lin} et~al.}{2021}]{2021ATel14599....1L}
{Lin} D.,  {Irwin} J.~A.,   {Berger} E.,  2021, The Astronomer's Telegram, \href {https://ui.adsabs.harvard.edu/abs/2021ATel14599....1L} {14599, 1}

\bibitem[\protect\citeauthoryear{{Lin}, {Irwin}, {Berger}  \& {Nguyen}}{{Lin} et~al.}{2022}]{2022ApJ...927..211L}
{Lin} D.,  {Irwin} J.~A.,  {Berger} E.,   {Nguyen} R.,  2022, \mn@doi [\apj] {10.3847/1538-4357/ac4fc6}, \href {https://ui.adsabs.harvard.edu/abs/2022ApJ...927..211L} {927, 211}

\bibitem[\protect\citeauthoryear{{Lyman} et~al.,}{{Lyman} et~al.}{2017}]{2017MNRAS.467.1795L}
{Lyman} J.~D.,  et~al., 2017, \mn@doi [\mnras] {10.1093/mnras/stx220}, \href {https://ui.adsabs.harvard.edu/abs/2017MNRAS.467.1795L} {467, 1795}

\bibitem[\protect\citeauthoryear{{MacLeod}, {Guillochon}, {Ramirez-Ruiz}, {Kasen}  \& {Rosswog}}{{MacLeod} et~al.}{2016}]{2016ApJ...819....3M}
{MacLeod} M.,  {Guillochon} J.,  {Ramirez-Ruiz} E.,  {Kasen} D.,   {Rosswog} S.,  2016, \mn@doi [\apj] {10.3847/0004-637X/819/1/3}, \href {https://ui.adsabs.harvard.edu/abs/2016ApJ...819....3M} {819, 3}

\bibitem[\protect\citeauthoryear{{Magnier} et~al.,}{{Magnier} et~al.}{2020}]{2020ApJS..251....3M}
{Magnier} E.~A.,  et~al., 2020, \mn@doi [\apjs] {10.3847/1538-4365/abb829}, \href {https://ui.adsabs.harvard.edu/abs/2020ApJS..251....3M} {251, 3}

\bibitem[\protect\citeauthoryear{{Maguire}, {Eracleous}, {Jonker}, {MacLeod}  \& {Rosswog}}{{Maguire} et~al.}{2020}]{2020SSRv..216...39M}
{Maguire} K.,  {Eracleous} M.,  {Jonker} P.~G.,  {MacLeod} M.,   {Rosswog} S.,  2020, \mn@doi [\ssr] {10.1007/s11214-020-00661-2}, \href {https://ui.adsabs.harvard.edu/abs/2020SSRv..216...39M} {216, 39}

\bibitem[\protect\citeauthoryear{{Marsh}}{{Marsh}}{2019}]{2019ascl.soft07012M}
{Marsh} T.,  2019, {molly: 1D astronomical spectra analyzer} (\mn@eprint {ascl} {1907.012})

\bibitem[\protect\citeauthoryear{{Massey} \& {Gronwall}}{{Massey} \& {Gronwall}}{1990}]{1990ApJ...358..344M}
{Massey} P.,  {Gronwall} C.,  1990, \mn@doi [\apj] {10.1086/168991}, \href {https://ui.adsabs.harvard.edu/abs/1990ApJ...358..344M} {358, 344}

\bibitem[\protect\citeauthoryear{{Matheson} et~al.,}{{Matheson} et~al.}{2008}]{2008AJ....135.1598M}
{Matheson} T.,  et~al., 2008, \mn@doi [\aj] {10.1088/0004-6256/135/4/1598}, \href {https://ui.adsabs.harvard.edu/abs/2008AJ....135.1598M} {135, 1598}

\bibitem[\protect\citeauthoryear{{Matzner} \& {McKee}}{{Matzner} \& {McKee}}{1999}]{Matzner1999}
{Matzner} C.~D.,  {McKee} C.~F.,  1999, \mn@doi [\apj] {10.1086/306571}, \href {https://ui.adsabs.harvard.edu/abs/1999ApJ...510..379M} {510, 379}

\bibitem[\protect\citeauthoryear{{McMahon}, {Banerji}, {Gonzalez}, {Koposov}, {Bejar}, {Lodieu}, {Rebolo}  \& {VHS Collaboration}}{{McMahon} et~al.}{2013}]{2013Msngr.154...35M}
{McMahon} R.~G.,  {Banerji} M.,  {Gonzalez} E.,  {Koposov} S.~E.,  {Bejar} V.~J.,  {Lodieu} N.,  {Rebolo} R.,   {VHS Collaboration} 2013, The Messenger, \href {https://ui.adsabs.harvard.edu/abs/2013Msngr.154...35M} {154, 35}

\bibitem[\protect\citeauthoryear{{Metzger} \& {Berger}}{{Metzger} \& {Berger}}{2012}]{2012ApJ...746...48M}
{Metzger} B.~D.,  {Berger} E.,  2012, \mn@doi [\apj] {10.1088/0004-637X/746/1/48}, \href {https://ui.adsabs.harvard.edu/abs/2012ApJ...746...48M} {746, 48}

\bibitem[\protect\citeauthoryear{{Metzger} \& {Piro}}{{Metzger} \& {Piro}}{2014}]{2014MNRAS.439.3916M}
{Metzger} B.~D.,  {Piro} A.~L.,  2014, \mn@doi [\mnras] {10.1093/mnras/stu247}, \href {https://ui.adsabs.harvard.edu/abs/2014MNRAS.439.3916M} {439, 3916}

\bibitem[\protect\citeauthoryear{{Metzger}, {Quataert}  \& {Thompson}}{{Metzger} et~al.}{2008}]{2008MNRAS.385.1455M}
{Metzger} B.~D.,  {Quataert} E.,   {Thompson} T.~A.,  2008, \mn@doi [\mnras] {10.1111/j.1365-2966.2008.12923.x}, \href {https://ui.adsabs.harvard.edu/abs/2008MNRAS.385.1455M} {385, 1455}

\bibitem[\protect\citeauthoryear{{Modjaz} et~al.,}{{Modjaz} et~al.}{2006}]{2006ApJ...645L..21M}
{Modjaz} M.,  et~al., 2006, \mn@doi [\apjl] {10.1086/505906}, \href {https://ui.adsabs.harvard.edu/abs/2006ApJ...645L..21M} {645, L21}

\bibitem[\protect\citeauthoryear{{Monet} et~al.,}{{Monet} et~al.}{2003}]{2003AJ....125..984M}
{Monet} D.~G.,  et~al., 2003, \mn@doi [\aj] {10.1086/345888}, \href {https://ui.adsabs.harvard.edu/abs/2003AJ....125..984M} {125, 984}

\bibitem[\protect\citeauthoryear{{Moran}, {Filippenko}  \& {Chornock}}{{Moran} et~al.}{2002}]{2002ApJ...579L..71M}
{Moran} E.~C.,  {Filippenko} A.~V.,   {Chornock} R.,  2002, \mn@doi [\apjl] {10.1086/345314}, \href {https://ui.adsabs.harvard.edu/abs/2002ApJ...579L..71M} {579, L71}

\bibitem[\protect\citeauthoryear{{Novara} et~al.,}{{Novara} et~al.}{2020}]{Novara2020}
{Novara} G.,  et~al., 2020, \mn@doi [\apj] {10.3847/1538-4357/ab98f8}, \href {https://ui.adsabs.harvard.edu/abs/2020ApJ...898...37N} {898, 37}

\bibitem[\protect\citeauthoryear{{Oke} et~al.,}{{Oke} et~al.}{1995}]{1995PASP..107..375O}
{Oke} J.~B.,  et~al., 1995, \mn@doi [\pasp] {10.1086/133562}, \href {https://ui.adsabs.harvard.edu/abs/1995PASP..107..375O} {107, 375}

\bibitem[\protect\citeauthoryear{{Peng} et~al.,}{{Peng} et~al.}{2010}]{2010ApJ...721..193P}
{Peng} Y.-j.,  et~al., 2010, \mn@doi [\apj] {10.1088/0004-637X/721/1/193}, \href {https://ui.adsabs.harvard.edu/abs/2010ApJ...721..193P} {721, 193}

\bibitem[\protect\citeauthoryear{{Phillips} et~al.,}{{Phillips} et~al.}{2020}]{Phillips2020}
{Phillips} M.~W.,  et~al., 2020, \mn@doi [\aap] {10.1051/0004-6361/201937381}, \href {https://ui.adsabs.harvard.edu/abs/2020A&A...637A..38P} {637, A38}

\bibitem[\protect\citeauthoryear{{Pickles}}{{Pickles}}{1998}]{1998PASP..110..863P}
{Pickles} A.~J.,  1998, \mn@doi [\pasp] {10.1086/316197}, \href {https://ui.adsabs.harvard.edu/abs/1998PASP..110..863P} {110, 863}

\bibitem[\protect\citeauthoryear{{Planck Collaboration} et~al.,}{{Planck Collaboration} et~al.}{2018}]{2018arXiv180706209P}
{Planck Collaboration} et~al., 2018, arXiv e-prints, \href {https://ui.adsabs.harvard.edu/abs/2018arXiv180706209P} {p. arXiv:1807.06209}

\bibitem[\protect\citeauthoryear{{Prentice} et~al.,}{{Prentice} et~al.}{2018}]{2018ApJ...865L...3P}
{Prentice} S.~J.,  et~al., 2018, \mn@doi [\apjl] {10.3847/2041-8213/aadd90}, \href {https://ui.adsabs.harvard.edu/abs/2018ApJ...865L...3P} {865, L3}

\bibitem[\protect\citeauthoryear{{Quirola-V{\'a}squez} et~al.,}{{Quirola-V{\'a}squez} et~al.}{2022}]{2022A&A...663A.168Q}
{Quirola-V{\'a}squez} J.,  et~al., 2022, \mn@doi [\aap] {10.1051/0004-6361/202243047}, \href {https://ui.adsabs.harvard.edu/abs/2022A&A...663A.168Q} {663, A168}

\bibitem[\protect\citeauthoryear{{Quirola-V{\'a}squez} et~al.,}{{Quirola-V{\'a}squez} et~al.}{2023}]{Quirola2023}
{Quirola-V{\'a}squez} J.,  et~al., 2023, \mn@doi [\aap] {10.1051/0004-6361/202345912}, \href {https://ui.adsabs.harvard.edu/abs/2023A&A...675A..44Q} {675, A44}

\bibitem[\protect\citeauthoryear{{Richardson}, {Jenkins}, {Wright}  \& {Maddox}}{{Richardson} et~al.}{2014}]{2014AJ....147..118R}
{Richardson} D.,  {Jenkins} Robert~L. I.,  {Wright} J.,   {Maddox} L.,  2014, \mn@doi [\aj] {10.1088/0004-6256/147/5/118}, \href {https://ui.adsabs.harvard.edu/abs/2014AJ....147..118R} {147, 118}

\bibitem[\protect\citeauthoryear{{Rosswog}, {Ramirez-Ruiz}  \& {Hix}}{{Rosswog} et~al.}{2009}]{2009ApJ...695..404R}
{Rosswog} S.,  {Ramirez-Ruiz} E.,   {Hix} W.~R.,  2009, \mn@doi [\apj] {10.1088/0004-637X/695/1/404}, \href {https://ui.adsabs.harvard.edu/\#abs/2009ApJ...695..404R} {695, 404}

\bibitem[\protect\citeauthoryear{{Saade}, {Brightman}, {Stern}, {Malkan}  \& {Garc{\'\i}a}}{{Saade} et~al.}{2022}]{2022ApJ...936..162S}
{Saade} M.~L.,  {Brightman} M.,  {Stern} D.,  {Malkan} M.~A.,   {Garc{\'\i}a} J.~A.,  2022, \mn@doi [\apj] {10.3847/1538-4357/ac88cf}, \href {https://ui.adsabs.harvard.edu/abs/2022ApJ...936..162S} {936, 162}

\bibitem[\protect\citeauthoryear{{Saxton}, {Komossa}, {Auchettl}  \& {Jonker}}{{Saxton} et~al.}{2020}]{2020SSRv..216...85S}
{Saxton} R.,  {Komossa} S.,  {Auchettl} K.,   {Jonker} P.~G.,  2020, \mn@doi [\ssr] {10.1007/s11214-020-00708-4}, \href {https://ui.adsabs.harvard.edu/abs/2020SSRv..216...85S} {216, 85}

\bibitem[\protect\citeauthoryear{{Schawinski} et~al.,}{{Schawinski} et~al.}{2008}]{Schawinski2008}
{Schawinski} K.,  et~al., 2008, \mn@doi [Science] {10.1126/science.1160456}, \href {https://ui.adsabs.harvard.edu/abs/2008Sci...321..223S} {321, 223}

\bibitem[\protect\citeauthoryear{{Schulze} et~al.,}{{Schulze} et~al.}{2021}]{2021ApJS..255...29S}
{Schulze} S.,  et~al., 2021, \mn@doi [\apjs] {10.3847/1538-4365/abff5e}, \href {https://ui.adsabs.harvard.edu/abs/2021ApJS..255...29S} {255, 29}

\bibitem[\protect\citeauthoryear{{Science Software Branch at STScI}}{{Science Software Branch at STScI}}{2012}]{2012ascl.soft07011S}
{Science Software Branch at STScI} 2012, {PyRAF: Python alternative for IRAF} (\mn@eprint {ascl} {1207.011})

\bibitem[\protect\citeauthoryear{{Smith} et~al.,}{{Smith} et~al.}{2020}]{2020PASP..132h5002S}
{Smith} K.~W.,  et~al., 2020, \mn@doi [\pasp] {10.1088/1538-3873/ab936e}, \href {https://ui.adsabs.harvard.edu/abs/2020PASP..132h5002S} {132, 085002}

\bibitem[\protect\citeauthoryear{{Soderberg} et~al.,}{{Soderberg} et~al.}{2008}]{2008Natur.453..469S}
{Soderberg} A.~M.,  et~al., 2008, \mn@doi [\nat] {10.1038/nature06997}, \href {https://ui.adsabs.harvard.edu/abs/2008Natur.453..469S} {453, 469}

\bibitem[\protect\citeauthoryear{{Stern} et~al.,}{{Stern} et~al.}{2012}]{2012ApJ...753...30S}
{Stern} D.,  et~al., 2012, \mn@doi [\apj] {10.1088/0004-637X/753/1/30}, \href {https://ui.adsabs.harvard.edu/abs/2012ApJ...753...30S} {753, 30}

\bibitem[\protect\citeauthoryear{Stoppa, Cator  \& Nelemans}{Stoppa et~al.}{2023}]{2023arXiv230209308S}
Stoppa F.,  Cator E.,   Nelemans G.,  2023, \mn@doi [Monthly Notices of the Royal Astronomical Society] {10.1093/mnras/stad1938}, 524, 1061

\bibitem[\protect\citeauthoryear{{Sun}, {Zhang}  \& {Gao}}{{Sun} et~al.}{2017}]{2017ApJ...835....7S}
{Sun} H.,  {Zhang} B.,   {Gao} H.,  2017, \mn@doi [\apj] {10.3847/1538-4357/835/1/7}, \href {https://ui.adsabs.harvard.edu/abs/2017ApJ...835....7S} {835, 7}

\bibitem[\protect\citeauthoryear{{Taddia} et~al.,}{{Taddia} et~al.}{2019}]{2019A&A...621A..71T}
{Taddia} F.,  et~al., 2019, \mn@doi [\aap] {10.1051/0004-6361/201834429}, \href {https://ui.adsabs.harvard.edu/abs/2019A&A...621A..71T} {621, A71}

\bibitem[\protect\citeauthoryear{{Tody}}{{Tody}}{1986}]{1986SPIE..627..733T}
{Tody} D.,  1986, in {Crawford} D.~L.,  ed.,  Society of Photo-Optical Instrumentation Engineers (SPIE) Conference Series Vol. 627, Instrumentation in astronomy VI. p.~733, \mn@doi{10.1117/12.968154}

\bibitem[\protect\citeauthoryear{{Tonry} et~al.,}{{Tonry} et~al.}{2018}]{2018PASP..130f4505T}
{Tonry} J.~L.,  et~al., 2018, \mn@doi [\pasp] {10.1088/1538-3873/aabadf}, \href {https://ui.adsabs.harvard.edu/abs/2018PASP..130f4505T} {130, 064505}

\bibitem[\protect\citeauthoryear{{Uddin} et~al.,}{{Uddin} et~al.}{2020}]{2020ApJ...901..143U}
{Uddin} S.~A.,  et~al., 2020, \mn@doi [\apj] {10.3847/1538-4357/abafb7}, \href {https://ui.adsabs.harvard.edu/abs/2020ApJ...901..143U} {901, 143}

\bibitem[\protect\citeauthoryear{{Waxman} \& {Katz}}{{Waxman} \& {Katz}}{2017}]{2017hsn..book..967W}
{Waxman} E.,  {Katz} B.,  2017, {Shock Breakout Theory}.
p.~967, \mn@doi{10.1007/978-3-319-21846-5_33}

\bibitem[\protect\citeauthoryear{{Webb} et~al.,}{{Webb} et~al.}{2020}]{2020A&A...641A.136W}
{Webb} N.~A.,  et~al., 2020, \mn@doi [\aap] {10.1051/0004-6361/201937353}, \href {https://ui.adsabs.harvard.edu/abs/2020A&A...641A.136W} {641, A136}

\bibitem[\protect\citeauthoryear{{Wilms} et~al.,}{{Wilms} et~al.}{2020}]{2020ATel13416....1W}
{Wilms} J.,  et~al., 2020, The Astronomer's Telegram, \href {https://ui.adsabs.harvard.edu/abs/2020ATel13416....1W} {13416, 1}

\bibitem[\protect\citeauthoryear{{Xue} et~al.,}{{Xue} et~al.}{2019}]{Xue2019}
{Xue} Y.~Q.,  et~al., 2019, \mn@doi [\nat] {10.1038/s41586-019-1079-5}, \href {https://ui.adsabs.harvard.edu/abs/2019Natur.568..198X} {568, 198}

\bibitem[\protect\citeauthoryear{{Zhang}}{{Zhang}}{2013}]{2013ApJ...763L..22Z}
{Zhang} B.,  2013, \mn@doi [\apjl] {10.1088/2041-8205/763/1/L22}, \href {https://ui.adsabs.harvard.edu/abs/2013ApJ...763L..22Z} {763, L22}

\makeatother
\end{thebibliography}







\bsp	
\label{lastpage}
\end{document}